\documentclass[10pt,a4paper]{article}
\usepackage[utf8]{inputenc}
\usepackage{amsmath}
\usepackage{amsfonts}
\usepackage{amssymb}
\usepackage{hyperref}
\usepackage{verbatim}
\usepackage[overlay]{textpos}
\usepackage{MnSymbol,wasysym}
\usepackage{subfigure}
\usepackage{tikz}
\usepackage{graphics}
\usepackage{tkz-euclide}
\usepackage[toc,page]{appendix}
\usetikzlibrary{decorations.markings,arrows}
\usetikzlibrary{arrows,shapes,backgrounds}
\usetikzlibrary{decorations.pathreplacing,decorations.markings}
\usetikzlibrary{shapes.misc}
\usepackage{youngtab}
\usepackage{rotating}
\usepackage{graphicx}
\usepackage{slashed}
\usepackage{a4wide}
\usepackage{appendix}
\usepackage{eufrak}

\def\makeatletter{\catcode`\@=11}
\makeatletter
\def\mathbox#1{\hbox{$\m@th#1$}}%
\def\math@ccstyles#1#2#3#4#5#6#7{{\leavevmode
      \setbox0\mathbox{#6#7}%
      \setbox2\mathbox{#4#5}%
      \dimen@ #3%
      \baselineskip\z@\lineskiplimit#1\lineskip\z@
      \vbox{\ialign{##\crcr
             \hfil \kern #2\box2 \hfil\crcr
             \noalign{\kern\dimen@}%
             \hfil\box0\hfil\crcr}}}}
\def\mathaccstyles{\math@ccstyles\maxdimen}
\def\maththroughstyles{\math@ccstyles{-\maxdimen}}
\def\unity%
 {\maththroughstyles{.45\ht0}\z@\displaystyle {\mathchar"006C}\displaystyle 1}

\tikzset{
    arrowMe/.style={
        postaction=decorate,
        decoration={
            markings,
            mark=at position .5 with {\arrow[thick]{#1}}
        }
    }
}
\pgfarrowsdeclaredouble{<<s}{>>s}{stealth}{stealth}
\pgfarrowsdeclaretriple{<<<s}{>>>s}{stealth}{stealth}

\tikzset{cross/.style={cross out, draw=black, minimum size=2*(#1-\pgflinewidth), inner sep=0pt, outer sep=0pt},
cross/.default={4pt}}

\tikzset{
  double arrow/.style args={#1 colored by #2 and #3}{
    -stealth,line width=#1,#2, 
    postaction={draw,-stealth,#3,line width=(#1)/3,
                shorten <=(#1)/3,shorten >=2*(#1)/3}, 
  }
}

\begin{document}

\begin{titlepage}

\vskip 2cm

\begin{center}
{\Large \bfseries
Aspects of the moduli space of instantons on $\mathbb{C}P^2$ and its orbifolds
}

\vskip 1.2cm

Alessandro Pini\footnote{pinialessandro@uniovi.es} and
Diego Rodriguez-Gomez\footnote{d.rodriguez.gomez@uniovi.es}

\bigskip
\bigskip

\begin{tabular}{c}
Department of Physics, Universidad de Oviedo, \\
Avda.~Calvo Sotelo 18, 33007, Oviedo, Spain
\end{tabular}

\vskip 1.5cm

\textbf{Abstract}
\end{center}
We study the moduli space of self-dual instantons on $\mathbb{C}P^2$. These are described by an ADHM-like construction which allows to compute the Hilbert series of the moduli space. The latter has been found to be blind to certain compact directions. In this paper we probe these, finding them to correspond to a Grassmanian, upon considering appropriate ungaugings. Moreover, the ADHM-like construction can be embedded into a $3d$ gauge theory with a known gravity dual. Using this, we realize in $AdS_4/CFT_3$ (part of) the instanton moduli space providing at the same time further evidence supporting the $AdS_4/CFT_3$ duality. Moreover, upon orbifolding, we provide the ADHM-like construction of instantons on $\mathbb{C}P^2/\mathbb{Z}_n$ as well as compute its Hilbert series. As in the unorbifolded case, these turn out to coincide with those for instantons on $\mathbb{C}^2/\mathbb{Z}_n$.
\medskip
\noindent

\bigskip
\vfill
\end{titlepage}

\setcounter{tocdepth}{2}

\tableofcontents

\section{Introduction}

It is well-known that instantons are very important configurations in gauge theory. For example, the partition function of gauge theories contains contributions from saddle points of all instanton numbers. This can be made fully precise in the case of supersymmetric gauge theories with eight supercharges, when the supersymmetric partition function can be computed exactly thanks to localization (see \cite{Pestun:2007rz} for a seminal contribution). One can then explicitly see that, in addition to purely perturbative saddle points, the partition function localizes on instantonic configurations, whose contribution one has to sum. On general grounds, such contributions are the one-loop determinants around each instanton saddle point, which can be computed by the so-called Nekrasov instanton partition function. In turn, in the case of pure gauge theories, the latter coincides with the Hilbert series of the instanton moduli space (see \textit{e.g.} \cite{Keller:2011ek,Rodriguez-Gomez:2013dpa}). Therefore, the construction of instanton moduli spaces, as well as the computation of their associated Hilbert series, is of the greatest importance (of course, the reasons alluded before are just a very limited subset of those making of the instanton moduli space a very interesting object).

In the case of instantons on $\mathbb{C}^2$ --or its conformal compactification $S^4$-- the problem of constructing instantons of pure gauge theories\footnote{We will concentrate on instantons in pure gauge theories with 8 supercharges throughout all the paper.} with gauge group $A,\,B,\,C,\,D$ was solved long ago by the ADHM construction \cite{adhm}. Moreover, it turns out that the ADHM construction has a natural embedding into string theory as it arises as the Higgs branch of the Dp-Dp+4 brane system \cite{Witten:1994tz,Witten:1995gx,Douglas:1995bn,Douglas:1996sw}. In this paper we will be interested on the parallel story but for the case of $\mathbb{C}P^2$. As opposed to $S^4$, $\mathbb{C}P^2$ is a K\"ahler manifold. This naturally induces a preferred orientation which distinguishes self-dual (SD) from anti-self-dual (ASD) 2-forms. As a result, the construction of gauge connections with ASD and SD curvatures is intrinsically different. In this paper we will concentrate on SD connections on $\mathbb{C}P^2$. In the mathematical literature an ADHM-like construction for such gauge bundles has been developed long ago \cite{Buchdal1,Buchdal2,Buchdal3,King,Sanders}. Very recently, it has been shown that such construction can be embedded into a gauge field theory which, moreover, admits a string/M theory interpretation \cite{Mekareeya:2014kca}. Surprisingly, the gauge theories engineering the ADHM construction for instantons on $\mathbb{C}P^2$ are $3d$ gauge theories with $\mathcal{N}=2$ supersymmetry --that is, 4 supercharges--. Nevertheless, as shown in \cite{NY1} (see also \cite{NY2,NY3}, and \cite{Mekareeya:2014kca} for a discussion in the physics context), the Hilbert series and other properties do indeed satisfy properties compatible with the expected hyperK\"ahler condition of the moduli space.

In this paper we study several aspects of these moduli spaces for SD instantons on $\mathbb{C}P^2$. As we briefly review, these are relevant in the computation of the partition function of the twisted gauge theory on $\mathbb{C}P^2$. The corresponding Hilbert series were computed in \cite{Mekareeya:2014kca}, where it was shown that they coincide with the Hilbert series of a ``parent" instanton on $\mathbb{C}^2$. Nevertheless, being $\mathbb{C}P^2$ a topologically non-trivial space, it is natural to expect that our instantons are described by extra topological data. Indeed, the dimension of the moduli space seen by the Hilbert series is smaller than the dimension of the actual moduli space. In this paper we explore the ``extra directions", associated to these extra topological data. With hindsight, the Hilbert series misses these directions as they correspond to a non-compact geometry. Indeed, in the case of unitary instantons, the theory describing these directions is a $3d$ version of the theory in  \cite{Hanany:1997vm} whose moduli space is a (compact) Grassmanian manifold. Upon appropriately ungauging $U(1)$ groups we turn them into non-compact by considering the complex cone over the compact base. In this modified scenario the Hilbert series probes the extra directions finding agreement with the expectations. 

The gauge theory containing the ADHM construction of unitary instantons admits a large $N$ limit where it is dual to an $AdS_4$ geometry. It is then natural to study the instanton moduli space in the gravity dual. This provides an interesting cross-check of our results as well as more non-trivial evidence of the proposed $AdS_4/CFT_3$ dualities.

The ADHM construction for instantons on a given space can be used to find the corresponding construction on related spaces obtained by orbifold projections. In this manner, we find the ADHM construction, as well as the Hilbert series, for moduli spaces of instantons on $\mathbb{C}P^2/\mathbb{Z}_n$.

The structure of this paper is as follows: in section \ref{SUSYZ} we briefly review the relevance of SD instantons on $\mathbb{C}P^2$ in the computation of the partition function for the topologically twisted gauge theory. In particular, we show how SD instantons on $\mathbb{C}P^2$ arise as the minima of the localization action, as well as (very briefly) review some relevant aspects of the ADHM construction in the mathematical literature. In section \ref{U} we study unitary instantons on $\mathbb{C}P^2$, considering in particular the resolution of the extra directions upon ungauging $U(1)$'s as well as the $AdS/CFT$ description of (part of) the instanton moduli space. In section \ref{sec:orbifolding} we consider the construction of unitary instantons on the orbifold space. In section \ref{sec:sp} we turn to the symplectic case, finding the ADHM construction of their moduli space on $\mathbb{C}P^2/\mathbb{Z}_n$. In section \ref{sec:son} we turn to orthogonal instantons, analyzing, very much like in the unitary case, the compact extra directions associated to the non-trivial topology. Moreover, we provide the construction of orthogonal instantons on the orbifolded space. We provide a short summary of the highlights as well as some conclusions in section \ref{conclusions}. Finally, we describe some exotic cases as well as compile some figures (relegated to the appendix in order not to clutter the text) in the appendices.

\section{Self-dual instanton contributions to supersymmetric gauge theory on $\mathbb{C}P^2$} \label{SUSYZ}

We are interested on pure gauge theories on $\mathbb{C}P^2$. Hence our first task would be the construction of the supersymmetric lagrangian for the theory on the curved manifold. To that matter we follow the approach in \cite{Festuccia:2011ws}, which amounts to consider the combined system of supergravity plus the gauge theory of interest. Then, a rigid limit freezes the gravitational dynamics so that we are automatically left with the supersymmetric gauge theory on the curved space. Since we are interested on $\mathcal{N}=2$ gauge theories, we will use conformal supergravity as in \cite{Klare:2013dka}. 

Recently, the partition function of supersymmetric gauge theories on $\mathbb{C}P^2$ was considered in \cite{Rodriguez-Gomez:2014eza}. However, in this paper we will be interested on a different version of the gauge theory. Recall that, in order to find the supersymmetric theory, we need to solve the gravitino variation as well as the auxiliary condition in \cite{Klare:2013dka}. These provide both the background fields as well as the Killing spinors for the gauge theory on the curved space. A natural solution to these equations is the topological twist \cite{Karlhede:1988ax}. On general grounds, this amounts to redefining the Lorentz group --generically locally $SO(4)\sim SU(2)_{\rm left}\times SU(2)_{\rm right}$-- by twisting either $SU(2)_{{\rm left},\,{\rm right}}$ with $SU(2)_R$.  Nevertheless, as described in \textit{e.g.} \cite{Witten:1994ev}, since for K\"ahler manifolds the holonomy is really $SU(2)_{{\rm right}}\times U(1)_{{\rm left}}$, a second version exists whereby one twists the $U(1)_{{\rm left}}$ by the Cartan of the $SU(2)_R$ (note that in this case one chirality is privileged over the other by the orientation naturally induced by the K\"ahler form). While in \cite{Rodriguez-Gomez:2014eza} this later choice was considered, in this paper we will focus on the former version of the topological twist, which can be performed both for positive and negative  chiralities of the background Killing spinors. 

Setting, to begin with, all supergravity fields other than metric and $SU(2)_R$ gauge field to zero, the equations defining the supersymmetric backgrounds are defined by the conformal Killing spinor equation \cite{Klare:2013dka} (we refer to this reference for details)

\begin{equation}
\mathcal{D}_{\mu}\epsilon_{\pm}^i-\frac{1}{4}\,\gamma_{\mu}\,\slashed{D}\epsilon_{\pm}^i=0\, ,
\end{equation}
where the covariant derivative acting on the background Killing spinors is

\begin{equation}
\mathcal{D}_{\mu}\epsilon_{\pm}^i=\nabla_{\mu}\epsilon_{\pm}^i+(\mathcal{A}_{\mu})^i_j\,\epsilon_{\pm}^j\, ,
\end{equation}
while $\mathcal{A}_{\mu}$ is the $SU(2)_R$ gauge field and $\nabla_{\mu}$ is the covariant derivative acting on spinors including the spin connection. Moreover the metric of the $\mathbb{C}P^2$ is

\begin{equation}
\label{metric}
ds_{\mathbb{C}P^2}=d\rho^2+\frac{\sin^2\rho}{4}\,\Big[ d\theta^2+\sin^2\theta\,d\phi^2+\cos^2\rho\,(d\psi+\cos\theta\,d\phi)^2\Big]\, ,\,\, \rho\in[0,\,\frac{\pi}{2}]\, ,\, \psi\in[0,\,4\,\pi]\, ,\, \theta\in[0,\,\pi]\, ,\, \phi\in[0,\,2\,\pi]\, .
\end{equation}

With hindsight, in this paper we will be interested on keeping the positive chirality spinors. Choosing then

\begin{equation}
(\mathcal{A}_{\mu})^i_j=-\frac{i}{4}\,\eta_{Iab}\,\omega_{\mu ab}\,(\sigma^I)^i_j\, ,
\end{equation}
where $\eta_{I ab} $ is the 't Hooft symbol and the $\sigma^I$ are the Pauli matrices, we have that the spin connection part in the covariant derivative is cancelled, so that the Killing spinors are simply\footnote{We choose a chiral representation for the Dirac algebra, so that $\Gamma_5={\rm diag}(\unity,\,-\unity)$.}

\begin{equation}
\label{spinors}
\epsilon_+^1=\left(\begin{array}{c} {\rm i} \alpha \\ 0 \\ 0 \\ 0\end{array}\right)\, ,\quad \epsilon_+^2=\left(\begin{array}{c} 0 \\ {\rm i} \alpha \\ 0 \\ 0\end{array}\right)\, ,\qquad \alpha\,\in\,\mathbb{R}\, .
\end{equation}
Furthermore, one can check that the remaining supergravity equation is solved upon appropriately tuning the supergravity scalar \cite{Karlhede:1988ax}.
 
Following \cite{Klare:2013dka}, negative chirality spinors could be included choosing a Killing vector $v$ of $\mathbb{C}P^2$ as $\epsilon_-^i=i\,\slashed{v}\,\epsilon_+^i$ upon turning on $T^-=2\,dv|_-$. Let us stick however to the topological case. Then, since the theory is invariant under the supersymmetry generated by the above $\epsilon_+^i$, we could add to the action the $Q$-invariant term $-t\,\int \delta \mathcal{V}$, being $\delta\mathcal{V}=|\delta\Omega_+^i|^2+|\delta\Omega_-^i|^2$. The standard argument suggest then that the action is $t$-invariant. A straightforward calculation gives (we set $(\epsilon_+^i)^{\dagger}\epsilon_+^i=1$)

\begin{equation}
\label{eq_deltaV}
\delta\mathcal{V}=\frac{1}{64}\,(F^+)^2+|D\bar{\phi}|^2+\frac{1}{8}\,|Y^i\,_j|^2+|[\phi,\,\bar{\phi}]|^2\, ,
\end{equation}
where  we have imposed the reality condition $Y^i\,_j=(Y^j\,_i)^{\star}$ \cite{Rodriguez-Gomez:2014eza}. Since eq.(\ref{eq_deltaV}) is strictly positive, in the classical limit $t\rightarrow \infty$ the theory localizes on configurations such that the scalar in the vector multiplet is constant and lies along the Cartan of the gauge group while $F^+=0$. Note that, had we chosen to keep negative chirality spinors, we would have obtained $F^-=0$. Being more explicit, the condition $F^+=0$ is, in the conventions of \cite{Klare:2013dka}, equivalent to\footnote{Here $(\star F)_{ab}=\frac{1}{2}\,\epsilon_{abcd}F^{cd}$.}

\begin{equation}
F^+=\frac{1}{2}\,(F-\star F)=0\quad\leadsto\quad F=\star F\, ,
\end{equation}
 that is, $F$ must be self-dual (SD). Since, for the standard orientation of the $\mathbb{C}P^2$, the K\"ahler form is also self-dual, we have that the relevant gauge configurations in this case are instantons of the same duality type of the K\"ahler form. This is precisely the type of instantons described in \cite{Mekareeya:2014kca} using the King and Bryan-Sanders constructions in \cite{King,Sanders} elaborating on \cite{Buchdal1, Buchdal2, Buchdal3}.

\subsection{The construction of self-dual instantons on $\mathbb{C}P^2$}
\label{sec:review}

While we are interested on constructing self-dual instantons on $\mathbb{C}P^2$, it is however more convenient to regard them, upon orientation reversal of the base manifold, as anti-self dual (ASD) instantons on $\overline{\mathbb{C}P}^2$ (the opposite-oriented $\mathbb{C}P^2$). Then, we can directly borrow the construction of their moduli spaces from King \cite{King} and Bryan-Sanders \cite{Sanders}. Let us give a lightning overview of the relevant ingredients of the construction and defer to \cite{Buchdal1, Buchdal2, Buchdal3,King,Sanders} for the detailed account (see also \cite{Mekareeya:2014kca} for more references).

On very general grounds, there is a correspondence between the moduli space of instantons on projective algebraic surfaces and the moduli space of (stable) holomorphic bundles which goes under the name of Hitchin-Kobayashi correspondence. In this context, the ADHM construction can be regarded as a device to construct holomorphic bundles over the appropriate manifold.

An alternative version of the Hitchin-Kobayashi correspondence, more useful for our purposes, was proven by Donaldson by using the so-called Ward correspondence, which associates an anti-self-dual (ASD) connection --that is, a connection whose curvature is ASD-- on a (not complex) manifold $X$ to a holomorphic bundle on an related  manifold $X_{\rm holo}$. Roughly speaking, one regards $X$ as a conformal compactification of some underlying complex manifold $X_{\rm cplx}$. Since both the Yang-Mills equations and the self-duality constraints are conformally invariant, solutions with definite duality properties (say ASD) on $X_{\rm cplx}$ can be naturally extended into solutions on $X$. Note that, in doing this, the behaviour of the gauge field at the added point must be specified, that is, a framing must be chosen. In particular, we choose a trivial framing, where the gauge transformations become the identity at infinity. 

On the other hand, it is well-known that connections with an ASD curvature on a complex manifold $X_{\rm cplx}$ are in one-to-one correspondence with holomorphic bundles on $X_{\rm cplx}$.\footnote{Roughly speaking, this is due to the fact that the ASD condition on a connection $A$ is equivalent to the integrability condition $\bar{\partial}_A^2=0$ of $\bar{\partial}_A=\bar{\partial}+\bar{A}$, hence defining a holomorphic bundle on $X_{\rm cplx}$ through the Newlander-Nirenberg theorem. See \cite{Buchdal1, Buchdal2, Buchdal3,King,Sanders} and \cite{Mekareeya:2014kca} for more references.} Since the moduli space of the latter is a rather sick notion, being $X_{\rm cplx}$ a non-compact space, we can considering a holomorphic compactification of $X_{\rm cplx}$ into $X_{\rm holo}$ whereby we add the complex line at infinity $\ell_{\infty}$ and demand the holomorphic bundle to be trivial over there. Hence, all in all, the problem of constructing trivially framed ASD connections on $X$ is mapped to the construction of holomorphic bundles over $X_{\rm holo}$ trivial over $\ell_{\infty}$. The ADHM construction is precisely the device constructing such bundles. 

In the case at hand we consider $X_{\rm cplx}=\widehat{\mathbb{C}}^2$, the blow-up of $\mathbb{C}^2$ at a point defined as

 \begin{equation}
\widehat{\mathbb{C}}^2=\{(x_1,\,x_2)\times [z_1,\,z_2]\,\in\,\mathbb{C}^2\times \mathbb{C}P^1\,/\,x_1\,z_1=x_2\,z_2\}\, .
\end{equation}
Then, on one hand we can find a conformal compactification of $X_{\rm cplx}=\widehat{\mathbb{C}}^2$ into $X=\overline{\mathbb{C}P}^2$ --the opposite-oriented $\mathbb{C}P^2$-- as follows

\begin{equation}
\label{conformalcompactification}
\widehat{\mathbb{C}}^2\,\rightarrow\,\overline{\mathbb{C}P}^2\,:\,\Big((x_1,\,x_2)\times [z_1,\,z_2]\Big)\,\rightarrow\,\begin{cases} [|x|^2,\,x_1,\,x_2] \, , \\ [0,\,z_1,\,z_2]\, . \end{cases}
\end{equation}
Note that $\widehat{\mathbb{C}P}^2$ is not really a complex manifold, as the orientation does not follow from the K\"ahler form.

On the other hand, we can find a holomorphic compactification by adding $\ell_{\infty}$ which compactifies $\widehat{\mathbb{C}}^2$ into $X_{\rm holo}=\mathbb{C}P^2$ blown up at a point, that is, Hirzebruch's first surface $\mathbb{F}_1$. Hence we have that framed ASD connections over $\mathbb{C}P^2$ are in one-to-one correspondence with holomorphic bundles over $\mathbb{F}^1$ which are trivial over $\ell_{\infty}$. Since upon orientation reversal, ASD connections on $\overline{\mathbb{C}P}^2$ become SD connections on $\mathbb{C}P^2$, it follows that the desired moduli spaces are in correspondence with holomorphic bundles over $\mathbb{F}_1$. Then, the ADHM construction is precisely the device to construct such bundles. 

While here we will not dive into more details, an instrumental notion in arriving to the actual ADHM construction, from this point of view, is the associated twistor space, which takes into account the sphere bundle of compatible complex structures over $X_{\rm holo}$. Instead of delving into more intricacies, here we will describe the ADHM-like description of instantons for unitary, orthogonal and symplectic gauge groups embedded in a gauge theory as in  \cite{Mekareeya:2014kca}, and refere to \cite{Buchdal1, Buchdal2, Buchdal3,King,Sanders} for the details of their construction along the lines outlined here.

On word of caution is in order. Even though in the following we will loosely refere to instantons on $\mathbb{C}P^2$, the previous description of the precise construction should be borne in mind --that is, we are describing SD instantons on $\mathbb{C}P^2$ or equivalently ASD instantons on $\overline{\mathbb{C}P}^2$--.

\section{$U(N)$ instantons on $\mathbb{C}P^2$}\label{U}

\begin{figure}[h!]
\centering 
\includegraphics[scale=1]{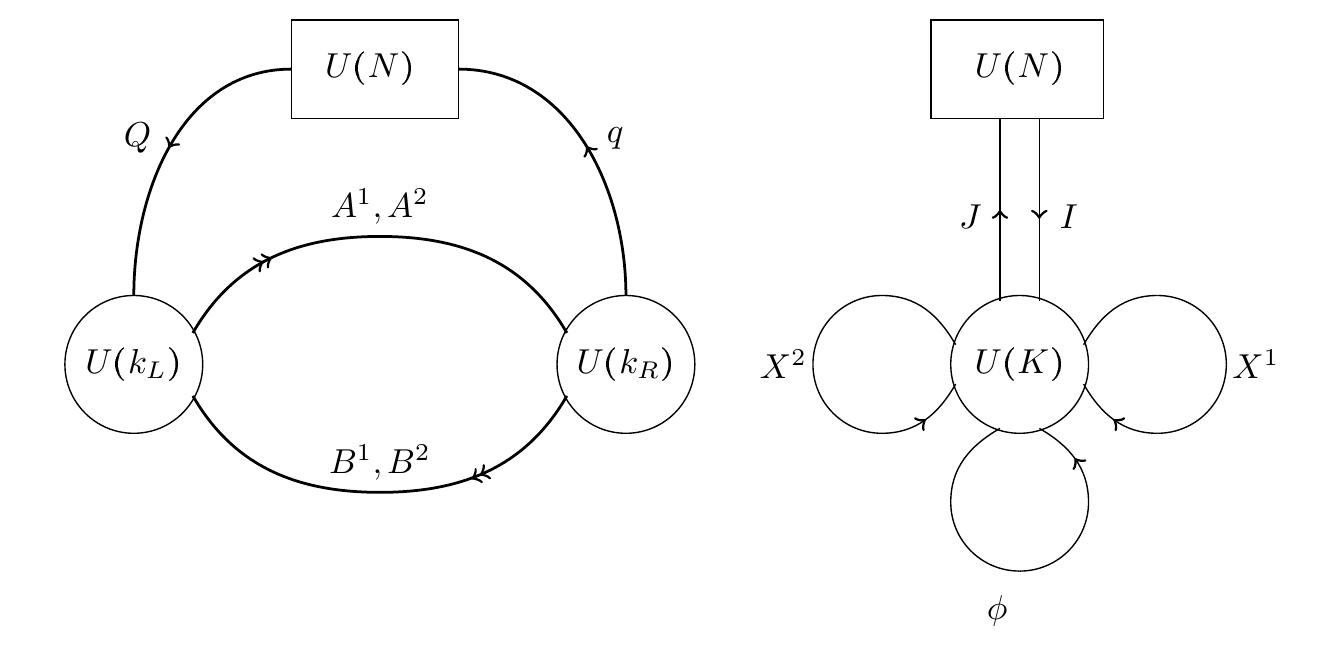} 
 
 \caption{Quiver diagram for $SU(N)$ instantons on 	$\mathbb{C}P^{2}$ (on the left) and for $SU(N)$ instantons on $\mathbb{C}^{2}$ (on the right). \label{fig:quiver1}}

\end{figure}

As described in \cite{Mekareeya:2014kca}, the King construction \cite{King} for unitary instantons on $\mathbb{C}P^2$ can be embedded into a $3d$ quiver gauge theory. The theory in question is a $3d$  $\mathcal{N}=2$ gauge theory whose quiver is in the left panel of fig.\ref{fig:quiver1}, supplemented with the superpotential 

\begin{equation}
\label{super}
W= \textrm{Tr}[A^{1}B^{1}A^{2}B^{2} -A^{1}B^{2}A^{2}B^{1} + qA^{1}Q]\ .
\end{equation}

Note that the chiral nature of the theory demands, because of the parity anomaly, the gauge nodes to have a non-vanishing Chern-Simons level $\frac{N}{2}+\mathbf{k}_L$ and $-\frac{N}{2}+\mathbf{k}_R$ respectively, where $\mathbf{k}_L,\,\mathbf{k}_R$ are integers including zero. In the following we will concentrate on the case $\mathbf{k}_L=\mathbf{k}_R=0$.

As a $3d$ gauge theory, it has been argued \cite{Davey:2009sr,Benini:2009qs} that the theory flows to an IR fixed point, where the charges of the fields are listed in table \ref{tab:1}.
\begin{table}
\centering
\renewcommand\arraystretch{1.3} 
\resizebox{10cm}{!}{
\begin{tabular}{c||cc||ccc||}
Fields  & $U(k_L)$  & $U(k_R)$ & $U(N)$ &  $SU(2)$    & $U(1)_R$    \\ 
 \hline 
 $A^{1}$ & $[1,0,...,0]_{+1}$ & $[0,...,0,1]_{-1}$ & $[\textbf{0}]$ & $[\textbf{0}]$  & 1/2 \\ 
\hline
 $A^{2}$ & $[1,0,...,0]_{+1}$ & $[0,...,0,1]_{-1}$ & $[\textbf{0}]$ & $[\textbf{0}]$  & 1/2 \\ 
\hline 
 $B^{1},B^{2}$ & $[0,...,0,1]_{-1}$ & $[1,0,...,0]_{+1}$ & $[\textbf{0}]$  & $[1]$ & 1/4 \\ 
\hline
 $q$ & $[\textbf{0}]$ & $[1,0,...,0]_{+1}$ & $[0,...,0,1]_{-1}$ & $[\textbf{0}]$  & 1-1/4$r$ \\ 
\hline
 $Q$ & $[0,...,0,1]_{-1}$ & $[\textbf{0}]$ & $[1,0,...,0]_{+1}$   & $[\textbf{0}]$&  1/4$r$ \\ 
\hline
 $F-$ term & $[0,...,0,1]_{-1}$ & $[1,0,...,0]_{+1}$ & $[\textbf{0}]$  &  $[\textbf{0}]$  & 1 \\

\end{tabular}  
}

\caption{Transformations of the fields for the $\mathbb{C}P^{2}$ quiver gauge theory. Here $r$ is an unknown real parameter whose value, nevertheless, does not affect subsequent results.\label{tab:1}}
\end{table}
For the paticular case $N=1$, as argued in \cite{Benini:2009qs}, the mesonic moduli space (excluding ``Higgs-like" directions where fundamental fields take a VEV) of the theory is the direct product of a conifold times the complex line. In general, as $N$ is increased, this geometric branch of the moduli space becomes an increasingly more involved toric manifold (see \cite{Benini:2009qs}). 

The instanton moduli space of interest is that of $G=U(N)$ instantons on $\mathbb{C}P^2$, denoted as $M^G_{\mathbb{C}P^2}$. It arises as a Higgs-like branch of the full moduli space of the gauge theory  dubbed as instanton branch where fundamental fields take a VEV. Note that the instanton gauge group appears as the flavor symmetry of the ADHM construction. Note as well that, in order to specify the instanton, in general a set of numbers $I$ including the instanton number is required. We will come back to this issue below.

More precisely, as described in \cite{Mekareeya:2014kca}, the instanton branch of the moduli space arises when we set $A^{1}$ (as well as all monopole operators, typically denoted by $T,\,\tilde{T}$) to zero. It is important to note that the truncation $A^1=T=\tilde{T}=0$ is consistent with the quantum constraint on the moduli space introduced in \cite{Benini:2009qs}. Then, the only relevant F-term arises from the superpotential and reads
\begin{equation}
\label{fcp2}
\partial_{A^{1}}W = B^{1}A^{2}B^{2} - B^{2}A^{2}B^{1} +qQ\, .
\end{equation}
Together with the field content and gauge groups of the $3d$ gauge theory, this constraint precisely realizes the King construction. Note that, even though the flavor symmetry is $U(N)$, the $U(1)$ part is really gauged. Hence we can think of our instantons as instantons of $SU(N)$ (even though, as we will review below, we should really think of $SU(N)/\mathbb{Z}_N$).

In the following we will be interested on the Hilbert series of the instanton moduli space. The ADHM construction just introduced (and the corresponding orthogonal and symplectic versions in addition to their orbifoldings to be described below) allows to compute it using by now standard methods as in \textit{e.g.} \cite{Mekareeya:2014kca,Benvenuti:2010pq,Hanany:2012dm,Dey:2013fea} (see also \cite{Nibbelink:2008qf} for the study of instantons on $\mathbb{C}^{2}/Z_{n}$). Let us pause to make a point on notation. Through all the paper we will denote the Hilbert series $H$ of the instantons moduli space as $H[I,G,M]$, being $I$ the integers characterizing instanton, which appears as the date of gauge group of the ADHM construction; $G$ those characterizing the instanton gauge group appearing as flavour group in the ADHM construction and $M$ the ambient manifold of the instanton. 

As anticipated, in order to specify a particular $G$ instanton on $\mathbb{C}P^2$ a set of quantum numbers $I$ is required. It is clear that one such integer is the instanton number. However, since $\mathbb{C}P^2$ is a topologically non-trivial manifold, it is natural to expect that instantons on $\mathbb{C}P^2$ might carry extra quantum numbers. Indeed, as reviewed in  \cite{Mekareeya:2014kca} following \cite{NY1}, we can characterize the instanton by its first Chern number $\hat{c}$ and its instanton number $\hat{k}$. Using the correspondence between ASD connections on $X$ and holomorphic bundles on $X_{holo}$, these can be written as

\begin{equation}
\langle c_1(E),\,[C]\rangle=-\hat{c}\, ,\qquad \langle c_2(E)-\frac{N-1}{2N}\,c_1(E)^2,\,[\mathbb{F}_1]\rangle=\hat{k}\, ,
\end{equation}
being $[C]$ the $\mathbb{C}P^1$ class inside $\mathbb{F}_1$ --recall that in this case $X=\overline{\mathbb{C}P^2}$ and $X_{holo}=\mathbb{F}_1$--. These, in turn, are related to the quiver data $k_L,\,k_R$ as follows

\begin{equation}
\hat{c}=k_R-k_L\, ,\qquad \hat{k}=\frac{1}{2}(k_L+k_R)-\frac{1}{2\,N}(k_L-k_R)^2\, .
\end{equation}

As an algebraic variety, $M^{SU(N)}_{\mathbb{C}P^2}$ can be mapped into the moduli space of a related instanton on $\mathbb{C}^2$ --described by the Higgs branch of the theory on the right panel of fig.\ref{fig:quiver1}--  in the following way
\begin{equation}
\label{mappi}
\pi :(A^{2},B^{1},B^{2},Q,q) \rightarrow (X^{1}=A^{2}B^{1}, X^{2}=A^{2}B^{2}, I=A^{2}q, J=Q)\, ,
\end{equation}
being $X^{1},X^{2},I,J$ the fields of the quiver diagram for $\mathbb{C}^{2}$ theory. Indeed, if we multiply the F-term relation (\ref{fcp2}) by $A^{2}$ and we apply the map (\ref{mappi}) we recover the F-term for $SU(N)$ instantons on $\mathbb{C}^{2}$
\begin{equation}
[X^{1},X^{2}] + I \cdot J = 0\ .
\end{equation}
In turn, the inverse map $\sigma$ can also be defined as

\begin{equation}
\sigma : (X^1,X^2,I,J) \rightarrow (A^2=\mathbf{1}_{K\times K}, B^1=X^1, B^2=X^2, q=I, Q=J)\, .
\end{equation}

Let us momentarily consider the case where $k_L=k_R$, which corresponds to $\hat{c}=0$ and $\hat{k}=k_L$. From the construction in eq.(\ref{mappi}), it is clear that the integer $K$ in the quiver on the right panel of fig.\ref{fig:quiver1} is identified with $k_L$. Thus, we have that as an algebraic variety, the moduli space of $k_L$ $SU(N)$ instantons on $\mathbb{C}P^2$ is identified with the moduli space of $k_L$ $SU(N)$ instantons on $\mathbb{C}^2$. Consistently, the Hilbert series of these instantons coincide, from which it follows that ${\rm dim}_{\mathbb{C}}M^{SU(N)}_{\mathbb{C}P^2}=2\,N\,k_L$.

In the general case $k_L\ne k_R$, one finds that the above construction still holds upon setting $K={\rm min}(k_L,k_R)$. Consistently, as described in \cite{Mekareeya:2014kca}, the Hilbert series corresponding to the instanton branch of the quiver on the left panel in fig.\ref{fig:quiver1} coincides with the Hilbert series of the Higgs branch of the quiver on the right panel of fig.\ref{fig:quiver1}, that is

\begin{equation}
\label{HSid}
H[(k_L,k_R),SU(N),\mathbb{C}P^{2}](t,x,\textbf{y}) = H[\textrm{min}(k_L,k_R),SU(N),\mathbb{C}^{2}](t^{3},x,\textbf{y})\, ,
\end{equation} 
where $t$ is the fugacity of the R-charge, $x$ the fugacity associated with the $SU(2)$ global symmetry and $\textbf{y}$ are the fugacities associated with the $U(N)$ global symmetry. Note that the fugacity associated to $R$-charge is re-scaled from $t$ in the $\mathbb{C}P^2$ case into $t^3$ in the $\mathbb{C}^2$ case.

Naively, eq.(\ref{HSid}) suggests that the dimension of the moduli space of unitary instantons on $\mathbb{C}P^2$ is

\begin{equation}
\label{eq:dim}
{\rm dim}_{\mathbb{C}}M^{SU(N)}_{\mathbb{C}P^2}=2\,N\,{\rm min}(k_L,\,k_R)\, \,.
\end{equation}
Note that, even though the quiver is specified by three integers $N,\,k_L,\,k_R$, eq.(\ref{eq:dim}) is only sensitive to two of them. However, it is possible to consider an extended notion of the moduli space where the extra directions associated to all the three quantum numbers specifying the instanton are taken into account. This is the so-called resolved (as the extra directions are discerned) moduli space, denoted as $\widehat{M}_{\mathbb{C}P^2}^{SU(N)}$, whose dimension is  \cite{NY1,NY2,NY3} 

\begin{equation}
\label{eq:dim_resolved}
{\rm dim}_{\mathbb{C}}\widehat{M}_{\mathbb{C}P^2}^{SU(N)}=2\,\hat{k}\,N={\rm dim}_{\mathbb{C}}M_{\mathbb{C}P^2}^{SU(N)}+\hat{c}\,(N-\hat{c})\, .
\end{equation}
Note that for $\hat{c}=0,\,N$ the dimension of $\widehat{M}_{\mathbb{C}P^2}^{SU(N)}$ is equal to the dimension of $M^{SU(N)}_{\mathbb{C}P^2}$. This suggests that $\hat{c}$ is really a modulo $N$ quantity corresponding to an instanton gauge group which is really $SU(N)/\mathbb{Z}_N$. We warn the reader that, while in the following we will not clutter notation by supresing the $\mathbb{Z}_N$, the global properties of the gauge group must be kept on mind.

\subsection{The resolved moduli space and the Grassmanian}\label{Grassmanian}

\begin{figure}[h!]
\centering
\includegraphics[scale=1]{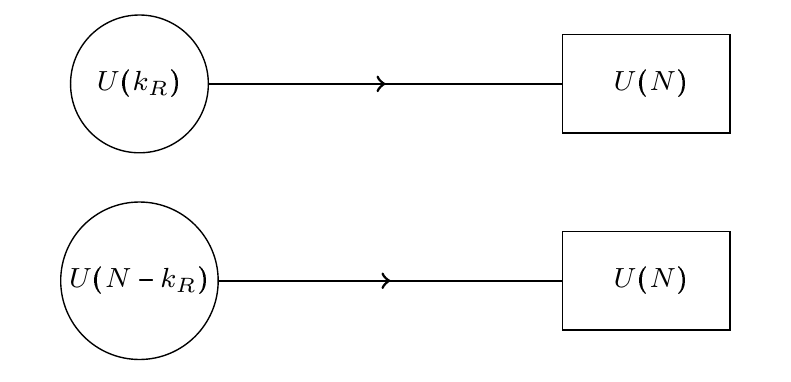}
 \caption{Quiver diagrams for Grassmanian (we show the dual pair --see text--). \label{fig:gras}}
\end{figure}

In order to explore the resolved moduli space it is instructive to first consider the simplest case where $k_L=0$. The theory simplifies into a one-noded quiver flavored only with fundamental fields (and not antifundamentals) shown in fig.\ref{fig:gras}. Recall that the CS level is adjusted so as to cancel the parity anomaly, and, furthermore, there is no superpotential.

The leftover theory in this particular case corresponds to a $3d$ version of the theory considered in \cite{Hanany:1997vm}. Then, as argued in that reference, the moduli space is a complex Grassmanian (compact) manifold, consistently with the expectations in \cite{NY1,NY2,NY3}.  

We can now understand why $M_{\mathbb{C}P^2}^{SU(N)}$ is insensitive to these extra directions, as, forming a compact Grassmanian manifold, the Hilbert series is blind to them. Indeed, since in the theory in fig.\ref{fig:gras} the gauge group is $U(k_R)$, the Higgs-like moduli space is empty, as no gauge-invariant can be constructed out of fundamental fields. Consistently, formula (\ref{eq:dim}) gives a zero-dimensional moduli space. However, as in \cite{Hanany:2014hia}, we can consider a version of the theory where only the non-abelian $SU(k_R)$ part of $U(k_R)$ is gauged, while the $U(1)$ is kept as a global baryonic symmetry (alternatively, we could think of this as the master space \cite{Forcella:2008bb} of the $U(k_R)$ theory). In this case we can form baryon-like gauge-invariant operators, thus finding a non-empty moduli space which in fact is a complex cone over the Grassmanian. It is straightforward to compute the Hilbert series and read-off the dimension of the moduli space from the pole at $t=1$, finding

\begin{equation}
\label{eq:dim_grassmanian}
{\rm dim}_{\mathbb{C}}M^{SU(N)}_{\mathbb{C}P^2}|_{\rm Grassmanian}=k_R\,(N-k_R)+1\ .
\end{equation}
Recalling that the $+1$ is due to the $U(1)$ over which we are not integrating over --resulting in moduli space which is a complex cone over the Grassmanian--, we find a result in accordance with eq.(\ref{eq:dim_resolved}).

Eq.(\ref{eq:dim_grassmanian}) is invariant under the exchange $k_R\leftrightarrow N-k_R$. Indeed, one can explicitly check that the Hilbert series of the theories with $SU(k_R)$ gauge group and $SU(N-k_R)$ are identical, thus suggesting a duality among these theories. Such duality is also suggested by the brane construction in \cite{Hanany:1997vm}. In that reference, in a IIA system consisting on an $NS$ brane and a $NS'$-$N$ D4 branes intersection, $k_R$ D2 branes are stretched along $x^6$ direction between the $NS$ and the $NS'$-D4 intersection. Then, the $N$ D4's can be broken on the $NS'$ and, say, the lower part of them can be sent to infinity. As argued in \cite{Hanany:1997vm}, the gauge theory on the D2's is precisely the $2d$ version of the gauge theory in the first panel in fig.\ref{fig:gras}. Upon T-duality along $x^2$, this system engineers the actual $3d$ gauge theory of interest, namely that in the first panel of fig.\ref{fig:gras}. Explicitly, the system contains

\begin{itemize}
\item An $NS$ brane along $012345$.
\item A braneweb with an $NS'$ brane along $012389$ meeting $N$ D5 branes along $012378$ and emanating a $(1,\,N)$ fivebrane.
\item $k_R$ D3 branes along $0126$, starting at the braneweb junction and ending on the $NS$.
\end{itemize}
Note that the $NS'$-D4 intersection in the IIA system becomes a braneweb in the IIB system, as D5-branes meeting an $NS'$ give rise to a $(1,\,N)$ fivebrane. In fact, it is precisely this bending what gives the expected CS level in the $3d$ gauge theory \cite{Kitao:1998mf,Bergman:1999na}. In this it is important to recall that the D3's meet the fivebranes right at the junction, as this is what makes the $3d$ theory to contain only fundamental (and not antifundamental) matter \cite{Hanany:1997vm} which in turn generates the $\frac{N}{2}$ CS level.

We can now imagine crossing the $NS$ to the other side. Then, due to the Hanany-Witten effect, the final configuration contains $N-k_R$ D3 branes but is otherwise identical, consistently with our finding that the two theories in fig.\ref{fig:gras} yield to the same Hilbert series (for a more detailed account of the duality in the $2d$ case we refere to \cite{Hanany:1997vm}).

Coming back to the general discussion, in view of the $k_L=0$ case it is natural to guess that ungauging the abelian part of the largest gauge symmetry will allow us to resolve the extra directions in $\widehat{M}$. To that matter, let us now consider the case $k_L=1$. Writing the remaining $U(k_R)$ gauge group as $U(1)\times SU(k_R)$, we can compute the Hilbert series upon integration only over the non-abelian $SU(k_R)$ part. Reading the dimension of the moduli space from the order of the pole at $t=1$, from explicit computations for $k_L=1$ and $k_R=2,\,3$ and $N=1,\,2,\,3$, we find

\begin{equation}
\label{dim}
\textrm{dim}_{\mathbb {C}}\hat{M}^{SU(N)}_{\mathbb{C}P^{2}} = 2k_LN + \hat{c}(N-\hat{c}) + 1\,, 
\end{equation}
which is precisely the expected result (\ref{eq:dim_resolved}). Unfortunately, explicitly checking higher rank cases is technically challenging. Nevertheless, it would be very interesting to perform further checks for higher ranks.

\subsection{Rank one and AdS/CFT}

In the particular case of $k_L=k_R$, upon setting $N=1$ and for $\mathbf{k}_L=\mathbf{k}_R=0$, the  theory engineering the moduli space of unitary instantons on $\mathbb{C}P^2$ becomes exactly that found in \cite{Benini:2009qs} to describe M2 branes probing $\mathcal{C}\times\mathbb{C}$, the direct product of a conifold times the complex line. The metric of the $CY_4$ cone can be written as

\begin{eqnarray}
ds_{\rm cone}^2&=&d\rho^2+\rho^2\,ds_{\mathcal{B}}^2\, , \\  ds_{\mathcal{B}}^2&=&d \alpha^{2} +\sin^{2}\alpha d\gamma^{2} +\frac{\cos^{2}\alpha}{9}\left(d\psi + \sum_{i=1}^2\cos\theta_{i}d\phi_{i}\right)^{2} + \sum_{i=1}^{2}\frac{\cos^{2}\alpha}{6}\left(d\theta_{i}^{2} +\sin^{2}\theta_{i}d\phi_{i}^{2}\right)\, . 
\end{eqnarray}
Then, on general grounds, the near-brane geometry for a stack of $k_L$ M2 branes probing this cone is $AdS_4\times \mathcal{B}$, which, in global coordinates, can be written as

\begin{equation}
ds^{2} = -\left(1+\frac{r^{2}}{L^{2}}\right)dt^{2}	+ \frac{dr^{2}}{(1+\frac{r^{2}}{L^{2}})} + r^{2}(\sin^{2}\theta d\theta^{2} +d\phi^{2}) + 4L^{2}ds^{2}_{\mathcal{B}}\,,
\end{equation}
being $L$  the radius of the $AdS_{4}$ space. Besides, there is a 6-form flux whose field strength integrates to $k_L$ on $\mathcal{B}$. Hence, in the large $k_L\,(=k_R)$ limit, the gauge theory is holographically dual to $AdS_4\times \mathcal{B}$ with $k_L$ units of flux through $\mathcal{B}$. It is thus natural to wonder whether, at least partially, the moduli space of unitary instantons on $\mathbb{C}P^2$ can be geometrically realized in this context.

As discussed in \cite{Benini:2009qs}, the gauge theory contains a mesonic branch of the moduli space which realizes the dual geometry. In general, it is natural to expect that the holographic dual captures gauge theory operators made out of bifundamental fields, while those corresponding to fundamental matter would require extra multiplets on top of the $AdS_4\times \mathcal{B}$ to account for the ``flavor brane open string" degrees of freedom. Hence, it is natural to expect that the subbranch of the instanton branch involving just $\{A^2,\,B^i\}$ fields is visible in the geometry. This is indeed analogous to the cases discussed in \cite{Dey:2013fea,Bergman:2012qh}, where only the ``closed string fields" in the quiver are captured by the gravity dual. 

More explicitly, following \cite{Dey:2013fea,Bergman:2012qh}, it is natural to expect that this subbranch of the instanton branch is captured by dual giant graviton branes moving in the appropriate subspace corresponding to the instanton branch. To that matter, we consider a probe M2 brane wrapping $(t,\Omega_{2})$, where $\Omega_2$ is the sphere inside  the $AdS_{4}$. Moreover we assume that $\psi = \psi(t)$ and $\phi_{2} = \phi_{2}(t)$, while 
\begin{equation}
\label{probe_brane}
\gamma, \alpha, \theta_{1}, \phi_{1}, \theta_{2}  = \textrm{constant}\, .
\end{equation}

The action for such probe brane is

\begin{equation}
S = -T_{2} \int \sqrt{-g} + T_{2} \int P[A^{(3)}]\, ,
\end{equation}
which becomes

\begin{equation*}
S = -T_{2}V_{2} \int dt \, r^{2}\left(\sqrt{\left(1 + \frac{r^{2}}{L^{2}} \right) - \frac{4L^2\cos^{2}\alpha}{9}\left(\dot{\psi}(t)  +\cos\theta_{2}\dot{\phi_{2}}(t)\right)^{2} - \frac{4L^2\cos^{2}\alpha \sin^{2}\theta_{2}}{6}\dot{\phi}_{2}(t)^{2}} - \frac{r^{3}}{L}\right)\, . 
\end{equation*}

It is easy to convince oneself that the equations of motion fix $\alpha = 0$ (for simplicity, from now on we set $\alpha=0$).  Then, Legendre transforming to the Hamiltonian $H = H(\theta_{2},r, P_{\psi},P_{\phi_{2}})$ we obtain 
\begin{equation*}
\label{h1}
H= \frac{1}{2L}\sqrt{\frac{r^2+L^2}{L^2}}\sqrt{\frac{3(5-\cos2\theta_{2})P_{\psi}^{2} -24\cos\theta_{2}P_{\psi}P_{\phi_{2}}+2(6P_{\phi_{2}}^{2}+4L^{2}r^{4}\sin^{2}\theta_{2} T_{2}^{2}V_{2}^{2})}{2\sin^{2}\theta_{2}}}- \frac{V_{2}T_{2}r^{3}}{L}\, .
\end{equation*}

The minimum energy configurations are

\begin{equation}
\label{sol}
\cos\theta_2=\frac{P_{\phi_2}}{P_{\psi}}\, ,
\end{equation}
for which 

\begin{equation}
r=0\qquad {\rm or}\qquad r=\frac{3\,P_{\psi}}{2\,L^2\,T_2\,V_2}\, .
\end{equation}
Both configurations are degenerated in energy, one corresponding to pointlike gravitons and the other to true dual giant gravitons. The energy is

\begin{equation}
\label{energy}
H=\frac{3\,P_{\psi}}{2\,L}\,.
\end{equation}

%

Coming back to the solution in eq.(\ref{sol}), we can parametrize the phase space of the spinning M2 as a dynamical system by the coordinates $Q^{A}= \{r,\alpha,\psi,\theta_{2},\phi_{2} \}$  and the conjugated momenta  $P_{A}= \{ P_{r},P_{\alpha},P_{\psi},P_{\theta_{2}},P_{\phi_{2}} \}$. Moreover the conjugated momenta $P_{A}$ must obey the following constraints
\begin{equation*}
f_{r} = P_{r}, \ \ f_{\alpha} = P_{\alpha}, \ \ f_{\theta_{2}}= P_{\theta_{2}}, \ \ f_{\psi} = P_{\psi} -\frac{2L^{2}T_{2}V_{2}r}{3}, \ \ f_{\phi_{2}} = P_{\phi_{2}} -\frac{2L^{2}T_{2}V_{2}r\cos\theta_{2}}{3}\ .
\end{equation*}
As usual, the matrix $M_{AB}= \{ f_{A},f_{B}\}_{PB}$ encodes the symplectic form associated to the phase space of our dynamical system as $\{Q^A,\,Q^B\}_{DB}=(M_{AB})^{-1}$ ($DB$ stands for Dirac brackets). Deleting the row and column corresponding to the trivial $\alpha$ coordinate, we find

\begin{equation*}
M^{AB} = \left( \begin{array}{cccc}
0 & \frac{2L^{2}T_{2}V_{2}}{3} & 0 & \frac{2L^{2}T_{2}V_{2}\cos\theta_{2}}{3} \\ 
\frac{-2LT_{2}V_{2}}{3} & 0 & 0 & 0 \\ 
0 & 0 & 0 & \frac{-2L^{2}r\sin\theta_{2}T_{2}V_{2}}{3} \\ 
\frac{-2L^{2}\cos\theta_{2}T_{2}V_{2}}{3} & 0 & \frac{2L^{2}r\sin\theta_{2}T_{2}V_{2}}{3} & 0
\end{array}  \right)\, .
\end{equation*}
Therefore the symplectic structure reads
\begin{equation*}
\omega = \frac{2L^{2}T_{2}V_{2}}{3} dr \wedge d\psi + \frac{2L^{2}T_{2}V_{2}\cos\theta_{2}}{3} dr \wedge d\phi_{2} -\frac{2L^{2}T_{2}V_{2}r\sin\theta_{2}}{3} d\theta_{2} \wedge d\phi_{2}\, . 
\end{equation*}
Integrating we obtain
\begin{equation}
\nu = \frac{2L^{2}T_{2}V_{2}r}{3} \left(d\psi + \cos\theta_{2}d\phi_{2} \right) \ \ \Rightarrow \ \ \omega=d\nu\, .
\end{equation}
Hence, upon introducing $\rho^{2}=4L^{2}T_{2}V_{2}r/3$, we just recover the data of $\mathbb{C}^{2}$. Following \cite{Dey:2013fea,Bergman:2012qh}, we can do symplectic quantization of this dynamical system. On general grounds, that amounts to identify the holomorphic functions on the phase space --in this case $\mathbb{C}^2$-- with the allowed wavefunctions. These can easily be counted, simply obtaining the Hilbert series for $\mathbb{C}^2$.

Let us now turn to the gauge theory. As discussed, we expect our probe branes to be dual to operators on the instanton branch not containing fundamental fields. These are of the schematic form 

\begin{equation}
\label{ops}
\mathcal{O}_{n,\,m}=(A^2\,B^1)^n\,(A^2\,B^2)^m\, .
\end{equation}
Note that the F-terms imply that the $B^i$ indices are completely symmetrized; that is, the operators $\mathcal{O}_{n,\,m}$ are in a spin $\frac{(n+m)}{2}$ representation of the $SU(2)$ global symmetry rotating the $B^i$'s. Hence, for a fixed $R$ charge $R[\mathcal{O}_{n,\,m}]=\frac{3}{4}\,(n+m)$, the number of operators is $(n+m)+1$, and the corresponding generating function is just $\sum_{j=0}^{\infty}(j+1)\,x^j=(1-x)^{-2}$, which is precisely the $\mathbb{C}^2$ Hilbert series --here $x$ is a generic fugacity--.

We can explicitly compare the gauge theory operators with our probe brane configurations in the gravity side. To that matter, let us first note that exactly the same configuration in the gravity side would have been obtained fixing $\theta_2=0,\,\pi$ and having our brane orbiting $\psi\pm\phi_1$ respectively. Hence, in all our formulas we can trade $\psi$ for $\tilde{\psi}=\psi\pm \phi_1$. In particular, eq.(\ref{energy}) becomes $H\,L=\frac{3}{2}\,P_{\tilde{\psi}}$. 

In order to compare our probe branes with the gauge theory operators we need to identify charges. It is reasonable to guess that the momentum along $\psi$ is proportional to the R-symmetry. Hence let us identify $P_{\psi}=r$, being $r$ (not to be confused with the arbitrary integer in table \ref{tab:1}) proportional to the charge $R$ under the $U(1)_R$ in way to which we will shortly come back. Moreover, in order to understand the $P_{\phi_{1,\,2}}$ momenta, it is instructive to consider momentarily removing the quarks from the gauge theory. It then exhibits an $SU(2)_A\times SU(2)_B$ global symmetry, rotating respectively the $A^i$ and $B^i$ fields. Then, the quark multiplets break the $SU(2)_A$ down to a $U(1)_A$, while the $SU(2)$ rotating the $B$'s remains as a global symmetry. We identify the $U(1)_A$ charge, denoted as $Q_A$, with $P_{\phi_1}$ as  $Q_A=P_{\phi_1}$. With no loss of generality, let's assume $Q_A[A^2]=\frac{1}{2}$, which corresponds to the choice $\theta_1=\pi$. Then $P_{\tilde{\psi}}=P_{\psi}-P_{\phi_1}$ translates into $P_{\tilde{\psi}}=r-Q_A$. Analogously, we identify $P_{\phi_2}$ with the Cartan of the $SU(2)_B$ denoted $Q_B$.

Note that eq.(\ref{sol}) translates into $Q_B=(r-Q_A)\,\cos\theta_2$, and therefore $Q_B\,\in\,[-(r-Q_A),\,(r-Q_A)]$. Let us compare this with the gauge theory operators (\ref{ops}). Using table \ref{tab:1}, the charges of the operators in the expression (\ref{ops}) are $R[\mathcal{O}_{n,\,m}]=\frac{3\,(n+m)}{4}$ and $Q_A[\mathcal{O}_{n,\,m}]=\frac{n+m}{2}$. As expected, being chiral operators, they satisfy the usual relation $\Delta=R$. Moreover, it is clear that $Q_B=\frac{n-m}{2}$, so that $Q_B\,\in\,[-\frac{2\,R}{3},\,\frac{2\,R}{3}]$. Comparing the ranges for $Q_B$ in gravity and field theory we find the identification 

\begin{equation}
\label{R}
R=\frac{3}{2}\,(r-Q_A)\, .
\end{equation}
Turning now to the energy for our branes, we find $H\,L=\frac{3}{2}\,(r-Q_A)$, which, upon using eq.(\ref{R}), becomes $\Delta=R$, precisely as expected for chiral operators.

Moreover, we can explicitly fix the value of $r$. To that matter, let us turn to the field theory operators and consider the highest $Q_B$ weight state, which corresponds to $m=0$. For this one $Q_A=Q_B=\frac{n}{2}$, while $R=\frac{3\,Q_A}{2}$. In turn, from the gravity side, the brane with highest $Q_B$ is $Q_B=r-Q_A$. Since this must correspond to $Q_B=Q_A$, we find $Q_A=2\,r$. Hence, this implies $r=\frac{4\,R}{3}$. 

We can offer an alternative test of our identifications. To that matter, let us consider metric fluctuations polarized along the internal manifold. On general grounds, these fluctuations correspond to operators of the schematic form $\mathcal{T}\,\mathcal{O}$, being $\mathcal{T}$ the stress-energy tensor of the theory. Note that, for the particular case when the inserted operator $\mathcal{O}$ is one of those in eq.(\ref{ops}), we expect that the dimension is $3+\Delta$. In turn, these fluctuations satisfy the Klein-Gordon equation in $AdS_4\times \mathcal{B}$. For a $CY_4$ of the form $\mathbb{C}\times \mathcal{C}$ this problem was considered in \cite{RodriguezGomez:2009ae}, where it was shown that the dimension of the dual operators can be written in terms of the eigenvalues of the scalar laplacian on $\mathcal{C}$. In turn, borrowing the results from \cite{Klebanov:2007us}, the eigenvalues of the scalar laplacian on the conifold are

\begin{equation}
E_\mathcal{C}=6\,\Big(\ell_1\,(\ell_1+1)+\ell_2\,(\ell_2+1)-\frac{r^2}{8}\Big)\, ,
\end{equation}
where $\ell_{1,\,2}$ are respectively, the $SU(2)_A\times SU(2)_B$ total spin and $r$ the charge along the $\psi$ direction. For the operators in eq.(\ref{ops}) we have that $\ell_1=\ell_2=\ell$. In turn, the charge $r$ must satisfy $\frac{r}{2}\,\in\,(-\ell,\,\ell)$. Focusing on the highest weight state, we would require $r=2\,\ell$, which is nothing but $r=2\,Q_A$ as seen before. Then, using \cite{RodriguezGomez:2009ae}, 

\begin{equation}
\Delta=3+\frac{3}{2}\,\ell\, .
\end{equation}
This precisely coincides with our expectations upon identifying $\Delta=\frac{3}{2}\,\ell$. This can be written as $\Delta=\frac{3\,r}{4}$, which becomes $\Delta=R$ upon using the identification $r=\frac{4\,R}{3}$ advocated above.

So far we have considered the case $k_L=k_R$. It is natural to expect that $k_L\ne k_R$ can be accommodated into the gravity dual by adding non-vanishing flat $B_2$ over a 2-cycle in the internal manifold \cite{Aharony:2008gk}. Nevertheless, such modification of the background would not change our computation. Hence we would find the same result even for the case $k_L\ne k_R$, in agreement with the field theory result where the Hilbert series only depends on ${\rm min}(k_L,\,k_R)$.

\section{ $U(N)$  instantons on $\mathbb{C}P^2/\mathbb{Z}_n$}
\label{sec:orbifolding}

A natural generalization of the ADHM construction of instantons on $\mathbb{C}P^2$ is to consider orbifolding the ambient manifold upon quotienting by a subgroup of its symmetries. In particular, since $\mathbb{C}P^2$ is invariant under a $U(1)\times U(1)$ action corresponding to the $\phi,\,\psi$ coordinates in eq.(\ref{metric}), it is natural to consider quotienting such symmetry by some discrete subgroup of it. Note that the spinors in eq.(\ref{spinors}) are constant and morever annihilated by $e^{i\,\frac{2\,\pi}{k}\,(J_{12}-J_{34})}$ ($J_{ij}$ are the Lorentz generators in tangent space indices $J_{ij}=\frac{{\rm i}}{2}\,[\Gamma_i,\,\Gamma_j]$). Therefore we can consider a $\mathbb{Z}_n$ orbifold of the $\phi$ direction whereby we restrict $\phi\sim \phi+\frac{2\pi}{n}$. In the rest of the paper we will be interested on the ADHM construction of instantons on these orbifolded spaces. To that matter, we will take as starting point the ADHM construction in the unorbifolded case, on which we will implement the orbifold by standard methods  \cite{Douglas:1996sw}. 

Let us consider the case of unitary instantons presented above. In order to find the orbifolded theory we first need to identify the transformation properties of the fields. These read:

\begin{itemize}
\item The fields $A^{j}$ (with $j=1,2$) in the bifundamental representation.\\
\begin{equation}
\label{ta}
A^{j} \mapsto \gamma_{1}A^{j}\gamma_{2}^{-1}\, , 
\end{equation}
\item The fields $B_{1}$ and $B_{2}$  in the bifundamental representation.
\begin{equation}
\label{b1}
B^{1} \mapsto \omega_{n}^{-1}\gamma_{2}B^{1}\gamma_{1}^{-1}, \ \
\ \  B^{2} \mapsto \omega_{n}\gamma_{2}B^{2}\gamma_{1}^{-1}, \ \ \textrm{with} \ \ \omega_{n} = e^{2\pi i/n}\, ,
\end{equation}
\item The fields $Q$ and $q$.
\begin{equation}
\label{tq}
q \mapsto \gamma_{2}q\gamma_{3}^{-1}, \ \ Q \mapsto \gamma_{3}Q\gamma_{1}^{-1}\, ,
\end{equation}
\end{itemize}
where the matrices $\gamma_{1}$, $\gamma_{2}$ and $\gamma_{3}$ are given by
\begin{equation*}
\gamma_{1} = \textrm{diag}(\underbrace{1,...,1}_{k_{1} \  \textrm{times}},\underbrace{\omega_{n},...,\omega_{n}}_{k_{3} \ \textrm{times}},...,\underbrace{\omega_{n}^{n-1}...,\omega_{n}^{n-1}}_{k_{2n-1} \ \textrm{times}}) \ \ \ \textrm{with} \ \ \sum_{i \  \textrm{odd}}^{2n-1}k_{i} = k_L\, ,
\end{equation*}
\begin{equation*}
\gamma_{2} = \textrm{diag}(\underbrace{1,...,1}_{k_{2} \  \textrm{times}},\underbrace{\omega_{n},...,\omega_{n}}_{k_{4} \ \textrm{times}},...,\underbrace{\omega_{n}^{n-1}...,\omega_{n}^{n-1}}_{k_{2n} \ \textrm{times}}) \ \ \ \textrm{with} \ \ \ \sum_{i \  \textrm{even}}^{2n}k_{i} = k_R\, ,
\end{equation*}
\begin{equation*}
\gamma_{3} = \textrm{diag}(\underbrace{1,...,1}_{N_{1} \  \textrm{times}},\underbrace{\omega_{n},...,\omega_{n}}_{N_{2} \ \textrm{times}},...,\underbrace{\omega_{n}^{n-1}...,\omega_{n}^{n-1}}_{N_{n} \ \textrm{times}}) \ \ \ \textrm{with} \ \ \ \sum_{i=1}^{n}N_{i}  = N\, .
\end{equation*}

It's easy to check that the superpotential (\ref{super}) is invariant under the transformations (\ref{ta})-(\ref{tq}). In addition, the two gauge groups $U(k_L)$ and $U(k_R)$ of the initial theory and the flavor group $U(N)$ are broken to
\begin{equation*}
 U(k_L) \mapsto \bigotimes_{i \ \textrm{odd}}^{2n-1} U(k_{i}), \ \ \ U(k_R) \mapsto \bigotimes_{i \  \textrm{even}}^{2n}U(k_{i}), \ \ \ U(N) \mapsto \bigotimes_{i=1}^{n} U(N_{i})\, ,
\end{equation*}
and after the action of the transformations (\ref{ta})-(\ref{tq}) the various fields become

{\centering
\begin{equation*}
A^{1} =\left(\begin{array}{ccccc}
A^{1}_{11} & 0 & 0 & ... & 0 \\ 
0 & A^{1}_{22} & 0 & ... & 0 \\ 
0 & 0 & A^{1}_{33} & .... & 0 \\ 
... & ... & ... & ... & 0 \\ 
0 & 0 & 0 & 0 & A^{1}_{nn}
\end{array} 
\right), \ \  A^{2} =\left(\begin{array}{ccccc}
A^{2}_{11} & 0 & 0 & ... & 0 \\ 
0 & A^{2}_{22} & 0 & ... & 0 \\ 
0 & 0 & A^{2}_{33} & .... & 0 \\ 
... & ... & ... & ... & 0 \\ 
0 & 0 & 0 & 0 & A^{2}_{nn}
\end{array} 
\right)\, ,
\end{equation*}
\begin{equation*}
B^{1}=\left(\begin{array}{ccccc}
0 & 0 & 0 & ... & B^{1}_{1,n} \\ 
B^{1}_{21} & 0 & 0 & ... & 0 \\ 
0 & B^{1}_{32} & 0 & ... & 0 \\ 
... & ... & ... & 0 & 0 \\ 
0 & 0 & 0 & B^{1}_{n,n-1} & 0
\end{array} 
\right), \ \ B^{2}=\left(\begin{array}{ccccc}
0 & B^{2}_{12} & 0 & ... & 0 \\ 
0 & 0 & B^{2}_{23} & ... & 0 \\ 
0 & 0 & 0 & ... & ... \\ 
... & ... & ... & ... & B^{2}_{n-1,n} \\ 
B^{2}_{n,n-1} & 0 & 0 & 0 & 0
\end{array}\right)\, , 
\end{equation*}
\begin{equation*}
q =\left(\begin{array}{ccccc}
q_{11} & 0 & 0 & ... & 0 \\ 
0 & q_{22} & 0 & ... & 0 \\ 
0 & 0 & q_{33} & .... & 0 \\ 
... & ... & ... & ... & 0 \\ 
0 & 0 & 0 & 0 & q_{nn}
\end{array} 
\right), \ \  Q =\left(\begin{array}{ccccc}
Q_{11} & 0 & 0 & ... & 0 \\ 
0 & Q_{22} & 0 & ... & 0 \\ 
0 & 0 & Q_{33} & .... & 0 \\ 
... & ... & ... & ... & 0 \\ 
0 & 0 & 0 & 0 & Q_{nn}
\end{array} 
\right)\, .
\end{equation*}
\par
}

\subsection{Constructing $U(N)$ instantons on $\mathbb{C}P^2/\mathbb{Z}_n$ }

Let us now show the actual construction of unitary instantons on $\mathbb{C}P^2/\mathbb{Z}_n$.

\subsubsection{The $\mathbb{C}P^{2}/\mathbb{Z}_{2}$ case}
\label{subsec:z2case}
\begin{figure}[h!]
\centering
\includegraphics[scale=1]{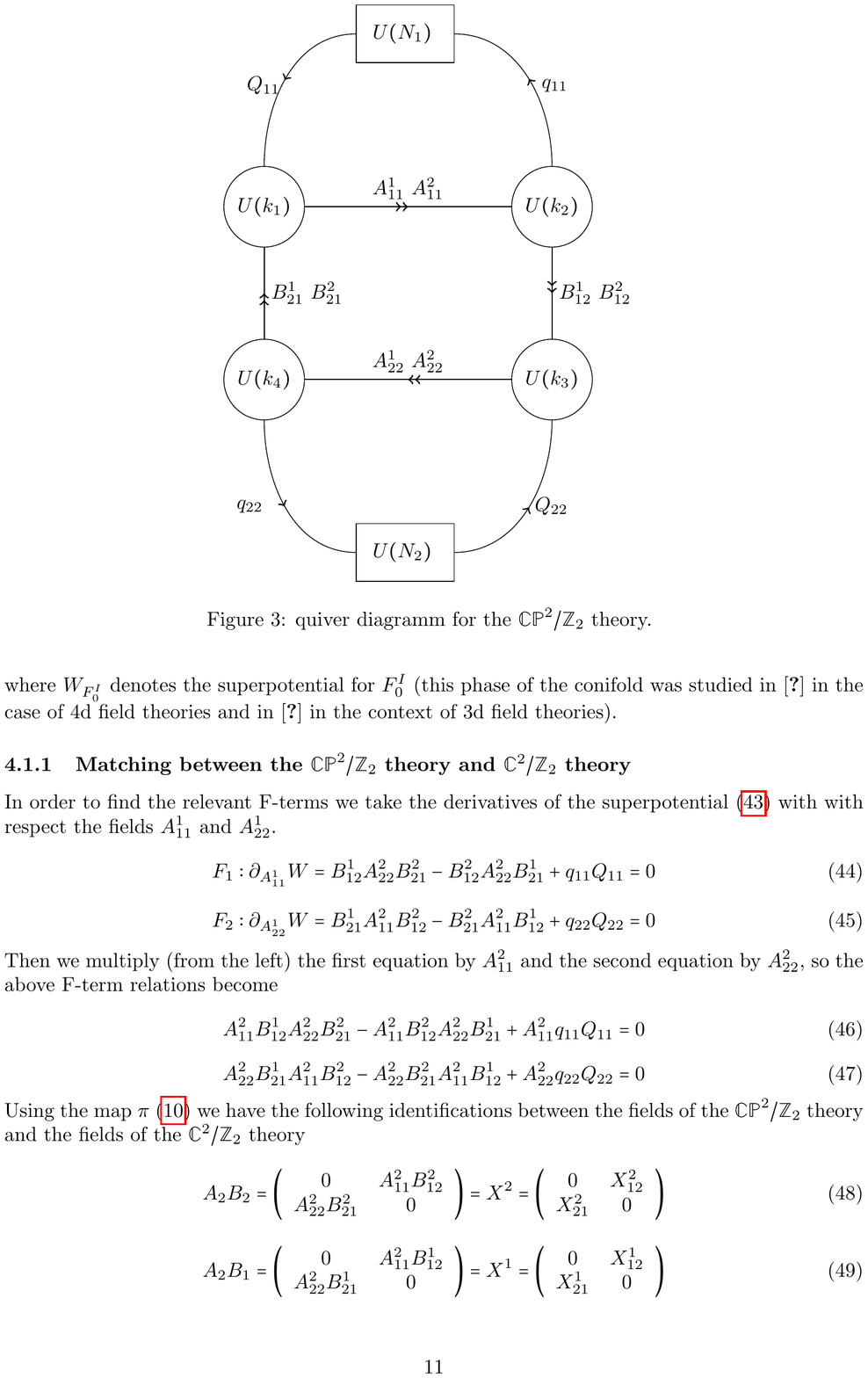} 
 \caption{Quiver diagram for the $\mathbb{C}P^{2}/\mathbb{Z}_{2}$ theory. \label{fig:quiverz2}}

\end{figure}

Let us consider the simplest case of the $\mathbb{Z}_{2}$ orbifold. Applying the rules above we obtain a theory whose quiver  is reported in fig.\ref{fig:quiverz2} together  with the superpotential (\ref{wz2}). Note that $W_{F_{0}^{I}}$ denotes the superpotential for $F_{0}^{I}$ (the first phase of the $F_0$ was studied in \cite{Butti:2007jv} in the case of $4d$ field theories and in \cite{Hanany:2008fj} in the context of $3d$ field theories). Moreover, for future reference, we compile the transformation properties of the fields and the F-terms under the various symmetry groups in table \ref{tab:orbifoldz2}.

\begin{table}[h!]
\renewcommand\arraystretch{1.3} 
\centering
\resizebox{15cm}{!}{
\begin{tabular}{c||cccc||cc||cc||}
Fields  & $U(k_{1})$ & $U(k_{2})$  & $U(k_{3})$  & $U(k_{4})$  & $U(N_{1})$ & $U(N_{2})$ & $SU(2)$  & $U(1)$   \\ 
 \hline 
 $A^{2}_{11}$ & $[1,0,...,0]_{+1}$ & $[0,...,0,1]_{-1}$ & $[\textbf{0}]_{0}$ & $[\textbf{0}]_{0}$ & $[\textbf{0}]_{0}$ & $[\textbf{0}]_{0}$ & $[0]$ & 1/2 \\ 
\hline
 $A^{2}_{22}$ & $[\textbf{0}]_{0}$ & $[\textbf{0}]_{0}$ & $[1,0,...,0]_{+1}$ & $[0,...,0,1]_{-1}$ & $[\textbf{0}]_{0}$ & $[\textbf{0}]_{0}$ & $[0]$ & 1/2 \\ 
\hline 
 $B^{1}_{12}$, $B^{2}_{12}$ & $[\textbf{0}]_{0}$ & $[1,0,...,0]_{+1}$ & $[0,...,0,1]_{-1}$ & $[\textbf{0}]_{0}$ & $[\textbf{0}]_{0}$ & $[\textbf{0}]_{0}$ & $[1]$ & 1/4 \\ 
\hline
 $B^{1}_{21}$,$B^{2}_{21}$ & $[0,...,0,1]_{-1}$ & $[\textbf{0}]_{0}$ & $[\textbf{0}]_{0}$ & $[1,0,...,0]_{+1}$ & $[\textbf{0}]_{0}$ & $[\textbf{0}]_{0}$ & $[1]$ & 1/4 \\ 
\hline
 $q_{11}$ & $[\textbf{0}]_{0}$ & $[1,0,...,0]_{+1}$ & $[\textbf{0}]_{0}$ & $[\textbf{0}]_{0}$ & $[0,...,0,1]_{-1}$ & $[\textbf{0}]_{0}$ & $[0]$ & 1-1/4$r$ \\ 
\hline
 $Q_{11}$ & $[0,...,0,1]_{-1}$ & $[\textbf{0}]_{0}$ & $[\textbf{0}]_{0}$ & $[\textbf{0}]_{0}$ & $[1,0,...,0]_{+1}$ &  $[\textbf{0}]_{0}$ & $[0]$ & 1/4$r$ \\ 
 \hline
$q_{22}$ & $[\textbf{0}]_{0}$ & $[\textbf{0}]_{0}$ & $[\textbf{0}]_{0}$ & $[1,0,...,0]_{+1}$ & $[\textbf{0}]_{0}$ & $[0,...,0,1]_{-1}$ & $[0]$ & 1-1/4$r$ \\ 
\hline
 $Q_{22}$ & $[\textbf{0}]_{0}$ & $[\textbf{0}]_{0}$ & $[0,...,0,1]_{-1}$ & $[\textbf{0}]_{0}$ & $[\textbf{0}]_{0}$ &  $[1,0,...,0]_{+1}$ & $[0]$ & 1/4$r$ \\ 
 \hline
 $F_{1}$ & $[0,...,0,1]_{-1}$ & $[1,0,...,0]_{+1}$ & $[\textbf{0}]_{0}$ & $[\textbf{0}]_{0}$ & $[\textbf{0}]_{0}$ & $[\textbf{0}]_{0}$ & $[0]$ & 1 \\ 
\hline 
 $F_{2}$ & $[\textbf{0}]_{0}$ & $[\textbf{0}]_{0}$ & $[0,...,0,1]_{-1}$ & $[1,0,...,0]_{+1}$ & $[\textbf{0}]_{0}$ & $[\textbf{0}]_{0}$ & $[0]$ & 1 \\ 
\end{tabular}  
}

\caption{Transformations of the fields and of the F-terms for the $\mathbb{C}P^{2}/\mathbb{Z}_{2}$ theory.\label{tab:orbifoldz2}}
\end{table}

\begin{equation}\begin{aligned}
\label{wz2}
 W = & \textrm{Tr}[A^{i}_{11}B^{j}_{12}A^{k}_{22}B^{l}_{21}\epsilon_{ik}\epsilon_{jl} + q_{11}A^{1}_{11}Q_{11} + q_{22}A^{1}_{22}Q_{22}]=\\
& = W_{F_{0}^{I}} + \textrm{Tr}[q_{11}A^{1}_{11}Q_{11} + q_{22}A^{1}_{22}Q_{22}]\, .
\end{aligned}\end{equation}

In the unorbifolded case the instanton branch appeared upon setting $A^1=0$. Therefore, in this case we need to impose $A^1_{11}=A^1_{22}=0$. Then, the only relevant F-terms are
\begin{equation}
\label{F1}
F_{1}: \partial_{A^{1}_{11}}W = B^{1}_{12}A^{2}_{22}B^{2}_{21} - B^{2}_{12}A^{2}_{22}B^{1}_{21} + q_{11}Q_{11} = 0\, ,
\end{equation}
\begin{equation}
\label{F2}
F_{2}: \partial_{A^{1}_{22}}W = B^{1}_{21}A^{2}_{11}B^{2}_{12} - B^{2}_{21}A^{2}_{11}B^{1}_{12} + q_{22}Q_{22} = 0\, .
\end{equation}
This describes the ADHM construction for instantons on $\mathbb{C}P^2/\mathbb{Z}_2$.

As we have reviewed above, in in the unorbifolded case it is possible to map instantons on $\mathbb{C}P^2$ into instantons on $\mathbb{C}^2$. Inherited from this we can find a mapping from the ADHM construction for instantons on the orbifolded space into that for instantons on the appropriate orbifold of $\mathbb{C}^2$. To see this, using the map $\pi$ in eq.(\ref{mappi}) we have the following identifications between the fields of the $\mathbb{C}P^{2}/\mathbb{Z}_{2}$ theory and the fields of the $\mathbb{C}^{2}/\mathbb{Z}_{2}$ theory
\begin{equation*}
A_{2}B_{2} = \left(\begin{array}{cc}
0 & A^{2}_{11}B^{2}_{12} \\ 
A^{2}_{22}B^{2}_{21} & 0
\end{array}\right) = \left(\begin{array}{cc}
0 & X^{2}_{12} \\ 
X^{2}_{21} & 0
\end{array}  \right) = X^{2}, \ \ A^{2}q = \left(\begin{array}{cc}
A^{2}_{11}q_{11} & 0 \\ 
0 & A^{2}_{22}q_{22}
\end{array} \right) = \left(\begin{array}{cc}
I_{11} & 0 \\ 
0 & I_{22}
\end{array}\right) = I,
\end{equation*}

{\centering
\begin{equation*}
A_{2}B_{1} = \left(\begin{array}{cc}
0 & A^{2}_{11}B^{1}_{12} \\ 
A^{2}_{22}B^{1}_{21} & 0
\end{array}\right) = \left(\begin{array}{cc}
0 & X^{1}_{12} \\ 
X^{1}_{21} & 0
\end{array}  \right) = X^{1}, \ \ Q=\left(\begin{array}{cc}
Q_{11} & 0 \\ 
0 & Q_{22}
\end{array}\right) =\left( \begin{array}{cc}
J_{11} & 0 \\ 
0 & J_{22}
\end{array}\right) = J\, .
\end{equation*}
\par}

Then, upon multiplication of the F-term relations (\ref{F1}) and (\ref{F2}) by $A_{11}^1$ and $A_{22}^2$ respectively, these can be rewritten as
\begin{equation}
\label{f11}
X^{1}_{12}X^{2}_{21} - X^{2}_{12}X^{1}_{21} + I_{11}J_{11} = 0\, ,
\end{equation}

\begin{equation}
\label{f22}
X^{1}_{21}X^{2}_{12} - X^{2}_{21}X^{1}_{12} +I_{22}J_{22} = 0\, ,
\end{equation}
which are the F-terms relations for the $\mathbb{C}^{2}/\mathbb{Z}_{2}$ theory \cite{Dey:2013fea}. Hence, we recover the analog to the unorbifolded case, namely that the moduli space (at least removing possible compact directions to which we will come back below) is biholomorphic to the moduli space of $\mathbb{C}^2/\mathbb{Z}_2$.

The Hilbert series of instantons described by the theory with flavor group $U(N_{1}) \times  U(N_{2})$ and gauge ranks $\textbf{k}=(k_{1},k_{2},k_{3},k_{4})$ \footnote{We will summarize the ranks of the various gauge groups with a vector $\textbf{k}$ and the ranks of the flavor groups with a vector $\textbf{N}$.} reads
\begin{equation}\begin{aligned}
\label{hsz2}
& H[\textbf{k},F, \mathbb{C}P^{2}/\mathbb{Z}_{2}](t,x,\textbf{y},\textbf{d})= \int d \mu_{U(k_{1})}(\textbf{u})\int d \mu_{U(k_{2})} (\textbf{w})\int d \mu_{U(k_{3})}(\textbf{z}) \times \\
& \int d \mu_{U(k_{4})}(\textbf{v})\textrm{PE}[\chi_{A^{2}_{11}}t^{2} +\chi_{A^{2}_{22}}t^{2}  + \chi_{B^{j}_{12}}t + \chi_{B^{j}_{21}}t + \chi_{q_{11}}t^{2} + \chi_{Q_{11}}t^{2} + \chi_{q_{22}}t^{2} + \\ 
& + \chi_{Q_{22}}t^{2} -\chi_{F_{1}}t^{4} -\chi_{F_{2}}t^{4}]\, ,
\end{aligned}\end{equation}
where we are using the following notation
\begin{itemize}
\item The fugacity $t$ is associated with the R-charge and keeps track of it in unit of one quarter.
\item The fugacities $\textbf{u},\textbf{w},\textbf{z}$ and $\textbf{v}$ are associated with the gauge groups $U(k_{1}),U(k_{2}),U(k_{3})$ and $U(k_{4})$ respectively.
\item The fugacities $x,\textbf{y}$ and $\textbf{d}$ are associated with the global symmetries $SU(2)$, $U(N_{1})$ and $U(N_{2})$ respectively.
\item The contribution of each field is given by
\begin{equation*}
\chi_{A^{2}_{11}} = \sum_{a=1}^{k_{1}}\sum_{b=1}^{k_{2}}u_{a}w_{b}^{-1}, \ \ \chi_{A^{2}_{22}} = \sum_{a=1}^{k_{3}}\sum_{b=1}^{k_{4}}z_{a}v_{b}^{-1}, \ \ \chi_{B_{12}^{j}} = \left(x+\frac{1}{x}\right)\sum_{a=1}^{k_{2}}\sum_{b=1}^{k_{3}}w_{a}z_{b}^{-1},
\end{equation*}
\begin{equation*}
\chi_{B^{j}_{21}} = \left(x +\frac{1}{x}\right)\sum_{a=1}^{k_{4}}\sum_{b=1}^{k_{1}}v_{a}u_{b}^{-1}, \ \ \chi_{F_{1}} = \sum_{a=1}^{k_{1}}\sum_{b=1}^{k_{2}}u_{a}^{-1}w_{b}, \ \ \chi_{F_{2}} = \sum_{a=1}^{k_{3}}\sum_{b=1}^{k_{4}}z_{a}^{-1}v_{b},
\end{equation*}
\begin{equation*}
\chi_{q_{11}} = \sum_{a=1}^{k_{2}}\sum_{b=1}^{N_{1}}w_{a}y_{b}^{-1}, \ \ \chi_{Q_{11}} = \sum_{a=1}^{N_{1}}\sum_{b=1}^{k_{1}}y_{a}u_{b}^{-1}, \ \  \chi_{q_{22}} = \sum_{a=1}^{k_{4}}\sum_{b=1}^{N_{2}}v_{a}d_{b}^{-1}, \ \ \chi_{Q_{22}} = \sum_{a=1}^{N_{2}}\sum_{b=1}^{k_{3}}d_{a}z_{b}^{-1}\, .
\end{equation*}
\item The Haar measure of each $U(k)$ gauge group is taken equal to
\begin{equation*}
\int d \mu_{U(k)}(\textbf{u}) = \frac{1}{k!}\left( \prod_{j=1}^{k}\oint_{\mid u_{j}\mid =1} \frac{du_{j}}{2\pi i u_{j}} \right) \prod_{1\leq i < j \leq k}(u_{i}-u_{j})(u_{i}^{-1}-u_{j}^{-1})
\end{equation*}
\end{itemize}
Explicit computation shows that the Hilbert series on the instanton branch for gauge group $G=U(k_1)\times U(k_2)\times U(k_3)\times U(k_4)$ with flavor group $U(N_{1}) \times U(N_{2})$ corresponding to instantons on $\mathbb{C}P^{2}/\mathbb{Z}_{2}$ is equal to the Hilbert series on the Higgs branch of the $A_1$ quiver with $U(K_1)\times U(K_2)$ gauge symmetry and global $U(N_{1}) \times U(N_{2})$ symmetry, corresponding to instantons on $\mathbb{C}^{2}/\mathbb{Z}_{2}$ \cite{Dey:2013fea}; where

\begin{equation}
K_{1} = \textrm{min}(k_{1},k_{2})\, ,\qquad K_{2}=\textrm{min}(k_{3},k_{4})\, .
\end{equation}

\begin{figure}[h!]
\centering
\includegraphics[scale=1]{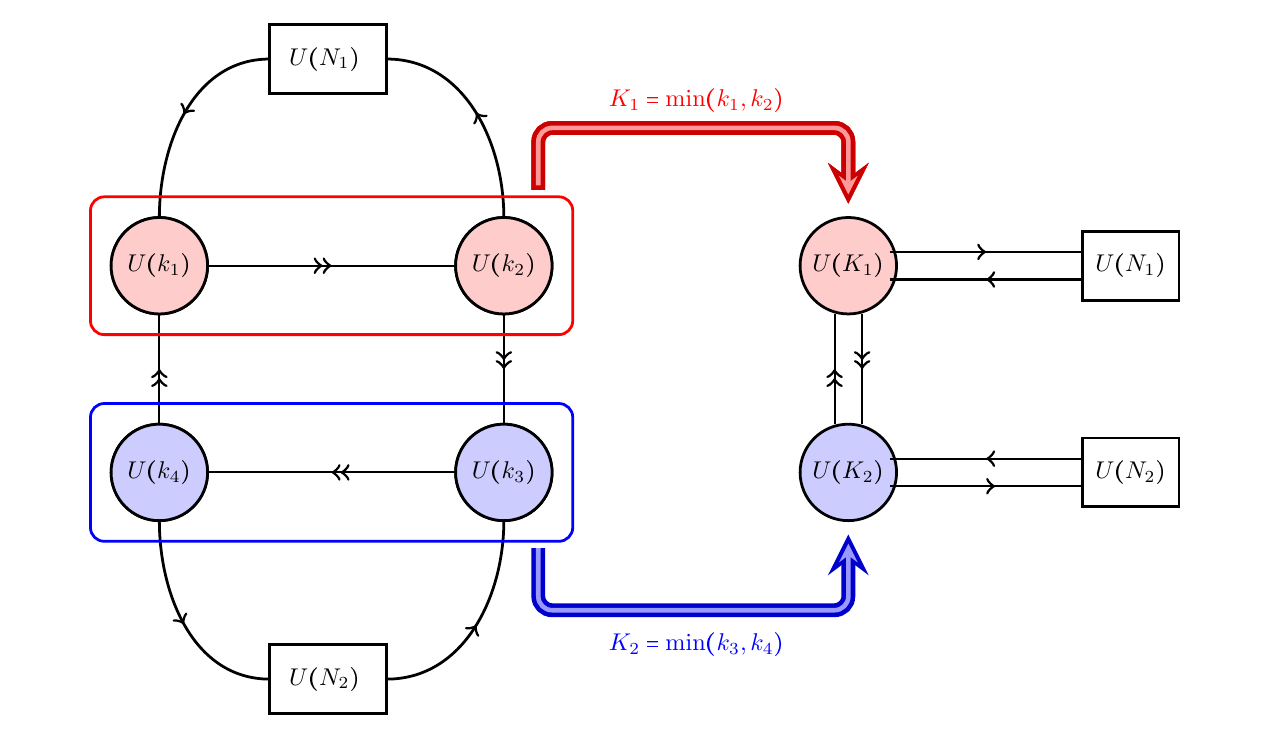}

\caption{Relation between the $\mathbb{C}P^{2}/\mathbb{Z}_{2}$ quiver gauge theory (on the left) and the corresponding $\mathbb{C}^{2}/\mathbb{Z}_{2}$ quiver gauge theory (on the right).  \label{fig:comparison}}

\end{figure}

In fig.\ref{fig:comparison} we graphically summarize the relation between the theory describing instantons on $\mathbb{C}P^2/\mathbb{Z}_2$ and that describing instantons on $\mathbb{C}^2/\mathbb{Z}_2$. Note that each flavor node flavors two adjacent nodes, which are precisely those ``merging" into a single node in the $\mathbb{C}^2/\mathbb{Z}_2$ cousin.

Let us turn to explicit examples supporting of our claim.\\
\\
\textbf{$U(N_1)$ instantons: $\textbf{k}=(1,1,1,1)$ and $\textbf{N}=(N_{1},0)$}
\\
\\
Using eq.(\ref{hsz2}) we have
\begin{equation*}\begin{aligned}
\label{HSn1}
& H[\textbf{k}=(1,1,1,1),\textbf{N}=(N_{1},0),\mathbb{C}P^{2}/\mathbb{Z}_{2}](t,x,\textbf{y}) = \frac{1}{(2\pi i)^{4}}\oint_{\mid u\mid=1} \frac{du}{u}\oint_{\mid w\mid=1} \frac{dw}{w}\oint_{\mid z\mid=1} \frac{dz}{z}\\
& \oint_{\mid v\mid=1} \frac{dv}{v} \times \textrm{PE}[\chi_{A^{2}_{11}}t^{2} +\chi_{A^{2}_{22}}t^{2}  + \chi_{B^{j}_{12}}t + \chi_{B^{j}_{21}}t + \chi_{q_{11}}t^{2} + \chi_{Q_{11}}t^{2} -\chi_{F_{1}}t^{4} -\chi_{F_{2}}t^{4}]\, ,
\end{aligned}\end{equation*}
where the various characters are given by
\footnote{We rewrite the flavor group $U(N_{1})$ as $U(1) \times SU(N_{1})$. We denote with $p$ the fugacity of the U(1) subgroup while we denote with $\vec{y}$ the fugacities of the $SU(N_{1})$ group.
} 
\begin{equation*}
\chi_{A^{2}_{11}} = uw^{-1}, \ \ \chi_{A^{2}_{22}} = zv^{-1}, \ \ \chi_{B_{12}^{j}} = \left(x+\frac{1}{x}\right)wz^{-1}, \ \ \chi_{B^{j}_{21}} = \left(x +\frac{1}{x}\right)u^{-1}v\, ,
\end{equation*}
\begin{equation*}
\chi_{F_{1}} = u^{-1}w, \ \ \chi_{F_{2}} = z^{-1}v, \ \ \chi_{q_{11}} = wp^{-1}[0,...,0,1]_{\vec{y}}, \ \ \chi_{Q_{11}} = u^{-1}p[1,0,...,0]_{\vec{y}}\, . 
\end{equation*}
Integrating over $z$ and $v$ we obtain
\begin{equation*}
\frac{1}{(2\pi i)^{2}}\oint_{\mid u\mid=1} \frac{du}{u}\oint_{\mid w\mid=1} \frac{dw}{w} \frac{(1-t^{6})x^{2}(u+t^{4}w)}{(t^{2}u-w)(t^{4}w-x^{2}u)(u-t^{4}x^{2}w)} \times \textrm{PE}[\chi_{q_{11}}t^{2} + \chi_{Q_{11}}t^{2}]\, ,
\end{equation*}
then integrating over the second gauge group we find
\begin{equation*}
\frac{1+t^{6}}{(1-t^{6}/x^{2})(1-t^{6}x^{2})}\times \frac{1-t^{6}}{(2\pi i)}\oint_{\mid u\mid=1} \frac{du}{u} \textrm{PE}[up^{-1}t^{4}[0,...,0,1]_{\vec{y}} +u^{-1}pt^{2}[1,0,...0]_{\vec{y}}]\, .
\end{equation*}
We can reabsorb the fugacity $p$ of the $U(1)$ flavour as $u' = up^{-1}$. Therefore the previous integral becomes
\begin{equation*}
\frac{1+t^{6}}{(1-t^{6}/x^{2})(1-t^{6}x^{2})}\times \frac{1-t^{6}}{(2\pi i)}\oint_{\mid u'\mid=1} \frac{du'}{u'} \times \textrm{PE}[u't^{4}[0,...,0,1]_{\vec{y}} + t^{2}/u'[1,0,...0]_{\vec{y}}]\, ,
\end{equation*}
finally, doing $u' = u_{2}/t$, the previous expression becomes

\begin{equation*}
\label{HSf}
\frac{1+t^{6}}{(1-t^{6}/x^{2})(1-t^{6}x^{2})}\times \frac{1-t^{6}}{(2\pi i)}\oint_{\mid u_{2}\mid=1} \frac{du_{2}}{u_{2}} \textrm{PE}[u_{2}t^{3}[0,...,0,1]_{\vec{y}} + t^{3}u_{2}^{-1}[1,0,...0]_{\vec{y}}]\, .
\end{equation*}
This last expression coincides with the Hilbert series for one $SU(N_{1})$ instanton on $\mathbb{C}^{2}/\mathbb{Z}_{2}$ (it coincides with the eq.(2.15) of \cite{Dey:2013fea}).\\

\begin{itemize}
\item{\textbf{$U(1)$ instanton: $\textbf{k}=(2,1,1,1)$, and $\textbf{N}=(1,0)$}}\\
\\
Using eq.(\ref{hsz2}), we find that
\begin{equation*}
H[\textbf{k}=(2,1,1,1),\textbf{N}=(1,0),\mathbb{C}P^{2}/\mathbb{Z}_{2}](t,x) = \frac{1+t^{6}}{(1-t^{6}/x^{2})(1-t^{6}x^{2})}\, ,
\end{equation*}
which is the Hilbert series of one $U(1)$ instanton on $\mathbb{C}^{2}/\mathbb{Z}_{2}$.\\

\item{\textbf{$U(1)$ instanton: $\textbf{k}=(2,1,2,1)$, and $\textbf{N}=(1,0)$}}\\
\\
Using eq.(\ref{hsz2}), we find that
\begin{equation*}
H[\textbf{k}=(2,1,2,1),\textbf{N}=(1,0),\mathbb{C}P^{2}/\mathbb{Z}_{2}](t,x) = \frac{1+t^{6}}{(1-t^{6}/x^{2})(1-t^{6}x^{2})}\, ,
\end{equation*}
which is again the Hilbert series of one $U(1)$ instanton on $\mathbb{C}^{2}/\mathbb{Z}_{2}$.\\

\item{\textbf{$U(1)$ instanton: $\textbf{k}=(1,2,1,2)$, and $\textbf{N}=(1,0)$}}\\
\\
Using eq.(\ref{hsz2}), we find that
\begin{equation*}
H[\textbf{k}=(1,2,1,2),\textbf{N}=(1,0),\mathbb{C}P^{2}/\mathbb{Z}_{2}](t,x) = \frac{1+t^{6}}{(1-t^{6}/x^{2})(1-t^{6}x^{2})}\, ,
\end{equation*}
which is again the Hilbert series of one $U(1)$ instanton on $\mathbb{C}^{2}/\mathbb{Z}_{2}$.\\

\item{\textbf{$U(2)$ instanton: $\textbf{k}=(2,1,1,1)$ and $\textbf{N}=(2,0)$}}\\
\\
Using eq.(\ref{hsz2}), we find that
\begin{equation*}
H[\textbf{k}=(2,1,1,1),\textbf{N}=(2,0),\mathbb{C}P^{2}/\mathbb{Z}_{2}](t,x,y_{1},y_{2}) =\frac{(1 + t^6)^2 x^2 y_{1} y_{2}}{(t^6 - x^2) (1 - t^6 x^2) (t^6 y_{1} - 
   y_{2}) (y_{1} - t^6 y_{2})}\, ,
\end{equation*}
being $y_{1}$ and $y_{2}$ the fugacities of the flavor group. The previous expression coincides with the Hilbert series for one $U(2)$ instanton on $\mathbb{C}^{2}/\mathbb{Z}_{2}$.\\

\item{\textbf{$U(2)$ instanton: $\textbf{k}=(2,2,1,1)$, and $\textbf{N}=(2,0)$}}\\
\\
Using eq.(\ref{hsz2}) and unrefining for simplicity we find
\begin{equation*}
H[\textbf{k}=(2,2,1,1),\textbf{N}=(2,0),\mathbb{C}P^{2}/\mathbb{Z}_{2}](t,1,1,1) = \frac{1 + 3 t^6 + 11 t^{12} + 10 t^{18} + 11 t^{24} + 
 3 t^{30} + t^{36}}{(1 - t^6)^6 (1 + t^6)^3}\, ,
\end{equation*}
which is the unrefined Hilbert series for $\textbf{K}=(2,1)$ instantons with flavor group $\textbf{N}=(2,0)$ on $\mathbb{C}^{2}/\mathbb{Z}_{2}$.\\ 

\item{\textbf{$U(2)$ instanton: $\textbf{k}=(2,2,1,1)$, and $\textbf{N}=(0,2)$}}\\
\\
Using eq.(\ref{hsz2}) this time we find that
\begin{equation*}
H[\textbf{k}=(2,2,1,1),\textbf{N}=(0,2),\mathbb{C}P^{2}/\mathbb{Z}_{2}](t,x,y_{1},y_{2}) = \frac{(1 + t^6) (x^2 + t^6 x^2 + t^{18} x^2 - 
   t^{12} (1 + x^2 + x^4)) y_{1} y_{2}}{(t^6 - x^2) (1 - t^6 x^2) (t^6 y_{1} - 
   y_{2}) (y_{1} - t^6 y_{2})}\, ,
\end{equation*}
being $y_{1}$ and $y_{2}$ the fugacities of the $U(2)$ flavor group. The previous expression is the Hilbert series of $\textbf{K}=(2,1)$ instantons with $\textbf{N}=(0,2)$ on $\mathbb{C}^{2}/\mathbb{Z}_{2}$.

\end{itemize}

\subsubsection{The $\mathbb{C}P^{2}/\mathbb{Z}_{3}$ case}

\begin{figure}[h!]
\centering
\includegraphics[scale=1]{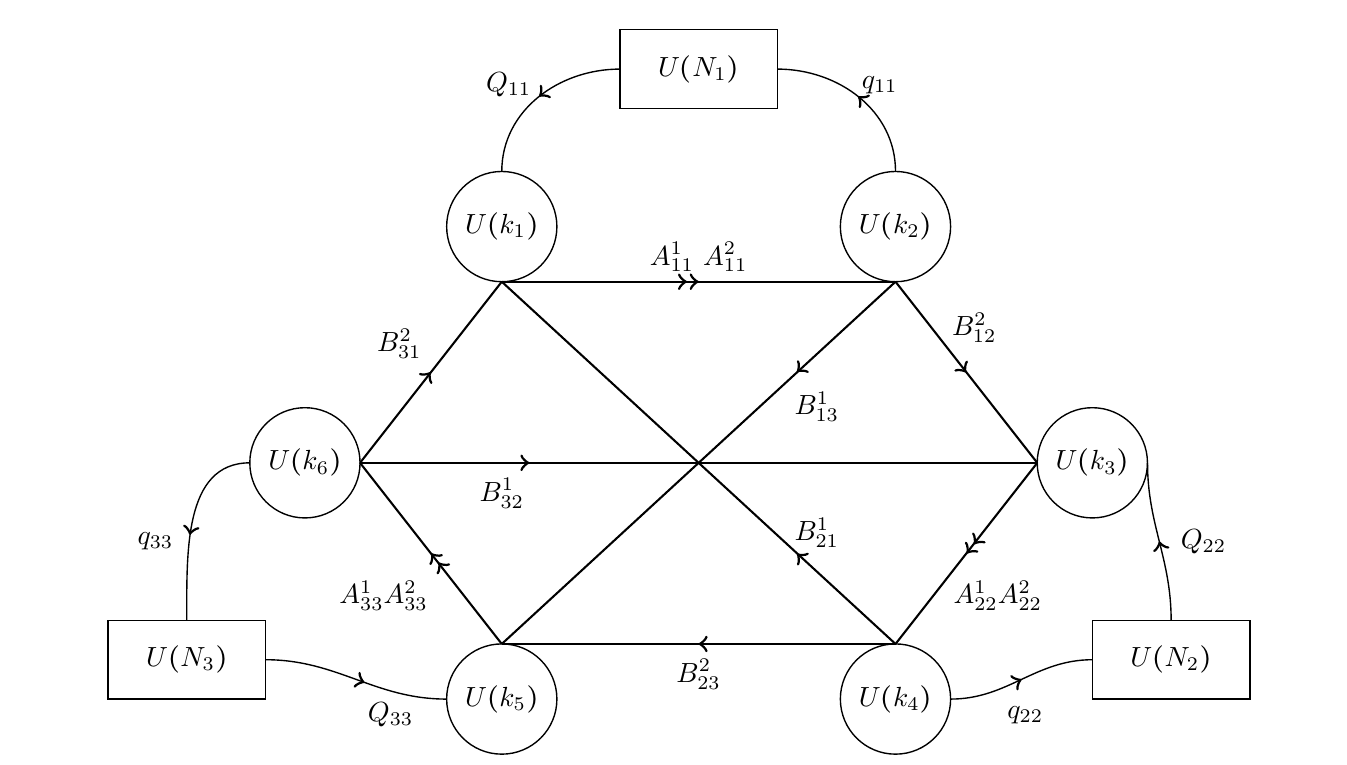}

\caption{The quiver diagram for the $\mathbb{C}P^{2}/\mathbb{Z}_{3}$  theory.  \label{fig:q2}}

\end{figure}

Let us now consider the case of $\mathbb{C}P^{2}/\mathbb{Z}_{3}$. Using the rules above, we find that the quiver describing the moduli space of instantons on the $\mathbb{C}P^{2}/\mathbb{Z}_{3}$ is fig.\ref{fig:q2}. We summarize the fields quantum numbers in table \ref{tab:2}.

The superpotential (\ref{super}) becomes

\begin{equation}\begin{split}
\label{wz3}
W = & \textrm{Tr}[A^{1}_{22}B^{1}_{21}A^{2}_{11}B^{2}_{12} - A^{1}_{11}B^{2}_{12}A^{2}_{22}B^{1}_{21} + A^{1}_{33}B^{1}_{32}A^{2}_{22}B^{2}_{23} - A^{1}_{22}B^{2}_{23}A^{2}_{33}B^{1}_{32} + \\
& - A^{1}_{33}B^{2}_{31}A^{2}_{11}B^{1}_{13} + A^{1}_{11}B^{1}_{13}A^{2}_{33}B^{2}_{31} + q_{11}A^{1}_{11}Q_{11} + q_{22}A^{1}_{22}Q_{22}  + q_{33}A^{1}_{33}Q_{33}]\, .
\end{split}\end{equation}
Now the instanton branch emerges upon setting $A^1_{ii}=0$. The relevant F-terms are

\begin{equation*}
F_{1}: \partial_{A^{1}_{11}}W = B^{1}_{13}A^{2}_{33}B^{2}_{31} - B^{2}_{12}A^{2}_{22}B^{1}_{21} + q_{11}Q_{11} = 0\, ,
\end{equation*}

\begin{equation*}
F_{2}: \partial_{A^{1}_{22}}W = B^{1}_{21}A^{2}_{11}B^{2}_{12} - B^{2}_{23}A^{2}_{33}B^{1}_{32} + q_{22}Q_{22} = 0\, ,
\end{equation*}

\begin{equation*}
F_{3}: \partial_{A^{1}_{33}}W = B^{1}_{32}A^{2}_{22}B^{2}_{23} - B^{2}_{31}A^{2}_{11}B^{1}_{13} + q_{33}Q_{33} = 0\, .
\end{equation*}
This defines the ADHM construction for instantons on $\mathbb{C}P^2/\mathbb{Z}_3$.

If we multiply the $F_{1}$,$F_{2}$ and $F_{3}$  respectively by $A^{2}_{11},A^{2}_{22}.A^{2}_{33}$ we  obtain
\begin{equation}
\label{e1}
A^{2}_{11}B^{1}_{13}A^{2}_{33}B^{2}_{31} - A^{2}_{11}B^{2}_{12}A^{2}_{22}B^{1}_{21} + A^{2}_{11}q_{11}Q_{11} = 0\, ,
\end{equation}

\begin{equation}
\label{e2}
A^{2}_{22}B^{1}_{21}A^{2}_{11}B^{2}_{12} - A^{2}_{22}B^{2}_{23}A^{2}_{33}B^{1}_{32} + A^{2}_{22}q_{22}Q_{22} = 0\, ,
\end{equation}

\begin{equation}
\label{e3}
 A^{2}_{33}B^{1}_{32}A^{2}_{22}B^{2}_{23} - A^{2}_{33}B^{2}_{31}A^{2}_{11}B^{1}_{13} + A^{2}_{33}q_{33}Q_{33} = 0\, .
\end{equation}
It is easy to check,  using the identification provided by the map $\pi$ in eq.(\ref{mappi}), 
 that the expressions (\ref{e1})-(\ref{e3}) match the corresponding F-terms of the $\mathbb{C}^{2}/\mathbb{Z}_{3}$ theory. Note that, as opposed to the unorbifolded and $\mathbb{Z}_2$ orbifold, the $SU(2)$ global symmetry rotating the $B_{i}$ fields is broken due to the orbifold action. This correlates with the fact that the moduli space of instantons on $\mathbb{C}P^2/\mathbb{Z}_n$ is biholomorphic to the moduli space of instantons on $\mathbb{C}^2/\mathbb{Z}_n$, which exhibits an $SU(2)$ symmetry for $n=1,\,2$ but not for higher $n$.

\begin{table}[h!]
\centering
\resizebox{15cm}{!}{
\renewcommand\arraystretch{1.3} 
\begin{tabular}{c||cccccc||ccc||c||}
Fields & $U(k_{1})$ & $U(k_{2})$ & $U(k_{3})$ & $U(k_{4})$ & $U(k_{5})$ & $U(k_{6})$ & $U(N_{1})$ & $ U(N_{2})$ & $U(N_{3})$ & $U(1)_{R}$ \\ 
\hline 
$A^{2}_{11}$ & $[1,0,...,0]_{+1}$ & $[0,...,0,1]_{-1}$ & $[\textbf{0}]_{0}$ & $[\textbf{0}]_{0}$ & $[\textbf{0}]_{0}$ & $[\textbf{0}]_{0}$ & $[\textbf{0}]_{0}$ & $[\textbf{0}]_{0}$ & $[\textbf{0}]_{0}$  & 1/2 \\ 
\hline 
$A^{2}_{22}$ & $[\textbf{0}]_{0}$ & $[\textbf{0}]_{0}$ & $[1,0,...,0]_{+1}$ & $[0,...,0,1]_{-1}$ & $[\textbf{0}]_{0}$ & $[\textbf{0}]_{0}$ & $[\textbf{0}]_{0}$ & $[\textbf{0}]_{0}$ & $[\textbf{0}]_{0}$  & 1/2 \\ 
\hline 
$A^{2}_{33}$ & $[\textbf{0}]_{0}$ & $[\textbf{0}]_{0}$ & $[\textbf{0}]_{0}$ & $[\textbf{0}]_{0}$ & $[1,0,...,0]_{+1}$ & $[0,...,0,1]_{-1}$ & $[\textbf{0}]_{0}$ & $[\textbf{0}]_{0}$ & $[\textbf{0}]_{0}$  & 1/2 \\ 
\hline 
$B^{1}_{13}$ & $[\textbf{0}]_{0}$ & $[1,0,...,0]_{+1}$ & $[\textbf{0}]_{0}$ & $[\textbf{0}]_{0}$ & $[0,...,0,1]_{-1}$ & $[\textbf{0}]_{0}$ & $[\textbf{0}]_{0}$ & $[\textbf{0}]_{0}$ & $[\textbf{0}]_{0}$  & 1/4 \\ 
\hline 
$B^{1}_{21}$ & $[0,...,0,1]_{-1}$ & $[\textbf{0}]_{0}$ & $[\textbf{0}]_{0}$ & $[1,0,...,0]_{+1}$ & $[\textbf{0}]_{0}$ & $[\textbf{0}]_{0}$ & $[\textbf{0}]_{0}$ & $[\textbf{0}]_{0}$ & $[\textbf{0}]_{0}$  & 1/4 \\ 
\hline 
$B^{1}_{32}$ & $[\textbf{0}]_{0}$ & $[\textbf{0}]_{0}$ & $[0,...,0,1]_{-1}$ & $[\textbf{0}]_{0}$ & $[\textbf{0}]_{0}$ & $[1,0,...,0]_{+1}$ & $[\textbf{0}]_{0}$ & $[\textbf{0}]_{0}$ & $[\textbf{0}]_{0}$  & 1/4 \\ 
\hline 
$B^{2}_{12}$ & $[\textbf{0}]_{0}$ & $[1,0,...,0]_{+1}$ & $[0,...,0,1]_{-1}$ & $[\textbf{0}]_{0}$ & $[\textbf{0}]_{0}$ & $[\textbf{0}]_{0}$ & $[\textbf{0}]_{0}$ & $[\textbf{0}]_{0}$ & $[\textbf{0}]_{0}$  & 1/4 \\ 
\hline 
$B^{2}_{23}$ & $[\textbf{0}]_{0}$ & $[\textbf{0}]_{0}$ & $[\textbf{0}]_{0}$ & $[1,0,...,0]_{+1}$ & $[0,...,0,1]_{-1}$ & $[\textbf{0}]_{0}$ & $[\textbf{0}]_{0}$ & $[\textbf{0}]_{0}$ & $[\textbf{0}]_{0}$  & 1/4 \\ 
\hline 
$B^{2}_{31}$ & $[0,...,0,1]_{-1}$ & $[\textbf{0}]_{0}$ & $[\textbf{0}]_{0}$ & $[\textbf{0}]_{0}$ & $[\textbf{0}]_{0}$ & $[1,0,...,0]_{+1}$ & $[\textbf{0}]_{0}$ & $[\textbf{0}]_{0}$ & $[\textbf{0}]_{0}$  & 1/4 \\ 
\hline 
$q_{11}$ & $[\textbf{0}]_{0}$ & $[1,0,...,0]_{+1}$ & $[\textbf{0}]_{0}$ & $[\textbf{0}]_{0}$ & $[\textbf{0}]_{0}$ & $[\textbf{0}]_{0}$ & $[0,...,0,1]_{-1}$ & $[\textbf{0}]_{0}$ & $[\textbf{0}]_{0}$  & $1-1/4r$ \\ 
\hline 
$Q_{11}$ & $[0,...,0,1]_{-1}$ & $[\textbf{0}]_{0}$ & $[\textbf{0}]_{0}$ & $[\textbf{0}]_{0}$ & $[\textbf{0}]_{0}$ & $[\textbf{0}]_{0}$ & $[1,0,...,0]_{+1}$ & $[\textbf{0}]_{0}$ & $[\textbf{0}]_{0}$  & $1/4r$ \\ 
\hline 
$q_{22}$ & $[\textbf{0}]_{0}$ & $[\textbf{0}]_{0}$ & $[\textbf{0}]_{0}$ & $[1,0,...,0]_{+1}$ & $[\textbf{0}]_{0}$ & $[\textbf{0}]_{0}$ & $[\textbf{0}]_{0}$  & $[0,...,0,1]_{-1}$ &$[\textbf{0}]_{0}$ & $1-1/4r$ \\ 
\hline 
$Q_{22}$ & $[\textbf{0}]_{0}$ & $[\textbf{0}]_{0}$ & $[0,...,0,1]_{-1}$ & $[\textbf{0}]_{0}$ & $[\textbf{0}]_{0}$ & $[\textbf{0}]_{0}$ & $[\textbf{0}]_{0}$  & $[1,0,...,0]_{+1}$ & $[\textbf{0}]_{0}$ & $1/4r$ \\ 
\hline 
$q_{33}$ & $[\textbf{0}]_{0}$ & $[\textbf{0}]_{0}$ & $[\textbf{0}]_{0}$ & $[\textbf{0}]_{0}$ & $[\textbf{0}]_{0}$ & $[1,0,...,0]_{+1}$ & $[\textbf{0}]_{0}$ & $[\textbf{0}]_{0}$ & $[0,...,0,1]_{-1}$ & $1-1/4r$ \\ 
\hline 
$Q_{33}$ & $[\textbf{0}]_{0}$ & $[\textbf{0}]_{0}$ & $[\textbf{0}]_{0}$ & $[\textbf{0}]_{0}$ & $[0,...,0,1]_{-1}$ & $[\textbf{0}]_{0}$ & $[\textbf{0}]_{0}$ & $[\textbf{0}]_{0}$ & $[1,0,...,0]_{+1}$ & $1/4r$ \\ 
\hline 
$F_{1}$ & $[0,...,0,1]_{-1}$ & $[1,0,...,0]_{+1}$ & $[\textbf{0}]_{0}$ & $[\textbf{0}]_{0}$ & $[\textbf{0}]_{0}$ & $[\textbf{0}]_{0}$ & $[\textbf{0}]_{0}$ & $[\textbf{0}]_{0}$ & $[\textbf{0}]_{0}$  & 1 \\ 
\hline 
$F_{2}$ & $[\textbf{0}]_{0}$ & $[\textbf{0}]_{0}$ & $[0,...,0,1]_{-1}$ & $[1,0,...,0]_{+1}$ & $[\textbf{0}]_{0}$ & $[\textbf{0}]_{0}$ & $[\textbf{0}]_{0}$ & $[\textbf{0}]_{0}$ & $[\textbf{0}]_{0}$  & 1 \\ 
\hline 
$F_{3}$ & $[\textbf{0}]_{0}$ & $[\textbf{0}]_{0}$ & $[\textbf{0}]_{0}$ & $[\textbf{0}]_{0}$ & $[0,....,0,1]_{-1}$ & $[1,0,...,0]_{+1}$ & $[\textbf{0}]_{0}$ & $[\textbf{0}]_{0}$ & $[\textbf{0}]_{0}$ &  1 \\ 
\end{tabular} 
}
\caption{Transformations of the fields and of the F-terms for the $\mathbb{C}P^{2}/\mathbb{Z}_{3}$ theory.\label{tab:2}}
\end{table}

The Hilbert series for $F=U(N_{1}) \times  U(N_{2}) \times U(N_{3})$ instantons on $\mathbb{C}P^{2}/\mathbb{Z}_{3}$ with the configuration $\textbf{k}=(k_{1},k_{2},k_{3},k_{4},k_{5},k_{6})$ reads
\begin{equation}\begin{aligned}
\label{hsz3}
& H[\textbf{k},F, \mathbb{C}P^{2}/\mathbb{Z}_{3}](t,\textbf{y},\textbf{d},\textbf{s})= \int d \mu_{U(k_{1})}(\textbf{u})\int d \mu_{U(k_{2})} (\textbf{w})\int d \mu_{U(k_{3})}(\textbf{z})\int d \mu_{U(k_{4})}(\textbf{v}) \times \\
& \int d \mu_{U(k_{5})}(\textbf{j})\int d \mu_{U(k_{6})}(\textbf{c}) \times \textrm{PE}[\chi_{A^{2}_{11}}t^{2} +\chi_{A^{2}_{22}}t^{2}  + \chi_{A^{2}_{33}}t^{2} + \chi_{B^{2}_{12}}t + \chi_{B^{2}_{23}}t +  \chi_{B^{2}_{31}}t + \chi_{B^{1}_{21}}t + \\
& + \chi_{B^{1}_{13}}t  + \chi_{B^{1}_{32}}t +   \chi_{q_{11}}t^{2} + \chi_{Q_{11}}t^{2} + \chi_{q_{22}}t^{2} + \chi_{Q_{22}}t^{2} +\chi_{q_{33}}t^{2} + \chi_{Q_{33}}t^{2}  -\chi_{F_{1}}t^{4} -\chi_{F_{2}}t^{4} - \chi_{F_{3}}t^{4}]\, ,
\end{aligned}\end{equation}
where the contributions of the F-terms and the various fields are given by
\begin{equation*} \ \ \chi_{F_{2}} = \sum_{a=1}^{k_{3}}\sum_{b=1}^{k_{4}}z_{a}^{-1}v_{b}, \ \
\chi_{F_{3}} = \sum_{a=1}^{k_{5}}\sum_{b=1}^{k_{6}}j_{a}^{-1}c_{b}, \ \
\chi_{q_{11}} = \sum_{a=1}^{k_{2}}\sum_{b=1}^{N_{1}}w_{a}y_{b}^{-1}, \ \ \chi_{Q_{11}} = \sum_{a=1}^{N_{1}}\sum_{b=1}^{k_{1}}y_{a}u_{b}^{-1}, \ \
\end{equation*}
\begin{equation*}
\chi_{q_{22}} = \sum_{a=1}^{k_{4}}\sum_{b=1}^{N_{2}}v_{a}d_{b}^{-1}, \ \
\chi_{Q_{22}} = \sum_{a=1}^{N_{2}}\sum_{b=1}^{k_{3}}d_{a}z_{b}^{-1}, \ \
\chi_{q_{33}} = \sum_{a=1}^{k_{6}}\sum_{b=1}^{N_{3}}c_{a}s_{b}^{-1}, \ \ \chi_{Q_{33}} = \sum_{a=1}^{N_{3}}\sum_{b=1}^{k_{5}}s_{a}j_{b}^{-1}\, ,
\end{equation*}
\begin{equation*}
\chi_{A^{2}_{11}} = \sum_{a=1}^{k_{1}}\sum_{b=1}^{k_{2}}u_{a}w_{b}^{-1}, \ \ \chi_{A^{2}_{22}} = \sum_{a=1}^{k_{3}}\sum_{b=1}^{k_{4}}z_{a}v_{b}^{-1}, \ \
\chi_{A^{2}_{33}} = \sum_{a=1}^{k_{5}}\sum_{b=1}^{k_{6}}j_{a}c_{b}^{-1}, \ \ \chi_{B_{12}^{2}} = \sum_{a=1}^{k_{2}}\sum_{b=1}^{k_{3}}w_{a}z_{b}^{-1}, \ \ \chi_{B_{23}^{2}} = \sum_{a=1}^{k_{4}}\sum_{b=1}^{k_{5}}v_{a}j_{b}^{-1},  
\end{equation*}
\begin{equation*}
\chi_{B_{31}^{2}} =\sum_{a=1}^{k_{6}}\sum_{b=1}^{k_{1}}c_{a}u_{b}^{-1},
\ \
\chi_{B^{1}_{21}} = \sum_{a=1}^{k_{4}}\sum_{b=1}^{k_{1}}v_{a}u_{b}^{-1},\\
\chi_{B^{1}_{13}} = \sum_{a=1}^{k_{2}}\sum_{b=1}^{k_{5}}w_{a}j_{b}^{-1},\\
\chi_{B^{1}_{32}} = \sum_{a=1}^{k_{6}}\sum_{b=1}^{k_{3}}c_{a}z_{b}^{-1},
\ \ \chi_{F_{1}} = \sum_{a=1}^{k_{1}}\sum_{b=1}^{k_{2}}u_{a}^{-1}w_{b}\, .
\end{equation*}

As in above, the Hilbert series on the instanton branch of the quiver describing instantons on $\mathbb{C}P^2/\mathbb{Z}_n$ with gauge group of $G=U(k_{1}) \times U(k_{2}) \times U(k_{3}) \times U(k_{4}) \times U(k_{5}) \times U(k_{6})$ and flavor group $U(N_{1}) \times U(N_{2}) \times U(N_{3})$ is equal to the Hilbert series of the Higgs branch describing the moduli space of instantons on $\mathbb{C}^2/\mathbb{Z}_3$ with flavor group $U(N_{1}) \times U(N_{2}) \times U(N_{3})$ instantons and gauge group $\textbf{K}=(K_{1},K_{2},K_{3})$ \cite{Dey:2013fea}, where

\begin{equation}
K_{1} = \textrm{min}(k_{1},k_{2}),  \ \ K_{2}=\textrm{min}(k_{3},k_{4}) \ \ \textrm{and} \ \ K_{3} =\textrm{min}(k_{5},k_{6})\, .
\end{equation}

\begin{figure}[h!]
\centering
\includegraphics[scale=1]{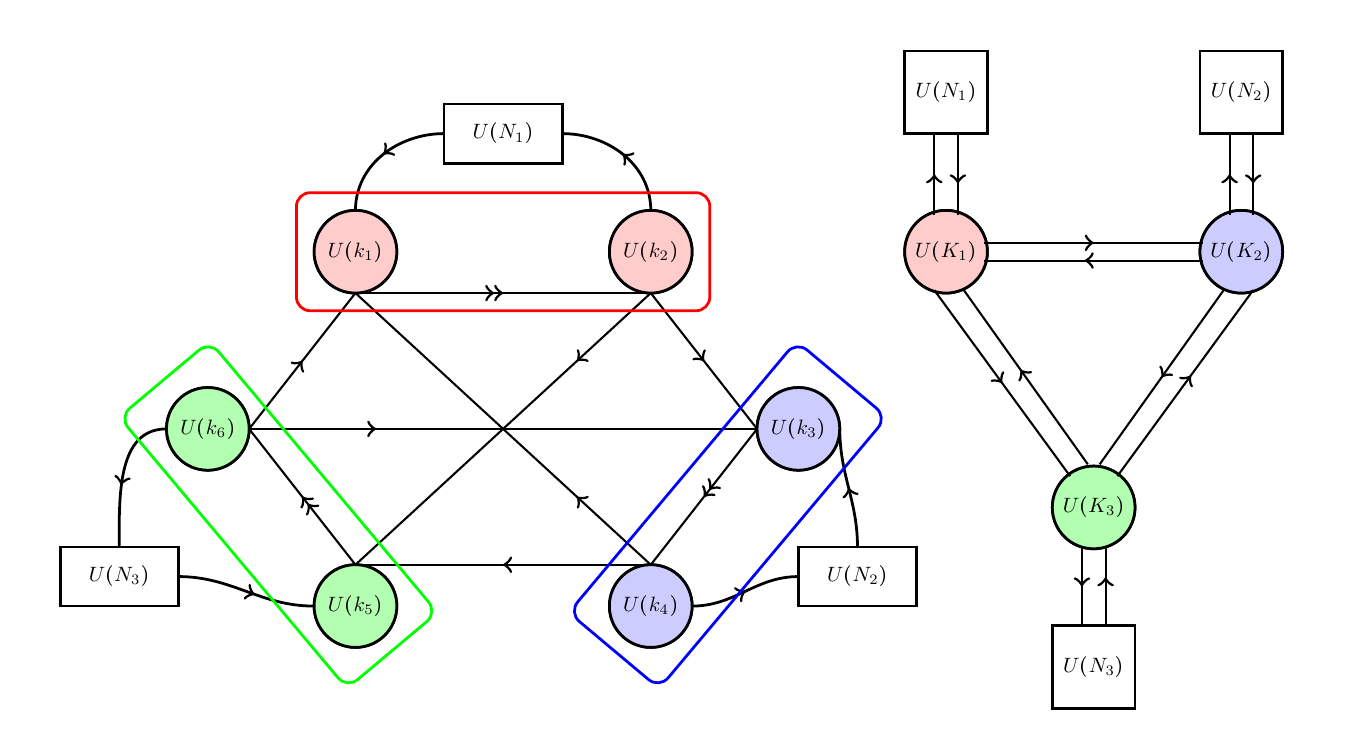}

\caption{Relation between the $\mathbb{C}P^{2}/\mathbb{Z}_{3}$ quiver gauge theory (on the left) and the corresponding $\mathbb{C}^{2}/\mathbb{Z}_{3}$ quiver gauge theory (on the right). \label{fig:comparisonz3}}

\end{figure}

We can again summarize graphically the relation between the theory describing $\mathbb{C}P^2/\mathbb{Z}_3$ instantons and its $\mathbb{C}^2/\mathbb{Z}_3$ cousin as in fig.\ref{fig:comparisonz3}. As in the $\mathbb{Z}_2$ orbifold case, each flavor node flavors a pair of gauge nodes which ``merge" into a single node in the cousin $\mathbb{C}^2/\mathbb{Z}_3$ theory.

Let us support our claim with explicit examples.
\\
\\
\textbf{$U(1)$ instanton: $\textbf{k}=(1,1,1,1,1,1)$, and $N=(1,0,0)$}\\
\\
Using eq.(\ref{hsz3}) we find that
\\
\begin{equation*}
 H[(1,1,1,1,1,1),(1,0,0), \mathbb{C}P^{2}/\mathbb{Z}_{3}](t)
 = \frac{1-t^{3}+t^{6}}{(1-t^{3})^{2}(1+t^{3}+t^{6})}\, ,
\end{equation*}
which is the Hilbert series for $\textbf{N}=(1,0,0)$ instantons and $\textbf{K}=(1,1,1)$ on $\mathbb{C}^{2}/\mathbb{Z}_{3}$.\\
\\
\textbf{$U(2)$ instanton: $\textbf{k}=(1,1,1,1,1,1)$ and $\textbf{N}=(1,1,0)$}\\
\\
Using eq.(\ref{hsz3}) and unrefining we find that 
\begin{equation*}
 H[(1,1,1,1,1,1),(1,1,0), \mathbb{C}P^{2}/\mathbb{Z}_{3}](t,1,1)
 = \frac{1 + t^6 + 2 t^9 + 2 t^{12} + 
 2 t^{15} + t^{18} + t^{24}}{(1 - t^3)^4 (1 + t^3)^2 (1 + t^6) (1 + t^3 + 
   t^6)^2}\, ,
\end{equation*}
which is the unrefined Hilbert series for $\textbf{N}=(1,1,0)$ instantons and $\textbf{K}=(1,1,1)$ on $\mathbb{C}^{2}/\mathbb{Z}_{3}$.\\
\\
\textbf{$U(1)$ instanton: $\textbf{k}=(2,1,1,1,1,1)$ and $\textbf{N}=(1,0,0)$}\\
\\
Using eq.(\ref{hsz3}) we find that
\begin{equation*}
H[(2,1,1,1,1,1),(1,0,0), \mathbb{C}P^{2}/\mathbb{Z}_{3}](t) = \frac{1-t^{3}+t^{6}}{(1-t^{3})^{2}(1+t^{3}+t^{6})}\, ,
\end{equation*}
which is again the Hilbert series for $\textbf{N}=(1,0,0)$ instantons and $\textbf{K}=(1,1,1)$ on $\mathbb{C}^{2}/\mathbb{Z}_{3}$.\\
\\
\textbf{$U(1)$ instanton: $\textbf{k}=(2,1,2,1,1,1)$, and $\textbf{N}=(1,0,0)$}\\
\\
Using eq.(\ref{hsz3}) we find that
\begin{equation*}
H[(2,1,2,1,1,1),(1,0,0), \mathbb{C}P^{2}/\mathbb{Z}_{3}](t) = \frac{1-t^{3}+t^{6}}{(1-t^{3})^{2}(1+t^{3}+t^{6})}\, ,
\end{equation*}
which is again the Hilbert series for $\textbf{N}=(1,0,0)$ instantons and $\textbf{K}=(1,1,1)$ on $\mathbb{C}^{2}/\mathbb{Z}_{3}$.\\
\\
\textbf{$U(2)$ instanton: $\textbf{k}=(2,1,1,1,1,1)$, and  $\textbf{N}=(2,0,0)$}\\
\\
Using eq.(\ref{hsz3}) we find that
\begin{equation*}
H[(2,1,1,1,1,1),(2,0,0), \mathbb{C}P^{2}/\mathbb{Z}_{3}](t,y_{1},y_{2}) = \frac{(1 - t^3 + 2 t^6 - t^9 + t^{12}) y_{1} y_{2}}{(1 - t^3)^2 (1 + t^3 + 
   t^6) (t^6 y_{1} - y_{2}) (t^6 y_{2} -y_{1})}\, ,
\end{equation*}
being $y_{1}$ and $y_{2}$ the fugacities of the flavor group $U(2)$. The previous expression is the Hilbert series for $\textbf{N}=(2,0,0)$ instantons and $\textbf{K}=(1,1,1)$ on $\mathbb{C}^{2}/\mathbb{Z}_{3}$.\\
\\
\textbf{$U(2)$ instanton: $\textbf{k}=(2,2,1,1,1,1)$, and $\textbf{N}=(2,0,0)$}\\
\\
Using eq.(\ref{hsz3}) and unrefining we find that
\begin{equation*}\begin{aligned}
& H[(2,2,1,1,1,1),(2,0,0), \mathbb{C}P^{2}/\mathbb{Z}_{3}](t,1,1) = \\ 
& = \frac{1 - t^3 + 2 t^6 - t^9 + 3 t^{12} + 2 t^{15} - t^{18} - t^{21} - 5 t^{27} + 
 2 t^{30} - 5 t^{33} - t^{39} - t^{42} + 2 t^{45} + 3 t^{48} - t^{51} + 
 2 t^{54} - t^{57} + t^{60}}{(1 - t^3)^4 (1 + t^3)^2 (1 + t^3 + t^6)(1 -
    t^{12})^2 (1 - t^{15})^2}\, ,
\end{aligned}\end{equation*}
which is the Hilbert series for $\textbf{N}=(2,0,0)$ instantons and $\textbf{K}=(2,1,1)$ on $\mathbb{C}^{2}/\mathbb{Z}_{3}$.\\
\\ 
\textbf{$U(2)$ instanton: $\textbf{k}=(2,1,2,1,1,1)$, and $\textbf{N}=(2,0,0)$}\\
\\
Using eq.(\ref{hsz3}) we find that
\begin{equation*}
H[(2,1,1,1,1,1),(2,0,0), \mathbb{C}P^{2}/\mathbb{Z}_{3}](t,y_{1},y_{2}) = \frac{(1 - t^3 + 2 t^6 - t^9 + t^{12}) y_{1} y_{2}}{(1 - t^3)^2 (1 + t^3 + 
   t^6) (t^6 y_{1} - y_{2}) (t^6 y_{2} -y_{1})}\, ,
\end{equation*}
being $y_{1}$ and $y_{2}$ the fugacities of the flavor group $U(2)$. The previous expression is the Hilbert series for $\textbf{N}=(2,0,0)$ instantons and $\textbf{K}=(1,1,1)$ on $\mathbb{C}^{2}/\mathbb{Z}_{3}$.

\subsubsection{The $\mathbb{C}P^{2}/\mathbb{Z}_{n}$ case ($n \geqslant 3$)}

\begin{figure}[h!]
\centering
\includegraphics[scale=1]{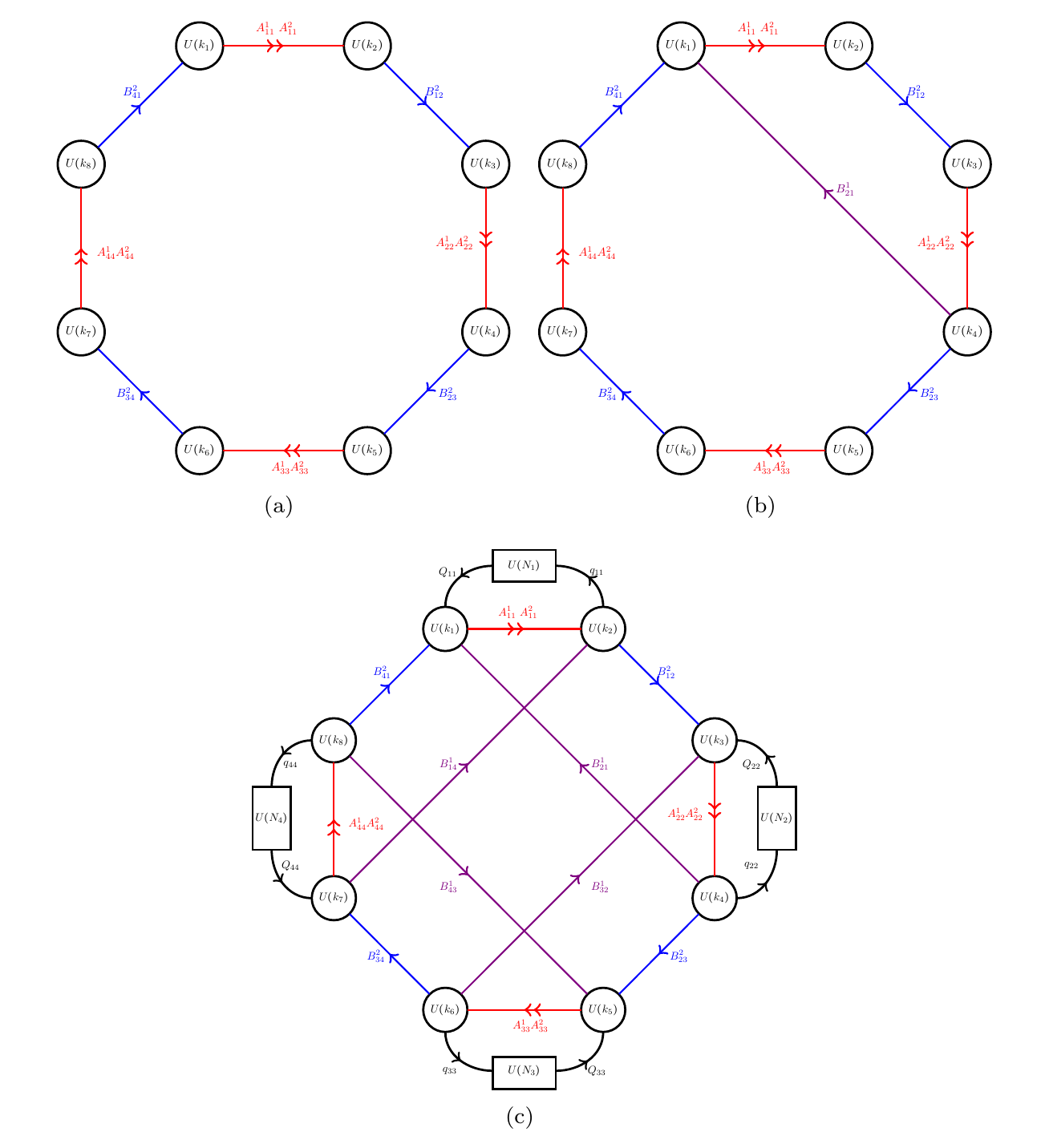}

\caption{Construction of the quiver diagram for the $\mathbb{C}P^{2}/\mathbb{Z}_{4}$ theory. \label{fig:quiverzn}}

\end{figure}

It is now easy to generalize the previous construction of $U(N)$ instantons to higher orbifolds of $\mathbb{C}P^2$. For a general $\mathbb{Z}_n$ orbifold, the resulting procedure is as follows (see fig.\ref{fig:quiverzn})

\begin{itemize}
\item The quiver has $2n$ circular nodes linked together in alternating way, i.e. a segment with fields $A^{1}_{ii}$ and $A^{2}_{ii}$ is alternated to a segment with field $B^{2}_{i,i+1}$ (see fig.\ref{fig:quiverzn} (a)).

\item Then we add the contribution due to the fields $B^{1}_{i+1,i}$. In order to do this we begin from one circular node (for example the one in which there is the gauge group $U(k_{1})$) and we move in clockwise sense counting three segments (in this case we will count the segment labelled by $A^{1}_{11}$, the segment labelled by $B^{2}_{12}$ and finally the segment labelled by $A^{1}_{22}$). When we reach the circular node at the end of the third segment we draw a line between this node and the initial circular node (in this case a line between the node $U(k_{4})$ and the initial node $U(k_{1})$). This line we will labelled by a $B^{1}_{i+1,i}$ field (in the case we are considering by the field $B^{1}_{2,1}$) (see fig.\ref{fig:quiverzn} (b)).

\item We apply the same procedure starting, this time, from the next circular node arising from the first gauge group $U(k_L)$ (in this case the one labelled by $U(k_{3})$) and we will continue to apply this algorithm up the end of circular nodes arising from the decomposition of the first gauge group. Finally we add the contributions due to the various flavour groups, and we obtain the quiver reported in fig.\ref{fig:quiverzn} (c).
 
\end{itemize}

Note that $N$ corresponds to the sum of the ranks of the flavor nodes. In turn, gauge ranks correspond to the instanton number as well as, together with relative flavor ranks, other quantum numbers describing the instanton (we will briefly come back to these issues below).

We can compute the Hilbert series on the instanton branch. In general, we find a correspondence between the Hilbert series for the moduli space of $\textbf{N}=(N_{1},...,N_{n})$ instantons with $\textbf{k}=(k_{1},k_{2},...,k_{2n})$ on $\mathbb{C}P^{2}/\mathbb{Z}_{n}$ and the Hilbert series for the moduli space of $\textbf{N}=(N_{1},...N_{n})$ instantons with $\textbf{K}=(K_{1},...,K_{n})$ on $\mathbb{C}^{2}/\mathbb{Z}_{n}$ upon identifying
\begin{equation}
K_{1}=\textrm{min}(k_{1},k_{2}), \ \ K_{2}=\textrm{min}(k_{3},k_{4}), \ \ ... \ \ K_{n}=\textrm{min}(k_{2n-1},k_{2n})\, .
\end{equation}
This can be easily proven in the particular case
\begin{equation*}
G = \bigotimes_{i=1}^{2n} U(1)_{i}, \ \ \ F = \bigotimes_{i=1}^{n} U(N_{i})\, .
\end{equation*} 
Moreover we denote with $z_{i}$ $i=1,...,2n$ the fugacities of the various $U(1)_{i}$ gauge groups and with $u_{i}$ and $\vec{y}_{i}$ the fugacities of each flavour group $U(N_{i})$ (being $u_{i}$ the fugacity of the $U(1)$ part while $\vec{y}_{i}$ are the fugacities associated with the $SU(N)$ part of the flavour group). 

The Hilbert series reads
\begin{equation}
\label{HSn}
\begin{aligned}
& H[(1,1,...,1),(N_{1},N_{2},...,N_{n}),\mathbb{C}P^{2}/\mathbb{Z}_{n}](t,u_{i},\vec{y}_{i}) = \prod_{i=1}^{2n} \frac{1}{2 \pi i} \oint_{\mid z_{i} \mid}\frac{dz_{i}}{z_{i}}\prod_{j=1}^{n} \chi_{A^{2}_{j,j}}(t,z_{2j-1},z_{2j}) \times\\ & \chi_{B^{2}_{j,j+1}}(t,z_{2j},z_{2j+1})
\chi_{B^{1}_{j+1,j}}(t,z_{2j},z_{2j-1})\chi_{F_{j}}(t,z_{2j-1},z_{2j})\chi_{q_{j,j}}(t,z_{2j},\vec{y_{j}},u_{j})\chi_{Q_{j,j}}(t,z_{2j-1},\vec{y_{j}},u_{j})\, .
\end{aligned}
\end{equation}

\begin{figure}
\centering
\includegraphics[scale=1]{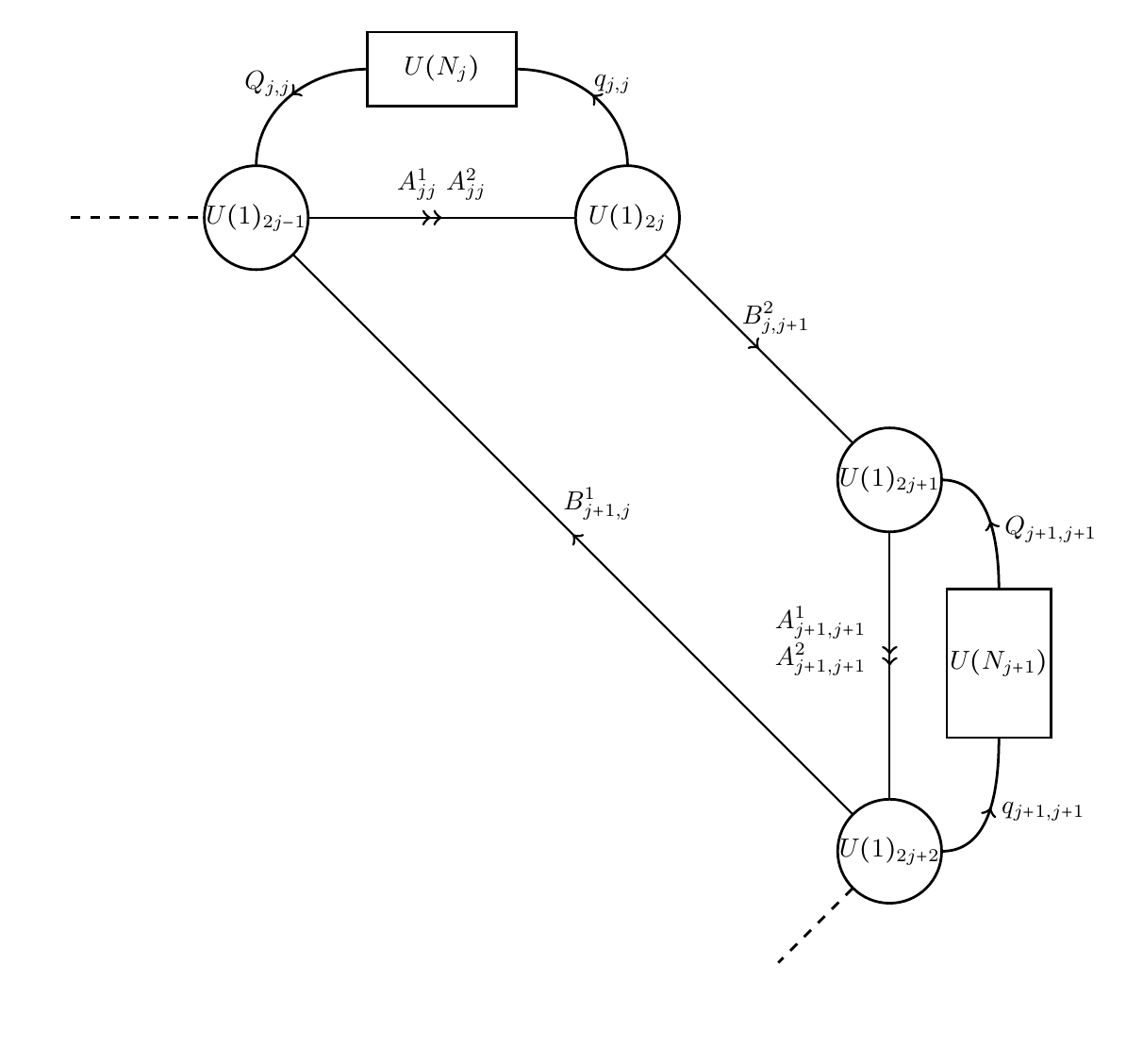}

\caption{Basic element of the quiver diagram for the  $\mathbb{C}P^{2}/\mathbb{Z}_{n}$ theory. \label{fig:pezzo}} 
\end{figure}

The contributions of the various fields are \footnote{See fig. \ref{fig:pezzo}.}
\begin{equation*}
\chi_{A^{2}_{j,j}}(t,z_{2j-1},z_{2j}) = \textrm{PE}[t^{2}z_{2j-1}z_{2j}^{-1}], \ \ \chi_{B^{2}_{j,j+1}}(t,z_{2j},z_{2j+1}) = \textrm{PE}[tz_{2j}z_{2j+1}^{-1}]\, ,
\end{equation*}
\begin{equation*}
\chi_{B^{1}_{j+1,j}}(t,z_{2j+2},z_{2j-1}) = \textrm{PE}[tz_{2j+2}z_{2j-1}^{-1}], \ \ \chi_{F_{j}}(t,z_{2j-1},z_{2j}) = \textrm{PE}[-t^{4}z_{2j-1}^{-1}z_{2j}]\, ,
\end{equation*}
\begin{equation*}
\chi_{Q_{j,j}}(t,z_{2j-1},\vec{y_{j}},u_{j})= \textrm{PE}[t^2z_{2j-1}^{-1}[1,0,...,0]_{\vec{y_{j}}}u_{j}], \ \ \chi_{q_{j,j}}(t,z_{2j},\vec{y_{j}},u_{j}) = \textrm{PE}[t^2z_{2j}[0,...,0,1]_{\vec{y_{j}}}u_{j}^{-1}]\, .
\end{equation*}

Therefore the Hilbert series (\ref{HSn}) becomes
\begin{equation*}
\prod_{i=1}^{2n} \frac{1}{2 \pi i} \oint_{\mid z_{i} \mid}\frac{dz_{i}}{z_{i}}\prod_{j=1}^{n}
\frac{\textrm{PE}[t^2z_{2j-1}^{-1}[1,0,...,0]_{\vec{y_{j}}}u_{j} + t^2z_{2j}[0,...,0,1]_{\vec{y_{j}}}u_{j}^{-1}](z_{2j-1}-t^{4}z_{2j})}{\frac{z_{2j-1}}{z_{2j}}(z_{2j}-t^{2}z_{2j-1})\left(1-\frac{tz_{2j}}{z_{2j+1}}\right)\left(1-\frac{tz_{2j+2}}{z_{2j-1}}\right)}\, .
\end{equation*}

It is important to note that we can integrate over the gauge group $U(1)_{i}$  with an even value of the index $i$. This is due to the fact that the only contribution to these integrals come from the poles located at $z_{2j}=t^{2}z_{2j-1}$. Therefore, performing the integrations, we obtain
\begin{equation*}
\prod_{i \ \textrm{odd}}^{2n} \frac{1}{2 \pi i} \oint_{\mid z_{i} \mid}\frac{dz_{i}}{z_{i}}\prod_{j=1}^{n}
\frac{\textrm{PE}[t^2z_{2j-1}^{-1}[1,0,...,0]_{\vec{y_{j}}}u_{j} + t^4z_{2j-1}[0,...,0,1]_{\vec{y_{j}}}u_{j}^{-1}](z_{2j-1}-t^{6}z_{2j-1})}{z_{2j-1}\left(1-\frac{t^3z_{2j-1}}{z_{2j+1}}\right)\left(1-\frac{t^3z_{2j+1}}{z_{2j-1}}\right)}\, ,
\end{equation*}
then we perform the change of variables $ z_{2j-1} \mapsto tz_{2j-1} $ 
\begin{equation*}
\prod_{i \ \textrm{odd}}^{2n} \frac{1}{2 \pi i} \oint_{\mid z_{i} \mid}\frac{dz_{i}}{z_{i}}\prod_{j=1}^{n}
\frac{\textrm{PE}[t^3z_{2j-1}^{-1}[1,0,...,0]_{\vec{y_{j}}}u_{j} + t^3z_{2j-1}[0,...,0,1]_{\vec{y_{j}}}u_{j}^{-1}](1-t^{6})}{\left(1-\frac{t^3z_{2j-1}}{z_{2j+1}}\right)\left(1-\frac{t^3z_{2j+1}}{z_{2j-1}}\right)}\, ,
\end{equation*}
finally we observe that, instead of consider only the odd numbers between 1 and $2n$, it is more useful to consider all the integers numbers between 1 and $n$. Therefore we can make the following replacements  $z_{2j-1} \mapsto z_{j}$   and $z_{2j+1} \mapsto z_{j+1}$ and we rewrite the previous integral as
\begin{equation*}
\prod_{i=1}^{n} \frac{1}{2 \pi i} \oint_{\mid z_{i} \mid}\frac{dz_{i}}{z_{i}}(1-t^{6})^n\prod_{j=1}^{n}
\textrm{PE}[t^3z_{j}^{-1}[1,0,...,0]_{\vec{y_{j}}}u_{j} + t^3z_{j}[0,...,0,1]_{\vec{y_{j}}}u_{j}^{-1}]\textrm{PE}[t^{3}z_{j}z_{j+1}^{-1}+t^{3}z_{j+1}z_{j}^{-1}]\, ,
\end{equation*}
which is the Hilbert series  for $\textbf{N}=(N_{1},N_{2},...,N_{n})$ instantons with $\textbf{K}=(1,1,...,1)$ on $\mathbb{C}^{2}/\mathbb{Z}_{n}$ (it coincides with the expression (2.41) of \cite{Dey:2013fea}).

Up to now we have deliberately postponed discussing the identification of the quantum numbers of the instanton. Recall that in the $\mathbb{C}^2/\mathbb{Z}_n$ case \cite{Dey:2013fea} the instanton is described by $n-1$ first Chern classes, one second Chern class and $n$ holonomies of the gauge field, all in all a total of $2\,n$ quantum numbers corresponding to the $2\,n$ integers specifying the $A_{n-1}$ quiver. 

In the case at hand, the quiver describing instantons on $\mathbb{C}P^2/\mathbb{Z}_n$ is specified by a total of $3\,n$ integers, corresponding to $2\,n$ gauge ranks and $n$ flavor ranks. In turn, we expect the instanton on $\mathbb{C}P^2/\mathbb{Z}_n$ to be described by $2\,n-1$ first Chern classes --corresponding to $n$ orbifold copies of the $\mathbb{C}P^2$ 2-cycle plus $n-1$ extra 2-cycles introduced by the orbifold--, one second Chern class and $n$ holonomies, hence totalling the expected $3\,n$ quantum numbers. While the exact identification of integers is not known, note that, from the examples above, the mapping of the $\mathbb{C}P^2/\mathbb{Z}_n$ quiver into the $\mathbb{C}^2/\mathbb{Z}_n$ one is such that one node of the latter arises from the ``merging" of two adjacent commonly flavored nodes of the former; in such a way that the common flavor group in the $\mathbb{C}P^2/\mathbb{Z}_n$ case becomes the flavor group in the $\mathbb{C}^2/\mathbb{Z}_n$ case. Hence it is natural to guess that the $n$ holonomies correspond to the $n$ flavor nodes. Moreover, the $n-1$ first Chern classes associated to the cycles arising from the orbifold are naturally associated to the diferences among the minima of the ranks of each pair of ``merging nodes". Obviously there are $n$ such nodes arising from ``merging", whose $n-1$ rank differences would correspond to first Chern classes. In turn, the relative rank between the ``merging nodes" is naturally associated with the $n$ remaining 2-cycles, orbifold copies of the original 2-cycle in $\mathbb{C}P^2$. Finally, the sum of ranks is naturally related to the second Chern class. Note that clearly, the identification of $N$ with the sum of the ranks of the flavor nodes is consistent.

As a small consistency check, let us consider the simple case of vanishing first Chern class associated to cycles introduced by the orbifold. This would correspond to a rank assignation of the form $(\cdots,\,k,\,q_n,\,k,\,q_{n+1},\,k,\,\cdots)$ with $q_i>k$, so that, among each ``merging pair" the minimum rank is $k$. Then all relative rank differences among the ``merged nodes" are 0 corresponding to a $\mathbb{C}^2/\mathbb{Z}_n$ instanton with zero first Chern classes. Moreover, let us consider the case of vanishing second Chern class from the $\mathbb{C}^2/\mathbb{Z}_n$ point of view, which demands $k=0$. This is analogous to the case $k_L=0$ in section \ref{Grassmanian}. We are then left with a gauge rank assignation of the form $(\cdots,\,0,\,q_n,\,0,\,q_{n+1},\,0,\,\cdots)$. According to our conjecture, these integers $q_i$ should correspond the first Chern classes on the $n$ 2-cycles coming from the orbifold images of the original 2-cycle. Indeed, if we consider just one of them, that is, we set all but one of the $q_i$'s to vanish, we simply recover the Grassmanian quiver above. Note that, as expected, indeed we have $n$ such possibilities corresponding to the $n$ 2-cycles coming from the orbifold images of the original 2-cycle.

\section{$Sp(N)$ instantons on $\mathbb{C}P^2/\mathbb{Z}_n$ }
\label{sec:sp}

\begin{figure}[h!]
\centering
\includegraphics[scale=1.25]{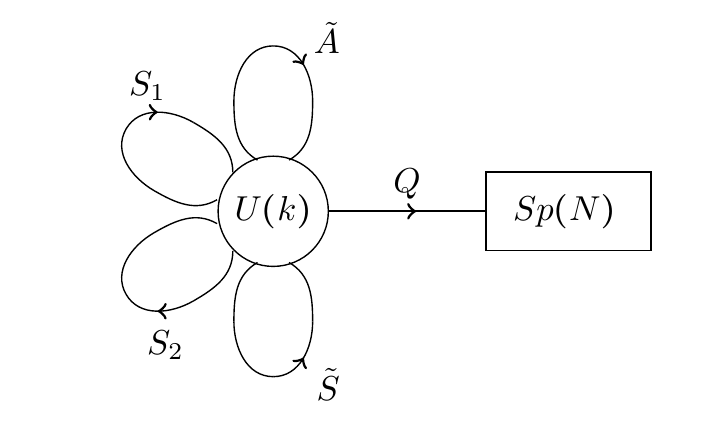}
\caption{Quiver diagram for $Sp(N)$ instantons on $\mathbb{C}P^{2}$.\label{SpADHM}}

\end{figure}

So far we have concentrated on the case of unitary instantons. Let us now turn to the case of instantons in the symplectic gauge group. The explicit ADHM construction of such instantons was introduced in \cite{Sanders}. As described in \cite{Mekareeya:2014kca}, it can be embedded into a $3d$ gauge theory upon restricting to the appropriate instanton branch. In $3d$ $\mathcal{N}=2$ notation, such theory contains one $U(k)$ vector multiplet coupled to one chiral multiplet $\tilde{A}$ in the second rank antisymmetric tensor representation of the gauge group and three chiral multiplets $S_1,\,S_2,\,\tilde{S}$ in the second rank symmetric tensor representation. In addition, there are a number of chiral multiplets in the fundamental representation with an $Sp(N)$ global symmetry. The corresponding quiver is reported in fig.\ref{SpADHM}.

In turn, the superpotential is

\begin{equation}
W=\epsilon^{\alpha\beta}\,(S_{\alpha})_{ab}\,\tilde{S}^{bc}\,(S_{\beta})_{cd}\,\tilde{A}^{da}+\tilde{A}^{ab}\,Q^i\,_a\,Q^j\,_b\,J_{ij}\, ,
\end{equation}
being $J$ the $Sp(N)$ symplectic matrix. As shown in \cite{Mekareeya:2014kca}, the instanton branch emerges upon setting $\tilde{A}$ --as well as the monopole operators-- to zero. 

As in the unitary case, it is possible to embed the $\mathbb{C}P^2$ symplectic instantons ADHM construction into the $\mathbb{C}^2$ symplectic ADHM construction and \textit{vice-versa} \cite{Mekareeya:2014kca}. It should be noted though that now the equivalent to the map $\pi$ in eq.(\ref{mappi}) is quadratic and hence does not define a proper mapping. Nevertheless, as a consequence, the Hilbert series for symplectic instantons on $\mathbb{C}P^2$ coincides with that of symplectic instantons on $\mathbb{C}^2$. We refere to \cite{Mekareeya:2014kca} for further details.

\subsection{Constructing $Sp(N)$ instantons on $\mathbb{C}P^2/\mathbb{Z}_n$ }

Just as in the case of unitary instantons, we can consider orbifolding the base $\mathbb{C}P^2$ manifold and study $Sp(N)$ instantons on $\mathbb{C}P^{2}/\mathbb{Z}_{n}$. It is then natural to engineer the ADHM-like construction by orbifolding the $\mathbb{C}P^2$ case, just as for unitary instantons. As guideline, let us compare with the case of instantons on $\mathbb{C}^2$ and its orbifolds \cite{Dey:2013fea}. The gauge theory realizing the ADHM construction for unitary instantons on $\mathbb{C}^2/\mathbb{Z}_n$ can be thought as the worldvolume theory on a D3-D7 system, where the transverse directions to the D3's inside the D7's wrap $\mathbb{C}^2/\mathbb{Z}_n$. Then, symplectic (and orthogonal) instantons can be constructed upon adding O7 planes of the appropriate charge. A comprehensive picture appears upon T-duality along the ALE space. Then, the D3 branes are mapped to D4 branes wrapping a circle. In turn, the D7 are mapped into D6 at fixed positions in the circle. Finally, $n$  NS5 branes on the circle arise from T-dualizing the ALE space. In this context, the construction of symplectic (alternatively orthogonal) instantons boils down to adding 2 identical --because they come from T-duality of a single O7-- O6 planes of the appropriate charge at opposite points in the circle such that each side of the circle mirrors --due to the orientifold projection-- the other side. This procedure highlights an obvious difference between the cases of even and odd orbifolds. As the distribution of NS5 branes must be symmetric on the circle, for an odd $n$ it is clear that one such NS5 must be stuck in an orientifold plane. In turn, in the case of even $n$ we can have a symmetric distribution by either sticking one NS5 at each O-plane or not sticking any NS5 on the O-planes. These possibilities lead, respectively, to the so-called no-vector-structure (NVS) and vector structure (VS) respectively. We refer to \cite{Dey:2013fea} and references therein for further explanations. Note that the T-duality construction suggests that the two O-planes are of the same type. Nevertheless, once in the IIA set-up, one might imagine other versions whereby the O-planes are of different type. These configurations were dubbed hybrid in \cite{Dey:2013fea}. We will briefly touch on the equivalent to these in the case at hand below, showing an explicit example in appendix \ref{app:A}.

In view of the $\mathbb{C}^2/\mathbb{Z}_n$ case, it is natural to proceed in a similar way in the case of instantons on the orbifolded $\mathbb{C}P^2$, that is, first consider orbifolding unitary instantons and then considering orientifolding. Note however that in this case the brane picture is much less clear. Nevertheless, as we will see, the results are qualitatively similar. Since we will set monopole operators to zero, formally the procedure is identical to the case of $4d$ gauge theories. Hence, we can borrow the  technology developed \cite{Franco:2007ii} and \cite{GarciaEtxebarria:2012qx} to construct the relevant theories.

As illustrated in \cite{Franco:2007ii} the orientifold field theory is obtained from the parent field theory performing a $\mathbb{Z}_{2}$ identification of the gauge groups, chiral multiplets and superpotential couplings. As explained in \cite{GarciaEtxebarria:2012qx}, this means that the O-plane involution defines a $\mathbb{Z}_{2}$ automorphism of the quiver diagram that reverses the directions of the arrows. Therefore the quiver of the parent theory has a $\mathbb{Z}_{2}$ symmetry, that can be visualized as a reflection through a fixed line once we embed the quiver diagram in $\mathbb{R}^{2}$. In the following we will follow the method used in \cite{GarciaEtxebarria:2012qx} that allows to obtain the orientifold theory starting directly from its quiver diagram. Of course, as can be verified, the application of the method of \cite{Franco:2007ii}, that acts on the dimer diagram of the theory, leads to the same results.

\begin{figure}[h!]\centering
\includegraphics{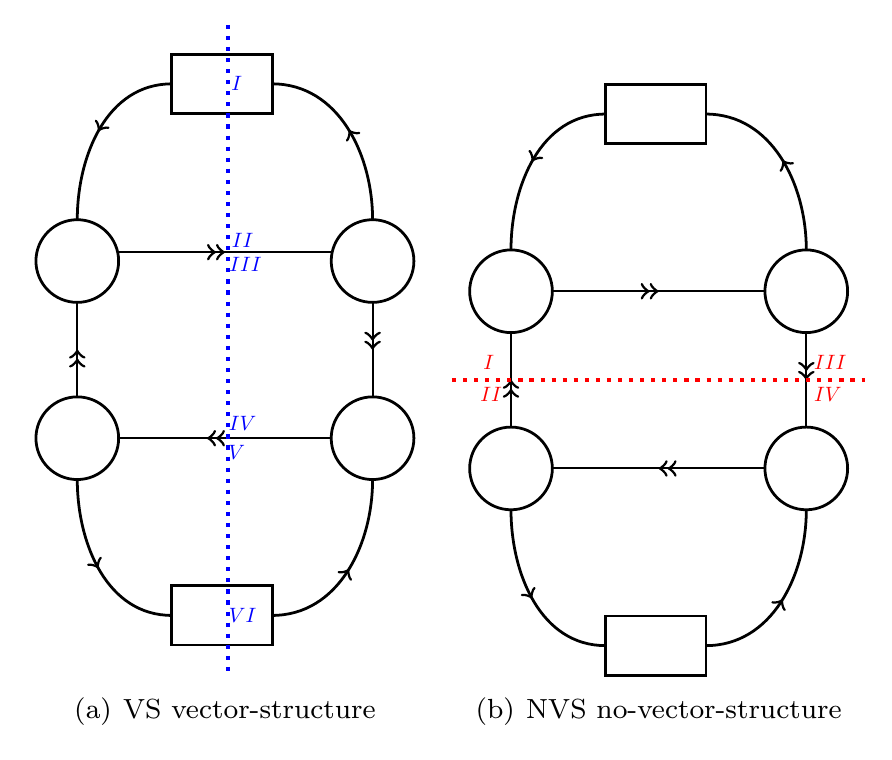}
\caption{The two inequivalent ways to obtain the $\mathbb{C}P^{2}/\mathbb{Z}_{2}$ orientifold theory. \label{fig:orientifolds}}

\end{figure}

In order to explain how this procedure works we apply it to the case of the $\mathbb{C}P^{2}/\mathbb{Z}_{2}$ theory and we refer to \cite{GarciaEtxebarria:2012qx} for the analysis of the general case. An inspection of the corresponding quiver diagram shows that there are two inequivalent ways to cut it with a line, such that the quiver displays  an arrows reversing symmetry with respect to this line (see fig.\ref{fig:orientifolds}).

In order to obtain the corresponding orientifold theory we label each node and each line intersecting perpendicularly the cutting line with a sign (denoted with a Roman number in the figure), that can be positive or negative. Then, the orientifold theory is constructed as follows. Each node untouched by the cutting line corresponds to a $U(k)$ group while each node touched by the line corresponds to a $SO(N)$ or $Sp(N)$ (for a positive or negative sign respectively) in the orientifold field theory. In the same way each edge of the quiver diagram away from the cutting line corresponds to bifundamental matter, while each edge crossing the cutting line perpendicularly corresponds to symmetric matter (positive sign) or antisymmetric matter (negative sign) in the orientifold field theory.
The values of the signs must be fixed requiring that the superpotential of the parent theory is invariant under the involution. 
Note that in general more than one choice is allowed. For example in the case of the quiver diagram in fig.\ref{fig:orientifolds} (b) we can choose the following values of the signs $(+,+,+,+), (-,+,+,-), (+,-,+,-), (+,+,-,-)$. In the following we will always fix the signs in order to obtain the theory whose Higgs branch describes the moduli space for $Sp(N)$ instantons (respectively $SO$) on $\mathbb{C}P^{2}/\mathbb{Z}_{n}$, which in the case at the hand means to select the $(+,+,+,+)$ configuration. The remaining allowed choices correspond to the ``hybrid configurations'' discussed in \cite{Dey:2013fea}. Even though we will not touch upon these further in this paper, we present an explicit example in appendix \ref{app:A}.

Therefore, as in \cite{Dey:2013fea} we have two different situations depending on whether the degree of the orbifold is even or odd.

\begin{itemize}
\item If \textbf{$n$ is odd} we have only one type of quiver diagram, corresponding to the fact that we have only one inequivalent way to cut it with a line.
\item If \textbf{$n$ is even} we have two types of quiver gauge theories corresponding to the two possible inequivalent ways to cut it with a fixed line. These two cases are just the equivalent of the vector-structure and no-vector-structure cases for $\mathbb{C}^{2}/\mathbb{Z}_n$ symplectic instantons. By analogy, in the following we will refer to theme as the \textit{VS} and the \textit{NVS} respectively.
\end{itemize}

Note that $N$ corresponds to the sum of the ranks of the flavor groups in the ADHM quiver. In turn, gauge group ranks correspond to instanton number (as well as to other possible quantum numbers labelling the instanton).

\subsubsection{$Sp(N)$ instantons on $\mathbb{C}P^{2}/\mathbb{Z}_{2}$ \ - \textit{VS}}

\begin{figure}[h!]
\centering
\includegraphics[scale=1]{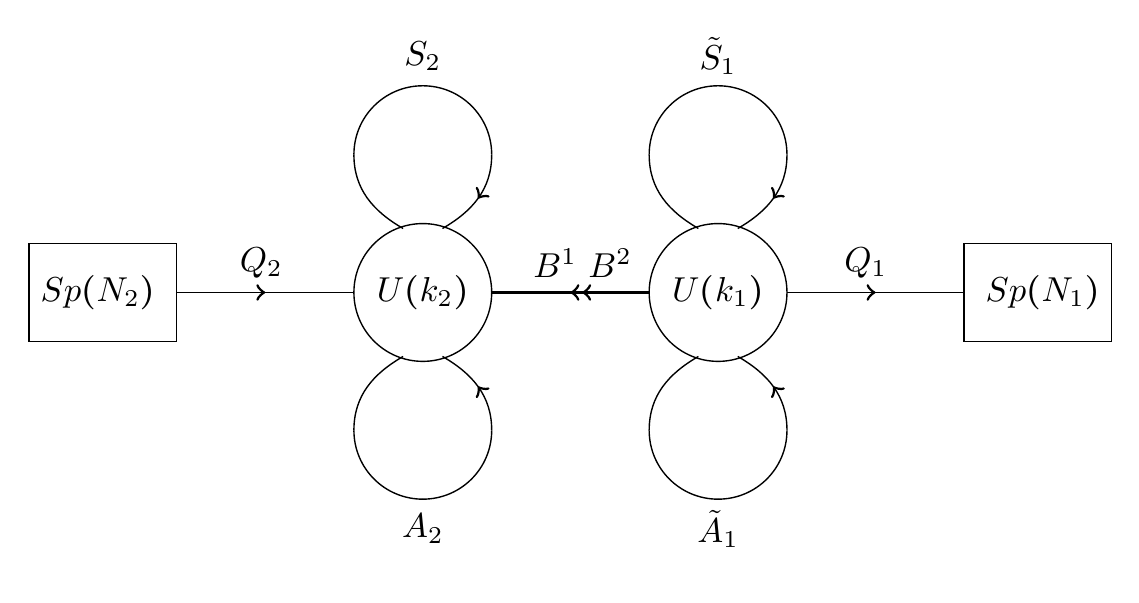}
\caption{Quiver diagram for VS symplectic instantons on $\mathbb{C}P^2/\mathbb{Z}_2$. \label{fig:spz2first}}
        
\end{figure}

Starting from the $\mathbb{C}P^{2}/\mathbb{Z}_{2}$ and applying the rules above we can obtain the VS theory for $Sp(N)$ instantons on $\mathbb{C}P^{2}/\mathbb{Z}_{2}$. The corresponding quiver diagram is reported in fig.\ref{fig:spz2first}, while we summarize the transformations of the fields under the different groups in table \ref{tab:tabusp}. Note that $N=N_1+N_2$.

\begin{table}[h!]
 \renewcommand\arraystretch{1.3} 

\centering
\resizebox{12cm}{!}{
\begin{tabular}{c||cc||cccc||} 
Fields & $U(k_{1})$ & $U(k_{2})$ & $Sp(N_{1})$ & $Sp(N_{2})$ & $SU(2)$ & $U(1)$ \\ 
\hline 
$\tilde{A}_{1}$ & $[0,1,0...,0]_{-2}$ & $[\textbf{0}]$ & $[\textbf{0}]$ & $[\textbf{0}]$ & $[0]$ & 1/2 \\ 
\hline 
$\tilde{S}_{1}$ & $[2,0,...,0]_{-2}$ & $[\textbf{0}]$ & $[\textbf{0}]$ & $[\textbf{0}]$ & $[0]$ & 1/2 \\ 
\hline 
$A_{2}$ & $[\textbf{0}]$ & $[0,1,0,...,0]_{+2}$ & $[\textbf{0}]$ & $[\textbf{0}]$ & $[0]$ & 1/2 \\ 
\hline 
$S_{2}$ & $[\textbf{0}]$ & $[2,0,...,0]_{+2}$ & $[\textbf{0}]$ & $[\textbf{0}]$ & $[0]$ & 1/2 \\ 
\hline 
$B^{1},B^{2}$ & $[1,0,...,0]_{+1}$ & $[0,...,0,1]_{+1}$ & $[\textbf{0}]$ & $[\textbf{0}]$ & $[1]$ & 1/4 \\ 
\hline 
$Q_{1}$ & $[1,0,...,0]_{+1}$ & $[\textbf{0}]$ & $[1,0,...,0]$ & $[\textbf{0}]$ & $[0]$ & 1/2 \\ 
\hline 
$Q_{2}$ & $[\textbf{0}]$ & $[0,...,0,1]_{+1}$ & $[\textbf{0}]$ & $[1,0,...,0]$ & $[0]$ & 1/2 \\ 
\hline 
$F_{1}$ & $[0,1,...,0]_{+2}$ & $[\textbf{0}]$ & $[\textbf{0}]$ & $[\textbf{0}]$ & $[0]$ & 1 \\ 
\hline 
$F_{2}$ & $[\textbf{0}]$ & $[0,1,...,0]_{-2}$ & $[\textbf{0}]$ & $[\textbf{0}]$ & $[0]$ & 1 \\ 
\end{tabular} 
}

\caption{Transformations of the fields for VS symplectic instantons on $\mathbb{C}P^{2}/\mathbb{Z}_2$. \label{tab:tabusp}}
\end{table}

The branch of the moduli space that can be identified with $Sp(N)$ instantons on $\mathbb{C}P^{2}/\mathbb{Z}_{2}$ is the one on which $\tilde{A}_{1}=0$ and $A_{2}=0$. Then, the Hilbert series of the instanton branch corresponding to the \textit{VS} theory with flavor symmetry $Sp(N_{1}) \times Sp(N_{2})$ and gauge ranks $\textbf{k}=(k_{1},k_{2})$ is

\begin{equation}\begin{aligned}
\label{spnhsz2}
& H[\textbf{k},F, \mathbb{C}P^{2}/\mathbb{Z}_{2}](t,x,\textbf{y},\textbf{d})= \int d \mu_{U(k_{1})}(\textbf{z})\int d \mu_{U(k_{2})}(\textbf{p}) \times \\
& \textrm{PE}[\chi_{S^{2}}t^{2} +\chi_{\tilde{S}^{1}}t^{2}  + \chi_{B^{j}}t   + \chi_{Q_{1}}t^{2}  + \chi_{Q_{2}}t^{2} -\chi_{F_{1}}t^{4} -\chi_{F_{2}}t^{4}]\, ,
\end{aligned}\end{equation}
where $\textbf{z}$ and $\textbf{p}$ are the fugacities of the $U(k_{1})$ and $U(k_{2})$ gauge groups respectively while $\textbf{y}$ and $\textbf{d}$ denote the fugacities of the $Sp(N_{1})$ and $Sp(N_{2})$ flavor groups respectively. Finally $x$ denotes the fugacity of the global $SU(2)$ symmetry rotating the $B_{1}$ and $B_{2}$ fields. The contribution of each field is given by
\begin{equation*}
\chi_{Q_{1}} = \sum_{i=1}^{N_{1}}\left(y_{i}+\frac{1}{y_{i}}\right)\sum_{a=1}^{k_{1}}z_{a}, \ \ \chi_{Q_{2}}=\sum_{
j=1}^{N_{2}}\left(d_{j}+\frac{1}{d_{j}}\right)\sum_{b=1}^{k_{2}}p_{b}^{-1}, \ \ \chi_{F_{1}}=\sum\nolimits_{1 \leq a < b \leq k_{1}} z_{a}z_{b}\, ,
\end{equation*}
\begin{equation*}
\chi_{S_{2}} = \sum\nolimits_{1 \leq a \leq b \leq k_{2}} p_{a}p_{b}, \ \ \chi_{\tilde{S}_{1}} = \sum\nolimits_{1 \leq a \leq b \leq k_{1}} z_{a}^{-1}z_{b}^{-1}, \ \ \chi_{B^{j}} = \left(x+\frac{1}{x}\right)\sum_{a=1}^{k_{1}}\sum_{b=1}^{k_{2}} z_{a}p_{b}^{-1} , \ \ \chi_{F_{2}}=\sum\nolimits_{1 \leq a < b \leq k_{2}} p_{a}^{-1}p_{b}^{-1}\, .
\end{equation*}

Explicit computation shows that the Hilbert series for the instanton branch of the VS theory with gauge group $G=U(k_{1})  \times U(k_{2})$ and flavor group $Sp(N_{1}) \times Sp(N_{2}) $ corresponding to the moduli space of instantons on $\mathbb{C}P^{2}/\mathbb{Z}_{2}$ turns out to be equal to the Hilbert series for $Sp(N)$ instantons on $\mathbb{C}^{2}/\mathbb{Z}_{2}$ with gauge group $G=O(K_{1}) \times O(K_{2})$ (see \cite{Dey:2013fea} for more details).  The two theories share the same flavor groups and the gauge groups are related as 

\begin{equation}
K_{1} = k_{1},\qquad K_{2} = k_{2}.
\end{equation}
Let us show some explicit examples supporting our claim.
\\
\\
\textbf{$Sp(2)$ instanton: $\textbf{k}=(1,1)$, and $\textbf{N}=(1,1)$}\\
\\
Using eq.(\ref{spnhsz2}) and unrefining we find that
\begin{equation*}
H[\textbf{k}=(1,1),Sp(1)\times Sp(1), \mathbb{C}P^{2}/\mathbb{Z}_{2}](t,1,1,1)= \frac{1-2t^3 +6t^6 -2t^9 +t^{12}}{(1-t^3)^{6}(1+t^3)^4}\, ,
\end{equation*}
which is the unrefined Hilbert series for $Sp(2)$ instantons on $\mathbb{C}^{2}/\mathbb{Z}_{2}$ with $\textbf{K}=(1,1)$ and $\textbf{N}=(1,1)$.\\
\\
\textbf{$Sp(3)$ instanton: $\textbf{k}=(1,1)$ and $\textbf{N}=(1,2)$}\\
\\
Using eq.(\ref{spnhsz2}) and unrefining we find that
\begin{equation*}
H[\textbf{k}=(1,1),Sp(1)\times Sp(2), \mathbb{C}P^{2}/\mathbb{Z}_{2}](t,1,1,1,1)= \frac{(1+t^6)(1-2t^3+10t^6-2t^9+t^{12})}{(1-t^3)^8(1+t^3)^6}\, ,
\end{equation*}
which is the unrefined Hilbert series for $Sp(3)$  instantons on $\mathbb{C}^{2}/\mathbb{Z}_{2}$ with $\textbf{K}=(1,1)$ and $\textbf{N}=(1,2)$.

\subsubsection{$Sp(N)$ instantons on $\mathbb{C}P^{2}/\mathbb{Z}_{2}$ \ - \textit{NVS}}

\begin{figure}[h!]
\centering
\includegraphics[scale=1]{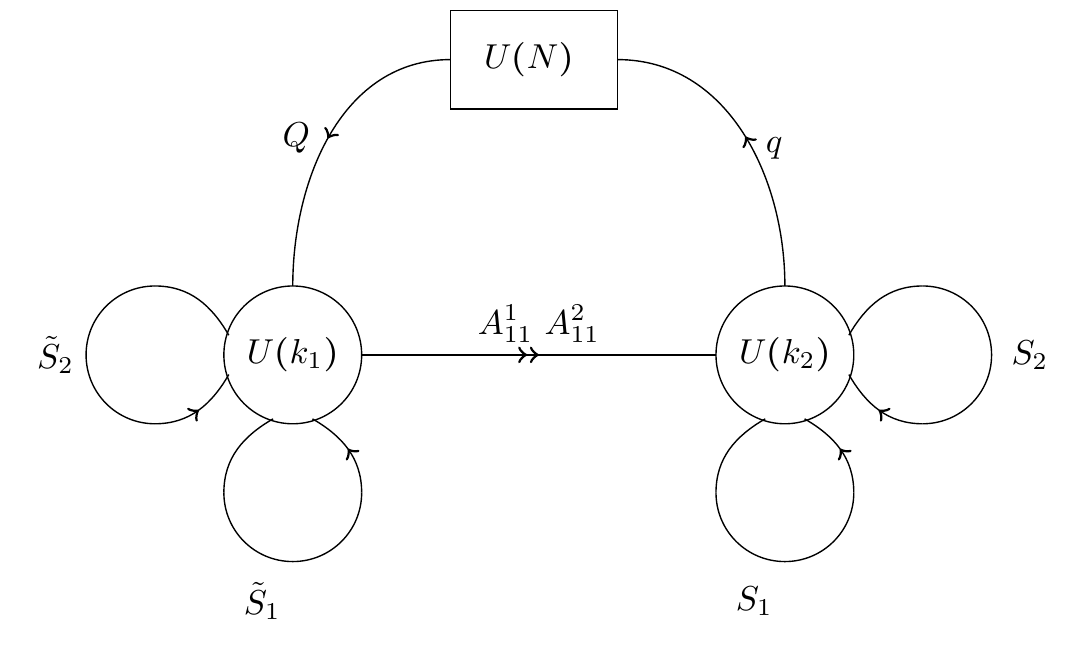}

 \caption{Quiver diagram for NVS symplectic instantons on $\mathbb{C}P^2/\mathbb{Z}_2$. \label{fig:spnz2second}}

\end{figure}

Let us now consider the second possible configuration, corresponding to the NVS case. The quiver diagram of the corresponding theory is reported in fig.\ref{fig:spnz2second}, while the transformations of the fields and of the F-term are summarized in table \ref{tab:spnz2second}. 

\begin{table}
 \renewcommand\arraystretch{1.3} 

\centering
\resizebox{12cm}{!}{
\begin{tabular}{c||cc||ccc||} 
Fields & $U(k_{1})$ & $U(k_{2})$ & $U(N)$  & $SU(2)$ & $U(1)$ \\ 
\hline 
$\tilde{S}_{1},\tilde{S}_{2}$ & $[2,0,...,0]_{-2}$ & $[\textbf{0}]$ & $[\textbf{0}]$  & $[1]$ & 1/4 \\ 
\hline 
$S_{1},S_{2}$ & $[\textbf{0}]$ & $[2,0,...,0]_{+2}$ & $[\textbf{0}]$  & $[1]$ & 1/4 \\ 
\hline 
$A^{2}_{11}$ & $ [1,0,...,0]_{+1}$ & $[0,0,...,1]_{+1}$ & $[\textbf{0}]$  & $[\textbf{0}]$ & 1/2 \\ 
\hline 
$q$ & $[\textbf{0}]$ & $[1,0,...,0]_{+1}$ & $[0,...,0,1]_{+1}$ & $[\textbf{0}]$ & 1/2 \\ 
\hline 
$Q$ & $[0,...,0,1]_{+1}$ & $[\textbf{0}]$ & $[1,0,...,0]_{+1}$ & $[\textbf{0}]$ & 1/2 \\ 
\hline 
$F$ & $[0,...,0,1]_{+1}$ & $[1,0,...,0]_{+1}$ & $[\textbf{0}]$ & $[\textbf{0}]$ & 1 \\ 
\end{tabular} 
}

\caption{Transformations of the fields for NVS symplectic instantons on $\mathbb{C}P^{2}/\mathbb{Z}_2$. \label{tab:spnz2second}}
\end{table}

\begin{figure}[h!]
\centering
\includegraphics[scale=1]{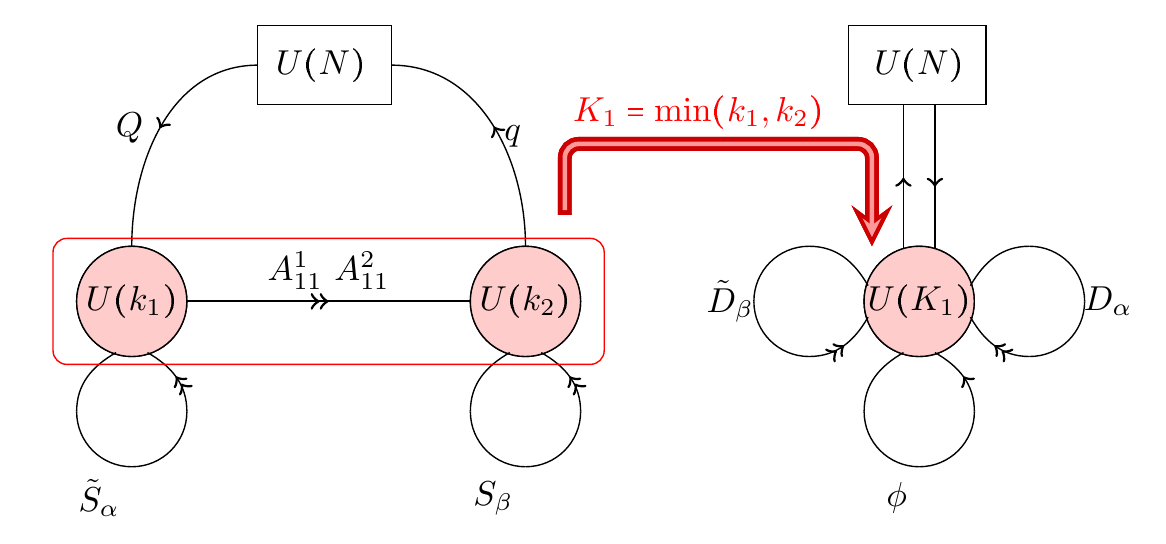}
\caption{Relation between the $\mathbb{C}P^{2}/\mathbb{Z}_{2}$ quiver gauge theory in the NVS case (on the left) and the $\mathbb{C}^{2}/\mathbb{Z}_{2}$ quiver gauge theory (on the right) . $\tilde{D}_{\beta}$ are two fields in the symmetric conjugate representation of the gauge group $U(K_{1})$. While $D_{\alpha}$ are two fields in the symmetric representation of the gauge group $U(K_{1})$ (see \cite{Dey:2013fea} for more details).\label{relationZ2symp}}

\end{figure}

The branch of the moduli space that can be identified with $Sp(N)$ instantons on $\mathbb{C}P^{2}/\mathbb{Z}_{2}$ is the one on which $A^{1}_{11}=0$. Then, the Hilbert series of the instanton branch corresponding to the NVS theory with flavor symmetry $U(N)$ and  gauge ranks $\textbf{k}=(k_{1},k_{2})$ is 

\begin{equation}\begin{aligned}
\label{spnhsz2sec1}
& H[\textbf{k},F, \mathbb{C}P^{2}/\mathbb{Z}_{2}](t,x,\textbf{y})= \int d \mu_{U(k_{1})}(\textbf{z})\int d \mu_{U(k_{2})}(\textbf{p}) \times \\
& \textrm{PE}[\chi_{S_{i}}t  +\chi_{\tilde{S}_{j}}t  + \chi_{A_{11}^{2}}t^{2}   + \chi_{Q}t^{2}  + \chi_{q}t^{2} -\chi_{F}t^{4}]\, ,
\end{aligned}\end{equation}
where $\textbf{z}$ and $\textbf{p}$ are the fugacities of the $U(k_{1})$ and $U(k_{2})$ gauge groups respectively while $\textbf{y}$  denote the fugacity of the $U(N)$ flavour group and $x$ denote the fugacity of the global $SU(2)$ symmetry acting separately on the two doublets $\tilde{S}_{\alpha}$ and $S_{\beta}$. The contribution of each field is given by
\begin{equation*}
\chi_{S_{j}} = \left(x+\frac{1}{x}\right)\sum_{1 \leq a \leq b \leq k_{2}} p_{a}p_{b}, \ \ \chi_{\tilde{S}_{i}} =\left(x+\frac{1}{x}\right) \sum_{1 \leq a \leq b \leq k_{1}} z_{a}^{-1}z_{b}^{-1}, \ \  
\end{equation*}
\begin{equation*}
\chi_{A^{2}_{11}} = \sum_{a=1}^{k_{1}}\sum_{b=1}^{k_{2}} z_{a}p_{b}^{-1}, \ \
\chi_{Q} = \sum_{i=1}^{N}\sum_{a=1}^{k_{1}}z_{a}^{-1}y_{i}, \ \ \chi_{q}=\sum_{
j=1}^{N}\sum_{b=1}^{k_{2}}p_{b}y_{j}^{-1}, \ \ \chi_{F}=\sum_{a=1}^{k_{1}}\sum_{b=1}^{k_{2}} z_{a}^{-1}p_{b}\, .
\end{equation*}
In this case, by explicit computation of the Hilbert series of the instanton branch of the NVS theory with gauge group $G=U(k_{1})  \times U(k_{2})$ and flavor group $U(N)$ for the moduli space of instantons on $\mathbb{C}P^{2}/\mathbb{Z}_{2}$ we find that it turns out to be equal to the Hilbert series for $Sp(N)$ instantons on $\mathbb{C}^{2}/\mathbb{Z}_{2}$ with gauge group $G=U(K_{1})$ (see \cite{Dey:2013fea} form more details). The two theories share the same flavor group and the gauge groups are related in the following way
\begin{equation}
K_{1} = \textrm{min}(k_{1},k_{2})\, .
\end{equation}
Let us explicitly show a few examples supporting our claim
\\
\\
\textbf{$Sp(1)$ instanton: $\textbf{k}=(1,1)$, and $N=1$}\\
\\
Using eq.(\ref{spnhsz2sec1}) and unrefining we find that
\begin{equation*}
 H[\textbf{k}=(1,1),U(1), \mathbb{C}P^{2}/\mathbb{Z}_{2}](t,1,1)=\frac{1+2t^{6}+2t^{9}+2t^{12}+t^{18}}{(1-t^3)^4(1+2t^3+2t^6+t^9)^2}\, ,
\end{equation*}
which is the Hilbert series for $Sp(1)$ instantons on $\mathbb{C}^{2}/\mathbb{Z}_{2}$ with $N=1$ and $K_{1}=1$.\\
\\
\textbf{$Sp(2)$ instanton: $\textbf{k}=(1,1)$, and $N=2$}\\
\\
Using eq.(\ref{spnhsz2sec1}) and unrefining we find that
\begin{equation*}
 H[\textbf{k}=(1,1),U(2), \mathbb{C}P^{2}/\mathbb{Z}_{2}](t,1,1,1)=\frac{1-t^3+5t^6+4t^9+4t^{12}+4t^{15}+5t^{18}-t^{21}+t^{24}}{(1-t^3)^6(1+t^3)^2(1+t^3+t^6)^3}\, ,
\end{equation*}
which is the Hilbert series for $Sp(2)$ instantons on $\mathbb{C}^{2}/\mathbb{Z}_{2}$ with $N=2$ and $K_{1}=1$.\\
\\
\textbf{$Sp(1)$ instanton: $\textbf{k}=(2,1)$, and $N=1$}\\
\\
Using eq.(\ref{spnhsz2sec1}) and unrefining we find that
\begin{equation*}
 H[\textbf{k}=(2,1),U(1), \mathbb{C}P^{2}/\mathbb{Z}_{2}](t,1,1)=\frac{1+2t^{6}+2t^{9}+2t^{12}+t^{18}}{(1-t^3)^4(1+2t^3+2t^6+t^9)^2}\, ,
\end{equation*}
which is again the Hilbert series for $Sp(1)$ instantons on $\mathbb{C}^{2}/\mathbb{Z}_{2}$ with $N=1$ and $K_{1}=1$.\\
\\
\textbf{$Sp(1)$ instanton: $\textbf{k}=(2,2)$, and $N=1$}\\
\\
Using eq.(\ref{spnhsz2sec1}) and unrefining we obtain
\begin{equation*}\begin{split}
 & H[\textbf{k}=(2,2),U(1), \mathbb{C}P^{2}/\mathbb{Z}_{2}](t,1,1) = \\
& = \frac{1}{(1-t^{3})^8(1+t^3)^4(1+t^6)^2(1+t^3+t^6)^2(1+t^3+t^6+t^9+t^{12})^2}(1+2t^6 +2t^9 +9t^{12} + 10t^{15} + \\
& 15t^{18} +18t^{21} +28t^{24} +26t^{27} +34t^{30} +26t^{33} + \textrm{palindrome} + t^{60})\, , 
\end{split}\end{equation*}
which is  the Hilbert series for $Sp(1)$ instantons on $\mathbb{C}^{2}/\mathbb{Z}_{2}$ with $N=1$ and $K_{1}=2$.
\\
\\
In the NVS case we can graphically summarize the relation between the parent $\mathbb{C}^2/\mathbb{Z}_2$ instanton and the $\mathbb{C}P^2/\mathbb{Z}_2$ one as in figure \ref{relationZ2symp}. Note that, as in the unitary instanton case, we again have a ``merging" of the flavored pair of gauge nodes into a single node with rank the minimum of the ``merged ones".

\subsubsection{$Sp(N)$ instantons on $\mathbb{C}P^{2}/\mathbb{Z}_{3}$ }

\begin{figure}[h!]
\centering
\includegraphics[scale=1]{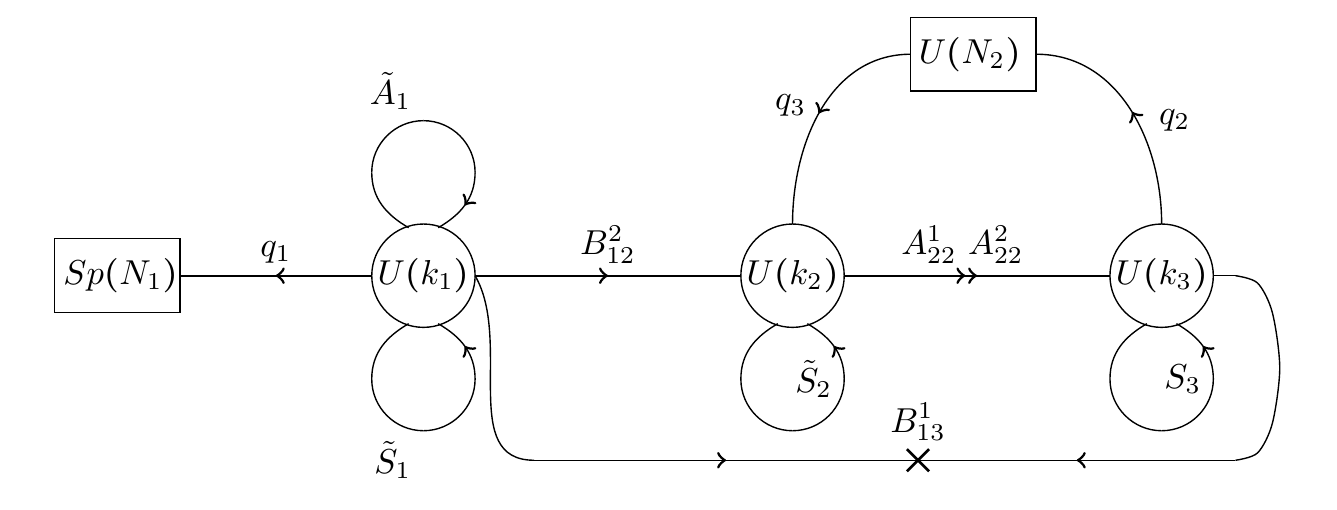}
\caption{Quiver diagram for symplectic instantons on $\mathbb{C}P^2/\mathbb{Z}_3$.\label{fig:orientifoldz3spn1}}

\end{figure}

For the case of odd orbifolds there is only one inequivalent choice. We report in fig.\ref{fig:orientifoldz3spn1} the quiver diagram of the corresponding field theory, while we summarize the fields and F-terms transformations in table \ref{tab:fn3}. Note that $N=N_1+N_2$.

\begin{table}\center
\resizebox{13cm}{!}{
 \renewcommand\arraystretch{1.3} 
\begin{tabular}{c||ccc||cccc||}

Fields & $U(k_{1})$ & $U(k_{2})$ & $U(k_{3})$ & $Sp(N_{1})$ & $U(N_{2})$ & $U(1)$ & $U(1)$ \\ 
\hline 
$q_{1}$ & $[1,0,...,0]_{+1}$ &  $[\textbf{0}]$ &  $[\textbf{0}]$ & $[1,0,...,0]$ &  $[\textbf{0}]$ &  $[\textbf{0}]$ & 1/2 \\ 
\hline 
$q_{2}$ &  $[\textbf{0}]$ &  $[\textbf{0}]$ & $[1,0,...0]_{+1}$ &  $[\textbf{0}]$ & $[0,...,0,1]_{+1}$ &  $[\textbf{0}]$ & 1/2 \\ 
\hline 
$q_{3}$ &  $[\textbf{0}]$ & $[0,...,0,1]_{+1}$ &  $[\textbf{0}]$ &  $[\textbf{0}]$ & $[1,0,...,0]_{+1}$ &  $[\textbf{0}]$ & 1/2 \\ 
\hline 
$B_{12}^{2}$ & $[1,0,...,0]_{+1}$ & $[0,...,0,1]_{+1}$ &  $[\textbf{0}]$ &  $[\textbf{0}]$ &  $[\textbf{0}]$ &  $[\textbf{0}]$ & 1/4 \\ 
\hline 
$A^{2}_{22}$ &  $[\textbf{0}]$ & $[1,0,...,0]_{+1}$ & $[0,...,0,1]_{+1}$ &  $[\textbf{0}]$ &  $[\textbf{0}]$ &  $[\textbf{0}]$ & 1/2 \\ 
\hline 
$B^{1}_{13}$ & $[1,0,...,0]_{+1}$ &  $[\textbf{0}]$ & $[1,0,...,0]_{+1}$ &  $[\textbf{0}]$ &  $[\textbf{0}]$ &  $[\textbf{0}]$ & 1/4 \\ 
\hline 
$\tilde{S}_{1}$ & $[2,0,...,0]_{-2}$ &  $[\textbf{0}]$ &  $[\textbf{0}]$ &  $[\textbf{0}]$ & $[\textbf{0}]$ &  $[\textbf{0}]$ & 1/2 \\ 
\hline 
$\tilde{S}_{2}$ &  $[\textbf{0}]$ & $[2,0,...,0]_{-2}$ &  $[\textbf{0}]$ &  $[\textbf{0}]$ &  $[\textbf{0}]$ & $1/x$ & 1/4 \\ 
\hline 
$S_{3}$ &  $[\textbf{0}]$ &  $[\textbf{0}]$ & $[2,0,...,0]_{+2}$ &  $[\textbf{0}]$ &  $[\textbf{0}]$ & $x$ & 1/4 \\ 
\hline 
$F_{1}$ & $[0,1,0,...,0]_{+1}$ &  $[\textbf{0}]$ &  $[\textbf{0}]$ &  $[\textbf{0}]$ &  $[\textbf{0}]$ &  $[\textbf{0}]$ & 1 \\ 
\hline 
$F_{2}$ &  $[\textbf{0}]$ & $[0,...,0,1]_{+1}$ & $[1,0,...,0]_{+1}$ &  $[\textbf{0}]$ &  $[\textbf{0}]$ &  $[\textbf{0}]$ & 1 \\ 
\end{tabular} 
}
\caption{Transformations of the fields for symplectic instantons on $\mathbb{C}P^{2}/\mathbb{Z}_{3}$. \label{tab:fn3}}

\end{table}

The branch of the moduli space that can be identified with $Sp(N)$ instantons on $\mathbb{C}P^{2}/\mathbb{Z}_{3}$ is the one on which $A^{1}_{22}=0$ and $\tilde{A}_{1}=0$. The Hilbert series of the instaton branch corresponding to the theory with flavor symmetry $Sp(N_{1}) \times U(N_{2})$ and gauge ranks $\textbf{k}=(k_{1},k_{2},k_{3})$ is

\begin{equation}\begin{aligned}
\label{spnhsz3}
& H[\textbf{k},F, \mathbb{C}P^{2}/\mathbb{Z}_{3}](t,x,\textbf{y},\textbf{d})= \int d \mu_{U(k_{1})}(\textbf{z})\int d \mu_{U(k_{2})}(\textbf{p}) \int d \mu_{U(k_{3})}(\textbf{w}) \times \\
& \textrm{PE}[\chi_{q_{1}}t^2  +\chi_{q_{2}}t^{2}  + \chi_{q_{3}}t^{2}   + \chi_{B_{12}^{2}}t  + \chi_{A^{2}_{22}}t^{2} +\chi_{B^{1}_{13}}t + \chi_{\tilde{S}_{1}}t^{2} +\chi_{\tilde{S}_{2}}t +\chi_{S_{3}}t -\chi_{F_{1}}t^{4} -\chi_{F_{2}}t^{4}]\, ,
\end{aligned}\end{equation}
where $\textbf{z}$, $\textbf{p}$ and $\textbf{w}$ are the fugacities of the $U(k_{1})$, $U(k_{2})$ and $U(k_{3})$ gauge groups respectively. While $\textbf{y}$  denote the fugacity of the $Sp(N_{1})$ flavor group and $\textbf{d}$ the fugacity of the $U(N_{2})$ flavor group. Finally $x$ is the fugacity of the $U(1)$ symmetry acting on $\tilde{S}_{2}$ and $S_{3}$ fields. The contribution of each field and of the F-terms are
\begin{equation*}
\chi_{\tilde{S}_{1}} = \sum\nolimits_{1\leq a \leq b \leq k_{1}} z_{a}^{-1}z_{b}^{-1}, \ \ \chi_{\tilde{S}_{2}} = \sum\nolimits_{1\leq a \leq b \leq k_{2}} p_{a}^{-1}p_{b}^{-1}x^{-1}, \ \ \chi_{S_{3}} = \sum\nolimits_{1\leq a \leq b \leq k_{3}} w_{a}w_{b}x\, ,
\end{equation*}
\begin{equation*}
\chi_{q_{1}} = \sum_{a=1}^{k_{1}}\sum_{i=1}^{N_{1}} z_{a}\left(y_{i} + \frac{1}{y_{i}} \right), \ \ \chi_{q_{2}} = \sum_{a=1}^{k_{3}}\sum_{j=1}^{N_{2}} w_{a}d_{j}^{-1}, \ \ \chi_{q_{3}} = \sum_{a=1}^{k_{2}}\sum_{j=1}^{N_{2}} p_{a}^{-1}d_{j}, \ \ \chi_{F_{2}} = \sum_{a=1}^{k_{2}}\sum_{b=1}^{k_{3}}p_{a}^{-1}w_{b}\, ,
\end{equation*}
\begin{equation*}
\chi_{B^{2}_{12}} = \sum_{a=1}^{k_{1}}\sum_{b=1}^{k_{2}} z_{a}p_{b}^{-1}, \ \
\chi_{A^{2}_{22}} = \sum_{a=1}^{k_{2}}\sum_{b=1}^{k_{3}}p_{a}w_{b}^{-1}, \ \ \chi_{B^{1}_{13}}=\sum_{
a=1}^{k_{1}}\sum_{b=1}^{k_{3}}z_{a}w_{b}, \ \ \chi_{F_{1}} = \sum\nolimits_{1\leq a < b \leq k_{1}} z_{a}z_{b}\, .
\end{equation*}

By explicit computation we find that the Hilbert series of the theory with gauge group $G=U(k_{1}) \times U(k_{2}) \times U(k_{3})$ and flavor group $Sp(N_{1}) \times U(N_{2})$ for the moduli space of instantons on $\mathbb{C}P^{2}/\mathbb{Z}_{3}$ coincides with the Hilbert series for the moduli space of $Sp(N)$ instantons on $\mathbb{C}^{2}/\mathbb{Z}_{3}$ with gauge group $G=O(K_{1})\times U(K_{2}) $ and flavor groups $Sp(N_{1}) \times U(N_{2})$ (see \cite{Dey:2013fea} for more details) upon identifying 
\begin{equation}
K_{1} = k_{1}, \ \ K_{2}= \textrm{min}(k_{2},k_{3})\, .
\end{equation}

Let us turn to explicit examples supporting our claim.
\\
\\
\textbf{ $Sp(1)$ instanton: $\textbf{k}=(1,1,1)$, and $\textbf{N}=(1,0)$}\\
\\ Using eq.(\ref{spnhsz3}) and unrefing we find that
\begin{equation*}
 H[\textbf{k}=(1,1,1),Sp(1), \mathbb{C}P^{2}/\mathbb{Z}_{3}](t,1,1) = \frac{(1 + t^6) (1 - t^3 + t^6)}{(1 - t^3)^4 (1 + t^3)^2 (1 + t^3 + t^6)}\, ,
\end{equation*}
which is the Hilbert series for $Sp(1)$ instantons on $\mathbb{C}^{2}/\mathbb{Z}_{3}$ with $\textbf{N}=(1,0)$ and $\textbf{K}=(1,1)$.\\
\\
\textbf{ $Sp(1)$ instanton: $\textbf{k}=(1,1,1)$, and $\textbf{N}=(0,1)$}\\
\\
Using eq.(\ref{spnhsz3}) and unrefing we find that
\begin{equation*}
H[\textbf{k}=(1,1,1),U(1), \mathbb{C}P^{2}/\mathbb{Z}_{3}](t,1,1)
= \frac{1 + t^6 + 2 t^9 + 2 t^{12} + 
 2 t^{15} + t^{18} + t^{24}}{(1 - t^3)^4 (1 + t^3)^2 (1 + t^6) (1 + t^3 + 
   t^6)^2}\, ,
\end{equation*}
which is the Hilbert series for $Sp(1)$ instantons on $\mathbb{C}^{2}/\mathbb{Z}_{3}$ with $\textbf{N}=(0,1)$ and $\textbf{K}=(1,1)$.\\
\\
\textbf{$Sp(2)$ instanton: $\textbf{k}=(1,1,1)$, and  $\textbf{N}=(1,1)$}\\
\\
Using eq.(\ref{spnhsz3}) and unrefing we find that
\begin{equation*}
H[\textbf{k}=(1,1,1),Sp(1) \times U(1), \mathbb{C}P^{2}/\mathbb{Z}_{3}](t,1,1,1)
= \frac{1 - 2 t^3 + 5 t^6 - 2 t^9 + 6 t^{12} - 2 t^{15} + 5 t^{18} - 
 2 t^{21} + t^{24}}{(1 - t^3)^6 (1 + t^6) (1 + 2 t^3 + 2 t^6 + t^9)^2}\, ,
\end{equation*}
which is the Hilbert series for $Sp(2)$ instantons on $\mathbb{C}^{2}/\mathbb{Z}_{3}$ with $\textbf{N}=(1,1)$ and $\textbf{K}=(1,1)$.\\
\\
\textbf{$Sp(1)$ instanton: $\textbf{k}=(1,2,1)$, and  $\textbf{N}=(1,0)$}\\
\\
Using eq.(\ref{spnhsz3}) and unrefing we find that
\begin{equation*}
H[\textbf{k}=(1,2,1),Sp(1), \mathbb{C}P^{2}/\mathbb{Z}_{3}](t,1,1) = \frac{(1 + t^6) (1 - t^3 + t^6)}{(1 - t^3)^4 (1 + t^3)^2 (1 + t^3 + t^6)}\, ,
\end{equation*}
which is again the Hilbert series for $Sp(1)$ instantons on $\mathbb{C}^{2}/\mathbb{Z}_{3}$ with $\textbf{N}=(1,0)$ and $\textbf{K}=(1,1)$.\\
\\
\textbf{$Sp(1)$ instanton: $\textbf{k}=(1,1,2)$, and $\textbf{N}=(1,0)$}\\
\\
Using eq.(\ref{spnhsz3}) and unrefing we find that
\begin{equation*}
H[\textbf{k}=(1,2,1),Sp(1), \mathbb{C}P^{2}/\mathbb{Z}_{3}](t,1,1) = \frac{(1 + t^6) (1 - t^3 + t^6)}{(1 - t^3)^4 (1 + t^3)^2 (1 + t^3 + t^6)}\, ,
\end{equation*}
which is again the Hilbert series for $Sp(1)$ instantons on $\mathbb{C}^{2}/\mathbb{Z}_{3}$ with $\textbf{N}=(1,0)$ and $\textbf{K}=(1,1)$.\\
\\
\textbf{$Sp(1)$ instanton: $\textbf{k}=(1,1,2)$, and $\textbf{N}=(0,1)$}\\
\\
Using eq.(\ref{spnhsz3}) and unrefing we find that
\begin{equation*}
H[\textbf{k}=(1,1,2),U(1), \mathbb{C}P^{2}/\mathbb{Z}_{3}](t,1,1) =\frac{1 + t^6 + 2 t^9 + 2 t^{12} + 
 2 t^{15} + t^{18} + t^{24}}{(1 - t^3)^4 (1 + t^3)^2 (1 + t^6) (1 + t^3 + 
   t^6)^2}\, ,
\end{equation*}
which is again the Hilbert series for $Sp(1)$ instantons on $\mathbb{C}^{2}/\mathbb{Z}_{3}$ with $\textbf{N}=(0,1)$ and $\textbf{K}=(1,1)$.\\
\\
\textbf{$Sp(1)$ instanton: $\textbf{k}=(1,2,1)$, and $\textbf{N}=(0,1)$}\\
\\
Using eq.(\ref{spnhsz3}) and unrefing we find
\begin{equation*}
H[\textbf{k}=(1,2,1),U(1), \mathbb{C}P^{2}/\mathbb{Z}_{3}](t,1,1) =\frac{1 + t^6 + 2 t^9 + 2 t^{12} + 
 2 t^{15} + t^{18} + t^{24}}{(1 - t^3)^4 (1 + t^3)^2 (1 + t^6) (1 + t^3 + 
   t^6)^2}\, ,
\end{equation*}
which is again the Hilbert series for $Sp(1)$ instantons on $\mathbb{C}^{2}/\mathbb{Z}_{3}$ with $\textbf{N}=(0,1)$ and $\textbf{K}=(1,1)$.\\
\\
\textbf{ $Sp(1)$ instanton: $\textbf{k}=(2,1,1)$, and $\textbf{N}=(1,0)$}\\
\\
Using eq.(\ref{spnhsz3}) and unrefing we find that
\begin{equation*}\begin{aligned}
& H[\textbf{k}=(2,1,1),Sp(1), \mathbb{C}P^{2}/\mathbb{Z}_{3}](t,1,1) = \\
& = \frac{1}{(1 - t^3)^6 (1 + t^3)^4 (1 + t^3 + 
   t^6) (1 + t^3 + 2 t^6 + 2 t^9 + 2 t^{12} + t^{15} + t^{18})^2}(1 + t^3 + \\
   & + 3 t^6 + 4 t^9 + 8 t^{12} + 14 t^{15} + 19 t^{18} + 23 t^{21} + 
 27 t^{24} + 26 t^{27}  + 27t^{30} + \textrm{palindrome} + t^{54})=\\
& = 1 + 4 t^6 + 2 t^9 + 13 t^{12} + 14 t^{15} + 33 t^{18} + 42 t^{21} + 80 t^{24} + 
 104 t^{27} + o(t^{27})\, , 
\end{aligned}\end{equation*}
which is the Hilbert series for $Sp(1)$ instantons on $\mathbb{C}^{2}/\mathbb{Z}_{3}$ with $\textbf{N}=(1,0)$ and $\textbf{K}=(2,1)$.
\\
\\

\begin{figure}[h!]
\centering
\includegraphics[scale=1]{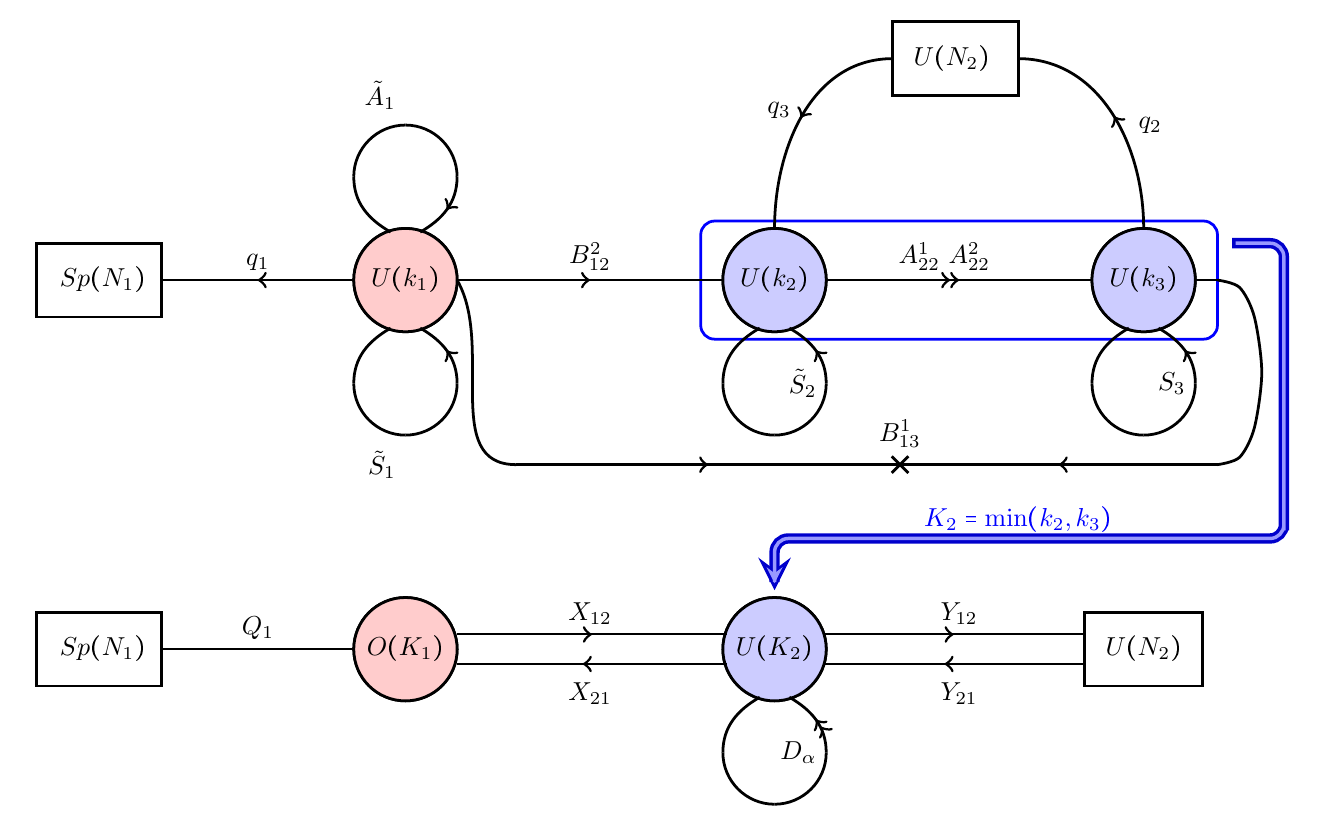}

 \caption{Relation between the quiver diagram for $Sp(N)$ instantons on $\mathbb{C}P^{2}/\mathbb{Z}_{3}$ and the quiver diagram for $Sp(N)$ instantons on $\mathbb{C}^{2}/\mathbb{Z}_{3}$. In the figure the symbol $D_{\alpha}$ denotes two fields transforming in the symmetric representation of the gauge group $ U(K_{2})$ (however see \cite{Dey:2013fea} for more details). \label{fig:orientifoldz3spn2}}

\end{figure}

As shown in fig.\ref{fig:orientifoldz3spn2}, we can graphically summarize the relation between the symplectic $\mathbb{C}P^2/\mathbb{Z}_3$ instanton and its cousin on $\mathbb{C}^2/\mathbb{Z}_3$ as the ``merging" of the flavored pair of gauge nodes into a single node whose rank is the minimum among the ``merging ones".

\subsubsection{$Sp(N)$ instantons on $\mathbb{C}P^{2}/\mathbb{Z}_{4}$ \ - \textit{VS}}
 
\begin{figure}[h!]
\centering
\includegraphics[scale=1]{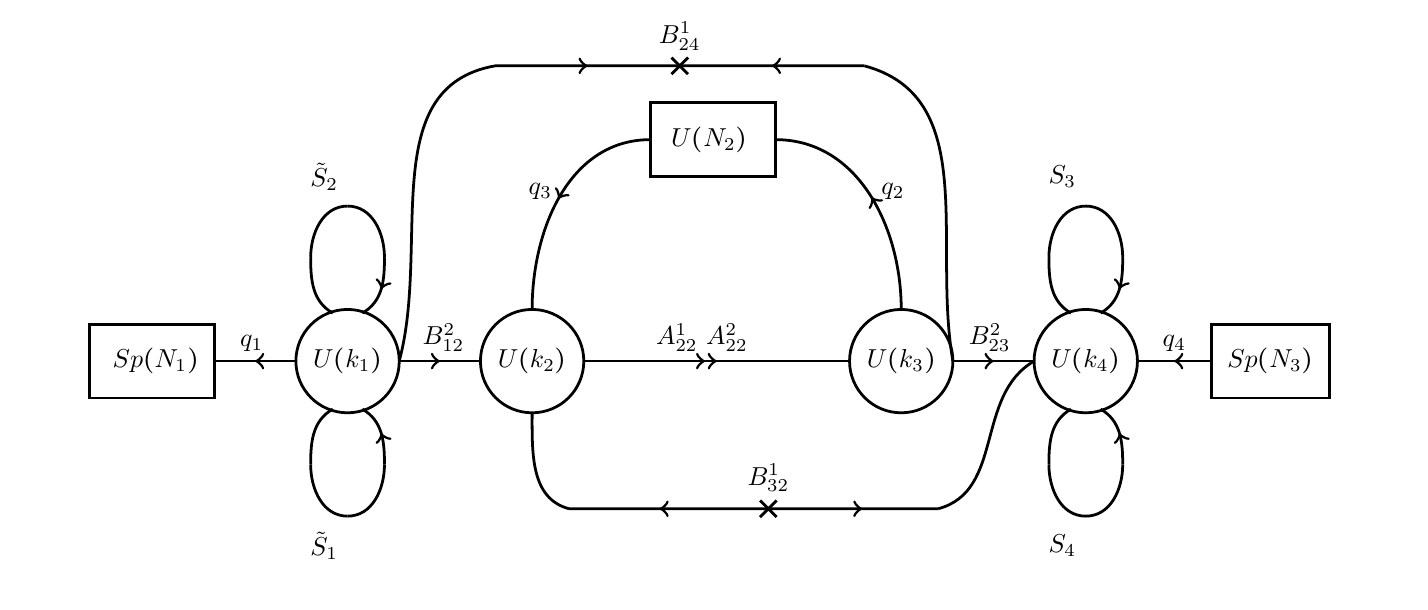}

\caption{Quiver diagram for VS symplectic instantons on $\mathbb{C}P^2/\mathbb{Z}_4$. \label{fig:orientifoldz4spn}}

\end{figure}

Starting from the theory whose instanton branch describes instantons on $\mathbb{C}P^{2}/\mathbb{Z}_{4}$ and applying the rules in \cite{Franco:2007ii} we obtain the theory for $Sp(N)$ instantons on $\mathbb{C}P^{2}/\mathbb{Z}_{4}$ in the VS case. The corresponding quiver diagram is reported in fig.\ref{fig:orientifoldz4spn}, while we summarize the transformations of the fields under the different groups in table \ref{tab:spnz41}.

\begin{table}
\renewcommand\arraystretch{1.3} 
\centering
\resizebox{14cm}{!}{
\begin{tabular}{c||cccc||cccc||} 
Fields & $U(k_{1})$ & $U(k_{2})$ & $U(k_{3})$ & $U(k_{4})$ & $Sp(N_{1})$ & $U(N_{2})$ & $Sp(N_{3})$ & $U(1)$ \\ 
\hline 
$B^{2}_{12}$ & $[1,0,...,0]_{+1}$ & $[0,...,0,1]_{+1}$ &  $[\textbf{0}]$ &  $[\textbf{0}]$ &  $[\textbf{0}]$ &  $[\textbf{0}]$ &  $[\textbf{0}]$ & 1/4 \\ 
\hline 
$A^{2}_{22}$ &  $[\textbf{0}]$ & $[1,0...,0]_{+1}$ & $[0,...,0,1]_{+1}$ &  $[\textbf{0}]$ &  $[\textbf{0}]$ &  $[\textbf{0}]$ &  $[\textbf{0}]$ & 1/2 \\ 
\hline 
$B^{2}_{23}$ &  $[\textbf{0}]$ &  $[\textbf{0}]$ & $[1,0...,0]_{+1}$ & $[0,...,0,1]_{+1}$ &  $[\textbf{0}]$ &  $[\textbf{0}]$ &  $[\textbf{0}]$ & 1/4 \\ 
\hline 
$\tilde{S}_{2}$ & $[2,0,...,0]_{-2}$ &  $[\textbf{0}]$ &  $[\textbf{0}]$ &  $[\textbf{0}]$ &  $[\textbf{0}]$ &  $[\textbf{0}]$ &  $[\textbf{0}]$ & 1/2 \\ 
\hline 
$S_{4}$ &  $[\textbf{0}]$ &  $[\textbf{0}]$ &  $[\textbf{0}]$ & $[2,0...,0]_{+2}$ &  $[\textbf{0}]$ &  $[\textbf{0}]$ &  $[\textbf{0}]$ & 1/2 \\ 
\hline 
$B^{1}_{24}$ & $[1,0,...,0]_{+1}$ &  $[\textbf{0}]$ & $[1,0,...,0]_{+1}$ &  $[\textbf{0}]$ &  $[\textbf{0}]$ &  $[\textbf{0}]$ &  $[\textbf{0}]$ & 1/4 \\ 
\hline 
$B^{1}_{32}$ &  $[\textbf{0}]$ & $[0,...,0,1]_{+1}$ &  $[\textbf{0}]$ & $[0,...,0,1]_{+1}$ &  $[\textbf{0}]$ &  $[\textbf{0}]$ &  $[\textbf{0}]$ & 1/4 \\ 
\hline 
$q_{1}$ & $[1,0,...,0]_{+1}$ &  $[\textbf{0}]$ &  $[\textbf{0}]$ & $[\textbf{0}]$ & $[1,0,....,0]$ &  $[\textbf{0}]$ &  $[\textbf{0}]$ & 1/2 \\ 
\hline 
$q_{3}$ &  $[\textbf{0}]$ & $[0,...,0,1]_{+1}$ &  $[\textbf{0}]$ &  $[\textbf{0}]$ &  $[\textbf{0}]$ & $[1,0,...,0]_{+1}$ & $[\textbf{0}]$  & 1/2 \\ 
\hline 
$q_{2}$ &  $[\textbf{0}]$ &  $[\textbf{0}]$ & $[1,0,...,0]_{+1}$ &  $[\textbf{0}]$ &  $[\textbf{0}]$ & $[0,...,0,1]_{+1}$ &  $[\textbf{0}]$ & 1/2 \\ 
\hline 
$q_{4}$ &  $[\textbf{0}]$ &  $[\textbf{0}]$ &  $[\textbf{0}]$ & $[0,...,0,1]_{+1}$ &  $[\textbf{0}]$ &  $[\textbf{0}]$ & $[1,0,...,0]$ & 1/2 \\ 
\hline 
$F_{1}$ & $[0,1,0...,0]_{+1}$ &  $[\textbf{0}]$ &  $[\textbf{0}]$ &  $[\textbf{0}]$ & $[\textbf{0}]$ &  $[\textbf{0}]$ &  $[\textbf{0}]$ & 1 \\ 
\hline 
$F_{2}$ &  $[\textbf{0}]$ & $[0,...,0,1]_{+1}$ & $[1,0,...,0]_{+1}$ &  $[\textbf{0}]$ &  $[\textbf{0}]$ &  $[\textbf{0}]$ &  $[\textbf{0}]$ & 1 \\ 
\hline 
$F_{3}$ &  $[\textbf{0}]$ &  $[\textbf{0}]$ &  $[\textbf{0}]$ & $[0,1,0,...,,0]_{-1}$ &  $[\textbf{0}]$ &  $[\textbf{0}]$ &  $[\textbf{0}]$ & 1 \\  
\end{tabular} 
}
\caption{Transformation of the fields for VS symplectic instantons on $\mathbb{C}P^{2}/\mathbb{Z}_{4}$. \label{tab:spnz41}}
\end{table}

The branch of the moduli space that can be identified with $Sp(N)$ instantons on $\mathbb{C}P^{2}/\mathbb{Z}_{4}$ is the one on which $A^{1}_{22}=0$, $\tilde{S}_{1}=0$ and $S_{3}=0$. The Hilbert series of the instanton branch corresponding to the VS theory with flavor symmetry  $Sp(N_{1}) \times U(N_{2}) \times Sp(N_{3})$ and gauge ranks $\textbf{k}=(k_{1},k_{2},k_{3},k_{4})$ is

\begin{equation}\begin{aligned}
\label{spnhsz4}
& H[\textbf{k},F, \mathbb{C}P^{2}/\mathbb{Z}_{4}](t,x,\textbf{y},\textbf{d},\textbf{u})= \int d \mu_{U(k_{1})}(\textbf{z})\int d \mu_{U(k_{2})}(\textbf{p}) \int d \mu_{U(k_{3})}(\textbf{w}) \\
& \int d \mu_{U(k_{4})}(\textbf{v})\times\textrm{PE}[\chi_{q_{1}}t^2  +\chi_{q_{2}}t^{2}  + \chi_{q_{3}}t^{2}   + \chi_{q_{4}}t^2 + \chi_{B_{12}^{2}}t  + \chi_{A^{2}_{22}}t^{2} + \chi_{B^{2}_{23}}t + \\
& + \chi_{B^{1}_{24}}t + \chi_{B^{1}_{32}}t + \chi_{\tilde{S}_{2}}t^{2} + \chi_{S_{4}}t^{2} -\chi_{F_{1}}t^{4} -\chi_{F_{2}}t^{4} -\chi_{F_{3}}t^{4}]\, ,
\end{aligned}\end{equation}
where $\textbf{z}$, $\textbf{p}$, $\textbf{w}$ and $\textbf{v}$  are the fugacities of the $U(k_{1})$, $U(k_{2})$,$U(k_{3})$ and $U(k_{4})$ gauge groups respectively while $\textbf{y}$,$\textbf{d}$ and $\textbf{u}$  denote the fugacities of the $Sp(N_{1})$ flavour group, the $U(N_{2})$ flavour group and the $Sp(N_{3})$ respectively. The contributions of the various fields are

\begin{equation*}
\chi_{B^{2}_{12}} = \sum_{a=1}^{k_{1}}\sum_{b=1}^{k_{2}}z_{a}p_{b}^{-1}, \ \  \chi_{A_{22}^{2}} = \sum_{a=1}^{k_{2}}\sum_{b=1}^{k_{3}}p_{a}w_{b}^{-1}, \ \ \chi_{B^{2}_{23}} = \sum_{a=1}^{k_{3}}\sum_{b=1}^{k_{4}}w_{a}v_{b}^{-1}\, , 
\end{equation*}
\begin{equation*}
\chi_{S_{4}} = \sum\nolimits_{1 \leq a \leq b \leq k_{4}}v_{a}v_{b}, \ \ \chi_{F_{1}}= \sum\nolimits_{1 \leq a < b \leq k_{1}} z_{a}z_{b},   \ \ \chi_{F_{3}} = \sum\nolimits_{1 \leq a < b \leq k_{4}} v_{a}^{-1}v_{b}^{-1}\, , 
\end{equation*}
\begin{equation*}
\chi_{B^{1}_{24}} = \sum_{a=1}^{k_{1}}\sum_{b=1}^{k_{3}}z_{a}w_{b}, \ \ \chi_{B^{1}_{32}} = \sum_{a=1}^{k_{2}}\sum_{b=1}^{k_{4}}p_{a}^{-1}v_{b}^{-1}, \ \ \chi_{\tilde{S_{2}}} = \sum\nolimits_{1 \leq a \leq b \leq k_{1}} z_{a}^{-1}z_{b}^{-1}, \ \ \chi_{F_{2}} = \sum_{a=1}^{k_{2}}\sum_{b=1}^{k_{3}} p_{a}^{-1}w_{b}\, , 
\end{equation*}
\begin{equation*}
\chi_{q_{1}} = \sum_{a=1}^{k_{1}}\sum_{j=1}^{N_{1}}z_{a}\left(y_{j}+\frac{1}{y_{j}}\right), \ \ \chi_{q_{3}}=\sum_{j=1}^{N_{2}}\sum_{b=1}^{k_{2}} d_{j}p_{b}^{-1}, \ \ \chi_{q_{2}}=\sum_{a=1}^{k_{3}}\sum_{i=1}^{N_{2}}w_{a}d_{i}^{-1},  \ \ \chi_{q_{4}}=\sum_{a=1}^{k_{4}}\sum_{i=1}^{N_{3}}v_{a}^{-1}\left(u_{i}+\frac{1}{u_{i}}\right)\, .
\end{equation*}

By explicit computation of the instanton branch Hilbert series for the theory with gauge group $G=U(k_{1}) \times U(k_{2}) \times U(k_{3}) \times U(k_{4})$ and flavor group $Sp(N_{1}) \times U(N_{2}) \times Sp(N_{3})$ we find that it is equal to the Hilbert series for $Sp(N)$ instantons on $\mathbb{C}^{2}/\mathbb{Z}_{4}$ with gauge group $G=O(K_{1})\times U(K_{2}) \times O(K_{3}) $ and flavor groups $Sp(N_{1}) \times U(N_{2}) \times Sp(N_{3})$ (see \cite{Dey:2013fea} for more details) upon identifying

\begin{equation}
K_{1} = k_{1}, \ \ K_{2}= \textrm{min}(k_{2},k_{3})\, ,\ \ K_{3}=k_{3}\, .
\end{equation}
Let us show some explicit examples supporting our claim.
\\
\\
\textbf{ $Sp(1)$ instanton:  $\textbf{k}=(1,1,1,1)$, and $\textbf{N}=(1,0,0)$}\\
\\
Using eq.(\ref{spnhsz4}) and unrefing we find that 
\begin{equation}
\label{h111}
H[\textbf{k}=(1,1,1,1),Sp(1), \mathbb{C}P^{2}/\mathbb{Z}_{4}](t,1) = \frac{1 + t^{12}}{(1 - t^6)^4}\, ,
\end{equation}
which is the Hilbert series for $Sp(1)$ instantons on $\mathbb{C}^{2}/\mathbb{Z}_{4}$ with $\textbf{N}=(1,0,0)$ and $\textbf{K}=(1,1,1)$.\\
\\
\textbf{$Sp(1)$ instanton: $\textbf{k}=(1,1,1,1)$, and $\textbf{N}=(0,1,0)$}\\
\\
Using eq.(\ref{spnhsz4}) and unrefing we find that
\begin{equation*}
H[\textbf{k}=(1,1,1,1),U(1), \mathbb{C}P^{2}/\mathbb{Z}_{4}](t,1) = \frac{1 + 4 t^{12} + t^{24}}{(1 - t^6)^4 (1 + t^6)^2}\, ,
\end{equation*}
which is the Hilbert series for $Sp(1)$ instantons on $\mathbb{C}^{2}/\mathbb{Z}_{4}$ with $\textbf{N}=(0,1,0)$ and $\textbf{K}=(1,1,1)$.\\
\\
\textbf{$Sp(1)$ instanton: $\textbf{k}=(1,2,1,1)$, and $\textbf{N}=(1,0,0)$}\\
\\
Using eq.(\ref{spnhsz4}) and unrefing we find again the expression (\ref{h111}).\\
\\
\textbf{$Sp(1)$ instanton: $\textbf{k}=(1,1,2,1)$, and $\textbf{N}=(1,0,0)$}\\
\\
Using eq.(\ref{spnhsz4}) and unrefing we find again the expression (\ref{h111}).\\
\\
\textbf{$Sp(1)$ instanton: $\textbf{k}=(2,1,1,1)$, and $\textbf{N}=(1,0,0)$}\\
\\
Using eq.(\ref{spnhsz4}) and unrefing we find that
\begin{equation*}
H[\textbf{k}=(2,1,1,1),Sp(1), \mathbb{C}P^{2}/\mathbb{Z}_{4}](t,1) =  \frac{1 + t^6 + 5 t^{12} + 8 t^{18} + 8 t^{24} + 8 t^{30} + 5 t^{36} + t^{42} + t^{48}}{(1 - t^6)^6 (1 + t^6) (1 + t^6 + t^{12})^2}\, ,
\end{equation*}
which is the Hilbert series for $Sp(1)$ instantons on $\mathbb{C}^{2}/\mathbb{Z}_{4}$ with $\textbf{N}=(1,0,0)$ and $\textbf{K}=(2,1,1)$.
\\
\\

\begin{figure}[h!]
\centering
\includegraphics[scale=0.8]{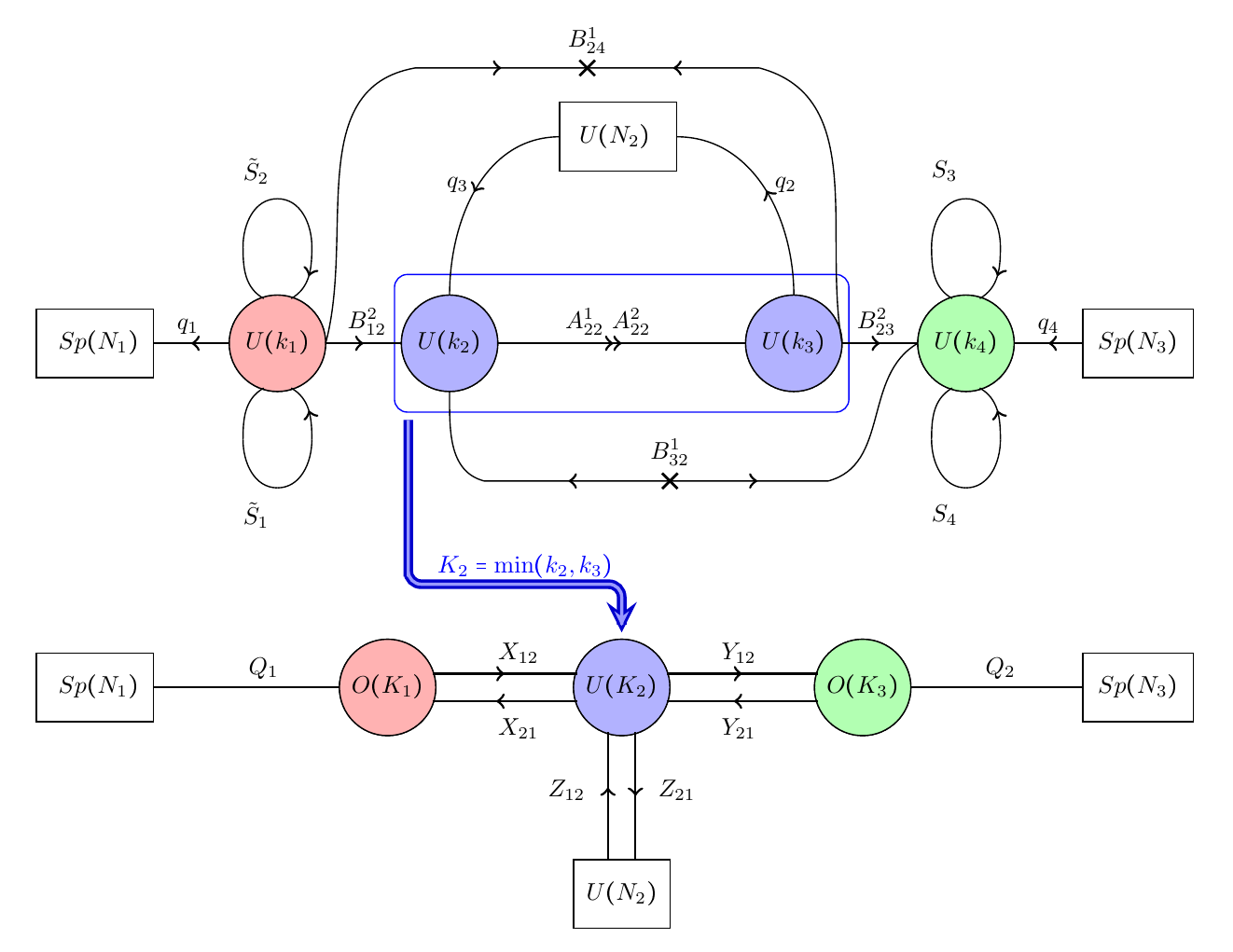}

\caption{Relation between the $\mathbb{C}P^{2}/\mathbb{Z}_{4}$ quiver  gauge theory in the VS case and the corresponding $\mathbb{C}^{2}/\mathbb{Z}_{4}$ quiver gauge theory. \label{fig:summaryorientifoldz4spn}}

\end{figure}

We can graphically relate the symplectic VS $\mathbb{C}P^2/\mathbb{Z}_4$ instantons with their cousin on $\mathbb{C}^2/\mathbb{Z}_4$ as in fig.\ref{fig:summaryorientifoldz4spn}.
\\
\\

\subsubsection{$Sp(N)$ instantons on $\mathbb{C}P^{2}/\mathbb{Z}_{4}$ \ - \textit{NVS}}

\begin{figure}[h!]
\centering
\includegraphics[scale=1]{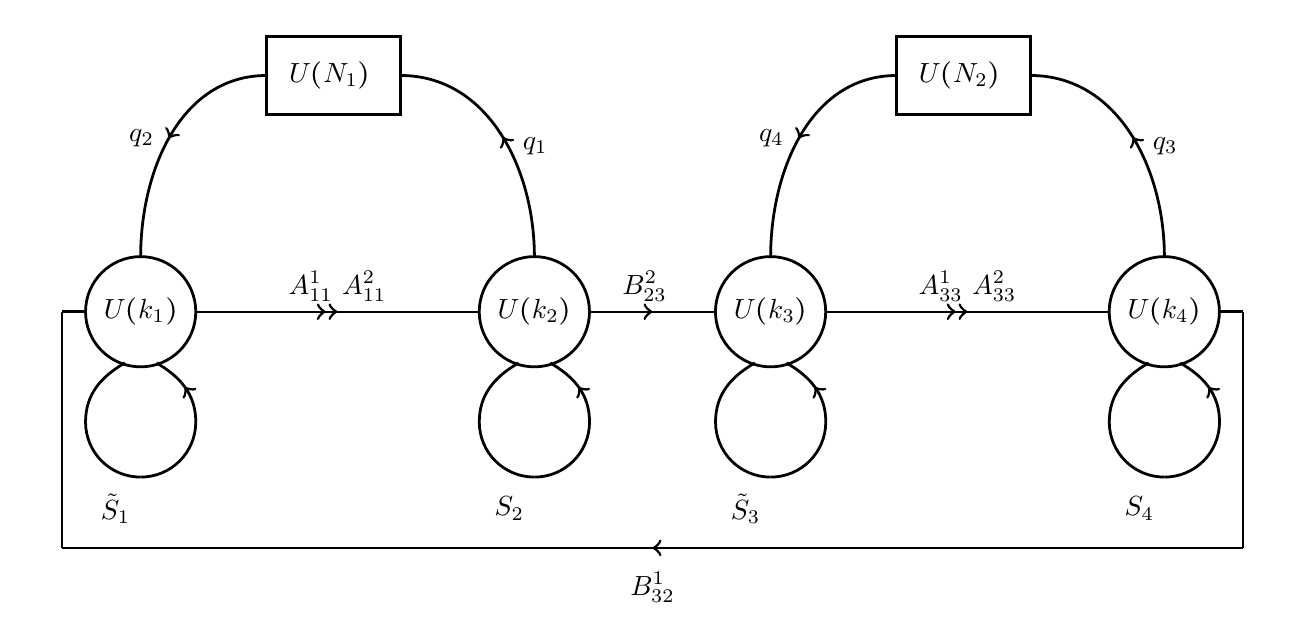}

 \caption{Quiver diagram for NVS symplectic instantons on $\mathbb{C}P^2/\mathbb{Z}_4$. \label{fig:z42}}

\end{figure}

Let us now consider the second configuration, leading to the NVS case. The quiver diagram of the corresponding theory is reported in fig.\ref{fig:z42}, while the transformations of the fields and of the F-terms are summarized in table \ref{tab:222}.

\begin{table}[h!]\centering

\renewcommand\arraystretch{1.3} 
\resizebox{14cm}{!}{
\begin{tabular}{c||cccc||ccc||}
Fields & $U(k_{1})$ & $U(k_{2})$ & $U(k_{3})$ & $U(k_{4})$ & $U(N_{1})$ & $U(N_{2})$ & $U(1)$\\ 
\hline 
$A^{2}_{11}$ & $[1,0,...,0]_{+1}$ & $[0,...,0,1]_{+1}$ & $[\textbf{0}]$  & $[\textbf{0}]$  & $[\textbf{0}]$  & $[\textbf{0}]$  & 1/2\\ 
\hline 
$B^{2}_{23}$ & $[\textbf{0}]$  & $[1,0,...,0]_{+1}$ & $[0,...,0,1]_{+1}$ & $[\textbf{0}]$  & $[\textbf{0}]$  & $[\textbf{0}]$  & 1/4\\ 
\hline 
$A^{2}_{33}$ & $[\textbf{0}]$  & $[\textbf{0}]$  & $[1,0,...,0]_{+1}$ & $[0,...,0,1]_{+1}$ & $[\textbf{0}]$  & $[\textbf{0}]$  & 1/2\\ 
\hline 
$B^{1}_{32}$ & $[0,...,0,1]_{+1}$ & $[\textbf{0}]$  & $[\textbf{0}]$  & $[1,0,...,0]_ {+1}$ & $[\textbf{0}]$  & $[\textbf{0}]$  & 1/4\\ 
\hline 
$\tilde{S}_{1}$ & $[2,0...,0]_{-2}$ & $[\textbf{0}]$  & $[\textbf{0}]$  & $[\textbf{0}]$  & $[\textbf{0}]$  & $[\textbf{0}]$  & 1/4\\ 
\hline 
$S_{2}$ & $[\textbf{0}]$  & $[2,0,...,0]_{+2}$ & $[\textbf{0}]$  & $[\textbf{0}]$  & $[\textbf{0}]$  & $[\textbf{0}]$  & 1/4\\ 
\hline 
$\tilde{S}_{3}$ & $[\textbf{0}]$  & $[\textbf{0}]$  & $[2,0,...,0]_{-2}$  &$[\textbf{0}]$  & $[\textbf{0}]$  & $[\textbf{0}]$  & 1/4\\ 
\hline 
$S_{4}$ & $[\textbf{0}]$  & $[\textbf{0}]$  & $[\textbf{0}]$  & $[2,0,...,0]_{+2}$ & $[\textbf{0}]$  & $[\textbf{0}]$  & 1/4\\ 
\hline 
$q_{1}$ & $[\textbf{0}]$  & $[1,0,...,0]_{+1}$ & $[\textbf{0}]$  & $[\textbf{0}]$  & $[0,...,0,1]_{+1}$ & $[\textbf{0}]$  & 1/2\\ 
\hline 
$q_{2}$ & $[0,...,0,1]_{+1}$ & $[\textbf{0}]$  & $[\textbf{0}]$  & $[\textbf{0}]$  & $[1,0,...,0]_{+1}$ & $[\textbf{0}]$  & 1/2\\ 
\hline 
$q_{3}$ & $[\textbf{0}]$  & $[\textbf{0}]$  & $[\textbf{0}]$  & $[1,0,...,0]_{+1}$ & $[\textbf{0}]$  & $[0,...,0,1]_{+1}$ & 1/2\\ 
\hline 
$q_{4}$ & $[\textbf{0}]$  & $[\textbf{0}]$  & $[0,...,0,1]_{+1}$ & $[\textbf{0}]$  & $[\textbf{0}]$ & $[1,0,...,0]_{+1}$  & 1/2 \\ 
\hline 
$F_{1}$ & $[0,...,0,1]_{+1}$ & $[1,0,...,0]_{+1}$ &  $[\textbf{0}]$  &  $[\textbf{0}]$  &  $[\textbf{0}]$  &  $[\textbf{0}]$  & 1 \\ 
\hline 
$F_{2}$ &  $[\textbf{0}]$  &  $[\textbf{0}]$  & $[0,...,0,1]_{+1}$ & $[1,0,...,0]_{+1}$ &  $[\textbf{0}]$  &  $[\textbf{0}]$  & 1\\ 
\end{tabular} 
}
\caption{Transformation of the fields for NVS symplectic instantons on $\mathbb{C}P^{2}/\mathbb{Z}_{4}$. \label{tab:222}}
\end{table}

The branch of the moduli space that can be identified with $Sp(N)$ instantons on $\mathbb{C}P^{2}/\mathbb{Z}_{4}$ in the NVS case is the one on which $A^{1}_{11}=0$  and $A_{33}^{1}=0$. The Hilbert series of the instanton branch corresponding to the NVS theory with flavor symmetry $U(N_{1}) \times U(N_{2})$ and gauge ranks $\textbf{k}=(k_{1},k_{2},k_{3},k_{4})$ is

\begin{equation}\begin{aligned}
\label{spnhsz42}
& H[\textbf{k},F, \mathbb{C}P^{2}/\mathbb{Z}_{4}](t,x,\textbf{y},\textbf{d})= \int d \mu_{U(k_{1})}(\textbf{z})\int d \mu_{U(k_{2})}(\textbf{p}) \int d \mu_{U(k_{3})}(\textbf{w}) \times \\
& \int d \mu_{U(k_{4})}(\textbf{v}) \times  \textrm{PE}[\chi_{q_{1}}t^2  +\chi_{q_{2}}t^{2}  + \chi_{q_{3}}t^{2}   + \chi_{q_{4}}t^2 + \chi_{B_{23}^{2}}t  + \chi_{A^{2}_{11}}t^{2} +\chi_{A^{2}_{33}}t^{2} +\\
& + \chi_{B^{1}_{32}}t  +\chi_{\tilde{S}_{1}}t + \chi_{S_{2}}t +\chi_{\tilde{S}_{3}}t +\chi_{S_{4}}t -\chi_{F_{1}}t^{4} -\chi_{F_{2}}t^{4}]\, ,
\end{aligned}\end{equation}
where $\textbf{z}$, $\textbf{p}$, $\textbf{w}$ and $\textbf{v}$  are the fugacities of the $U(k_{1})$, $U(k_{2})$, $U(k_{3})$ and $U(k_{4})$ gauge groups respectively, while $\textbf{y}$ and $\textbf{d}$ denote the fugacities of the $U(N_{1})$ flavor group and the $U(N_{2})$ flavor group respectively. The contributions of the various fields are given by
\begin{equation*}
\chi_{S_{4}} = \sum\nolimits_{ 1 \leq a \leq b \leq k_{4}}  v_{a}v_{b}, \ \
\chi_{F_{1}} = \sum_{a=1}^{k_{1}}\sum_{b=1}^{k_{2}} p_{b}z_{a}^{-1}, \ \ \chi_{F_{2}} = \sum_{a=1}^{k_{3}}\sum_{b=1}^{k_{4}} w_{a}^{-1}v_{b}\, ,
\end{equation*}
\begin{equation*}
 \ \ \chi_{\tilde{S}_{1}} = \sum\nolimits_{ 1 \leq a \leq b \leq k_{1}} z_{a}^{-1}z_{b}^{-1}, \ \ \chi_{S_{2}}=\sum\nolimits_{ 1 \leq a \leq b \leq k_{2}}  p_{a}p_{b}, \ \ \chi_{\tilde{S}_{3}} = \sum\nolimits_{1 \leq a\leq b \leq k_{3}} w_{a}^{-1}w_{b}^{-1}\, ,   
\end{equation*}
\begin{equation*}
\chi_{q_{1}} = \sum_{a=1}^{k_{2}}\sum_{i=1}^{N_{1}} p_{a}y_{i}^{-1}, \ \
\chi_{q_{2}} = \sum_{a=1}^{k_{1}}\sum_{i=1}^{N_{1}} z_{a}^{-1}y_{i}, \ \ \chi_{q_{3}} = \sum_{a=1}^{k_{4}}\sum_{j=1}^{N_{2}} v_{a}d_{j}^{-1}, \ \ \chi_{q_{4}} = \sum_{a=1}^{k_{3}}\sum_{j=1}^{N_{2}} w_{a}^{-1}d_{j}\, ,
\end{equation*}
\begin{equation*}
\chi_{A^{2}_{11}} = \sum_{a=1}^{k_{1}}\sum_{b=1}^{k_{2}}z_{a}p_{b}^{-1}, \ \ \chi_{B^{2}_{23}}=\sum_{a=1}^{k_{2}}\sum_{b=1}^{k_{3}} p_{a}w_{b}^{-1}, \ \ \chi_{A^{2}_{33}}=\sum_{a=1}^{k_{3}}\sum_{b=1}^{k_{4}} w_{a}v_{b}^{-1}, \ \ \chi_{B^{1}_{32}} = \sum_{a=1}^{k_{1}}\sum_{b=1}^{k_{4}} v_{b}z_{a}^{-1}\, .
\end{equation*}

Explicit computation of the instanton branch Hilbert series with gauge group $G=U(k_{1}) \times U(k_{2}) \times U(k_{3}) \times U(k_{4})$ and flavor group $U(N_{1}) \times U(N_{2})$ shows that it coincides with the Hilbert series for $Sp(N)$ instantons on $\mathbb{C}^{2}/\mathbb{Z}_{4}$ with gauge groups $G=U(K_{1})\times U(K_{2}) $ and flavor groups $U(N_{1}) \times U(N_{2})$ (see \cite{Dey:2013fea} for more details) upon the identification

\begin{equation}
K_{1} = \textrm{min}(k_{1},k_{2})\, ,\qquad K_{2}= \textrm{min}(k_{3},k_{4})\, .
\end{equation}
Let us show a few explicit examples.
\\
\\
\textbf{$Sp(1)$ instanton: $\textbf{k}=(1,1,1,1)$, and $\textbf{N}=(1,0)$}\\
\\Using eq.(\ref{spnhsz42}) and unrefining we find that
\begin{equation}
\label{hs333}
 H[\textbf{k}=(1,1,1,1),U(1),\mathbb{C}P^{2}/\mathbb{Z}_{4}](t,1)
 = \frac{1 - t^3 + 
 2 t^9 - t^{15} + t^{18}}{(1 - t^3)^4 (1 + t^3)^2 (1 + t^3 + t^6 + t^9 +
    t^{12})}\, ,
\end{equation}
which is the Hilbert series for $Sp(1)$ instantons on $\mathbb{C}^{2}/\mathbb{Z}_{4}$ with $\textbf{N}=(1,0)$ and $\textbf{K}=(1,1)$.\\
\\
\textbf{$Sp(2)$ instanton: $\textbf{k}=(1,1,1,1)$, and $\textbf{N}=(1,1)$}\\
\\
Using eq.(\ref{spnhsz42}) and unrefining we find that
\begin{equation*}\begin{aligned}
& H[\textbf{k}=(1,1,1,1),U(1) \times U(1),\mathbb{C}P^{2}/\mathbb{Z}_{4}](t,1,1)
 = \\ 
 & = \frac{1 + 2 t^6 + 3 t^9 + 8 t^{12} + 11 t^{15} + 13 t^{18} + 12 t^{21} + 13 t^{24} + 
 11 t^{27} + 8 t^{30} + 3 t^{33} + 
 2 t^{36} + t^{42}}{(1 - t^3)^6 (1 + t^3)^2 (1 + t^3 + t^6)^3 (1 + 
   t^3 + 2 t^6 + 2 t^9 + 2 t^{12} + t^{15} + t^{18})}\, ,
\end{aligned}\end{equation*}
which is the Hilbert series for $Sp(2)$ instantons on $\mathbb{C}^{2}/\mathbb{Z}_{4}$ with $\textbf{N}=(1,1)$ and $\textbf{K}=(1,1)$.\\
\\
\textbf{$Sp(1)$ instantons: $\textbf{k}=(1,2,1,1)$, and $\textbf{N}=(1,0)$}\\
\\
Using eq.(\ref{spnhsz42}) we find again the expression (\ref{hs333}).
\\
\\

\begin{figure}[h!]
\centering
\includegraphics[scale=0.8]{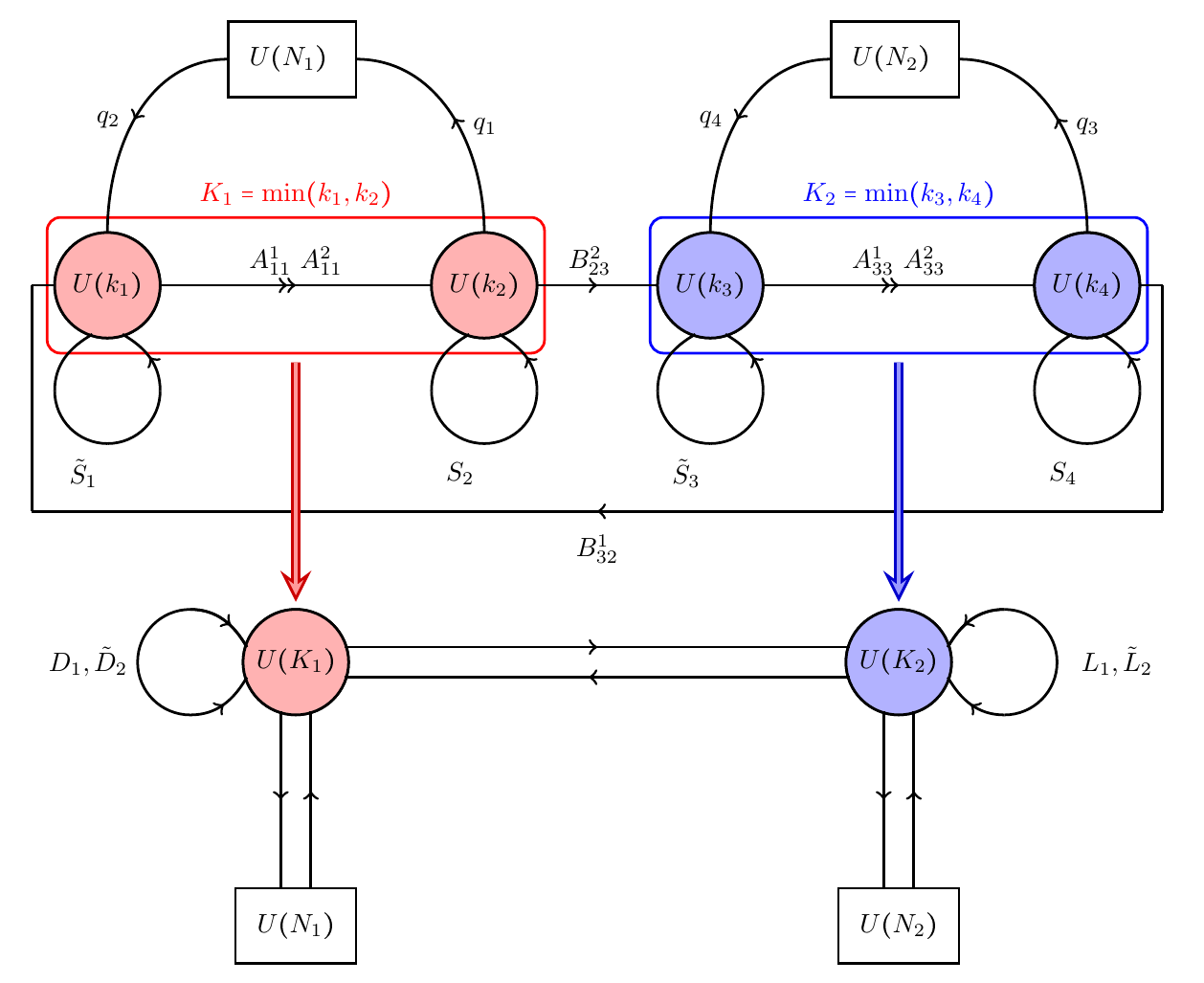}

\caption{Relation between the $\mathbb{C}P^{2}/\mathbb{Z}_{4}$ quiver gauge theory in the NVS case and the corresponding  $\mathbb{C}^{2}/\mathbb{Z}_{4}$ quiver gauge theory. Where $D_{1},\tilde{D}_{2}$ are two fields in the symmetric representation of the gauge group $U(K_{1})$, while $L_{1},\tilde{L}_{2}$ are two fields in the symmetric representation of the gauge group $U(K_{2})$ (however see \cite{Dey:2013fea} for more details regarding the $\mathbb{C}^{2}/\mathbb{Z}_{4}$ theory). \label{fig:z42relation1}}

\end{figure}

Finally, in fig.\ref{fig:z42relation1} we graphically show the relation between symplectic NVS instantons on $\mathbb{C}P^2/\mathbb{Z}_4$ and their cousins on $\mathbb{C}^2/\mathbb{Z}_4$.

\subsubsection{$Sp(N)$ instantons on $\mathbb{C}P^{2}/\mathbb{Z}_{n}$ with $n>4$}
\label{subsec:genarlsp}
Let us now consider the generic case of instantons on $\mathbb{Z}_n$ orbifolds of $\mathbb{C}P^2$ with $n >4$. Based on the previous examples, we can extract the generic pattern of both the quiver as well as the relation between the symplectic instanton on $\mathbb{C}P^2/\mathbb{Z}_n$ with its relative on $\mathbb{C}^2/\mathbb{Z}_n$.

Recall that $N$ is the sum of the ranks of the flavor groups in the ADHM quiver, while the ranks of the gauge groups is related to instanton number and, together with the relative flavor ranks, to other possible quantum numbers labelling the instanton. Unfortunately the precise identification between quiver data and instanton data is not know.
\\
\\
\textbf{$Sp(N)$ instantons on $\mathbb{C}P^{2}/\mathbb{Z}_{2n+1}$}
\\
\\
Elaborating on the previous examples, we conjecture that the theory describing symplectic instantons on $\mathbb{C}P^{2}/\mathbb{Z}_{2n+1}$ is related to its counterpart on $\mathbb{C}^2/\mathbb{Z}_{2n+1}$ as in fig.\ref{fig:cp22n+1}. Moreover, while the flavor groups continue to be the same, the ranks of the gauge groups are related in the following way 
\begin{equation}
K_{1} = k_{1}, \ \ K_{2}=\textrm{min}(k_{2},k_{3}), \ \ K_{3}=\textrm{min}(k_{4},k_{5}), \ \ ... \ \ K_{n+1}=\textrm{min}(k_{2n},k_{2n+1})\, .
\end{equation}
\\
\\

\noindent
\textbf{$Sp(N)$ instantons on $\mathbb{C}P^{2}/\mathbb{Z}_{2n}$- \textit{VS}}
\\
\\
Elaborating on the lowest $n$ cases, we can extrapolate both the quiver for VS symplectic instantons on $\mathbb{C}P^{2}/\mathbb{Z}_{2n}$ and their relation to their cousins (of course, VS) on $\mathbb{C}^2/\mathbb{Z}_{2n}$ as shown in fig.\ref{fig:cp22n+1first}. Moreover, while flavor nodes remain the same, the gauge rank identification is as follows

\begin{equation}
K_{1}=k_{1}, \ \ K_{2}=\textrm{min}(k_{2},k_{3}), \ \ ... \ \ K_{n-1}=\textrm{min}(k_{2n-2},k_{2n-1}),  \ \ K_{n}=k_{2n}\, . 
\end{equation}
\\
\\

\noindent
\textbf{$Sp(N)$ instantons on $\mathbb{C}P^{2}/\mathbb{Z}_{2n}$- \textit{NVS}}
\\
\\
Elaborating on the lowest $n$ cases, in this case we can extrapolate both the quiver for NVS symplectic instantons on $\mathbb{C}P^{2}/\mathbb{Z}_{2n}$ and their relation to their cousins (of course, NVS) on $\mathbb{C}^2/\mathbb{Z}_{2n}$ as shown in fig.\ref{fig:cp22n+1second}. Moreover, while flavor nodes remain the same, the gauge rank identification is as follows

\begin{equation}
 K_{1}=\textrm{min}(k_{1},k_{2}), \ \ K_{2}=\textrm{min}(k_{3},k_{4}), \ \ ... \ \ K_{n}=\textrm{min}(k_{2n-1},k_{2n})\, . 
\end{equation}
\\
\\
It is interesting to note that the ``merging nodes" are those going over, in the $\mathbb{C}^2/\mathbb{Z}_n$ parent, to unitary gauge groups. In turn, in the parent $\mathbb{C}^2/\mathbb{Z}_n$, these are the nodes admitting a blow-up mode through the FI parameter. It would be interesting to have a deeper understanding of these facts, as well as the topological data characterizing $Sp$ instantons on $\mathbb{C}P^2/\mathbb{Z}_n$.

\section{$SO(N)$ instantons on $\mathbb{C}P^2$ and its orbifolds}
\label{sec:son}

\begin{figure}[h!]
\centering
\includegraphics[scale=1.25]{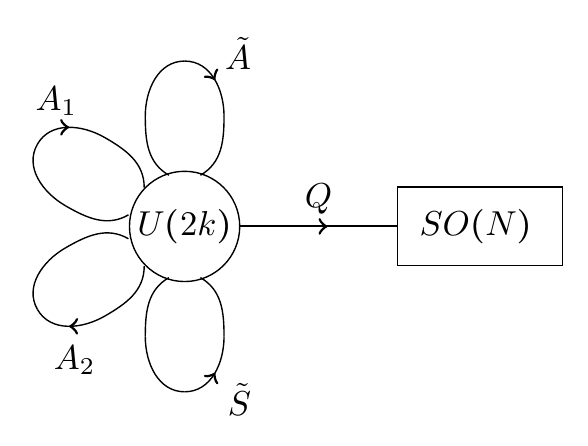}
\caption{Quiver diagram for $SO(N)$ instantons on $\mathbb{C}P^{2}$.\label{OinstantonsCP2}}

\end{figure}

We now turn to the case of orthogonal instantons on $\mathbb{C}P^2$ and its orbifolds. As described in \cite{Mekareeya:2014kca}, the ADHM construction for orthogonal instantons can be embedded into a $3d$ gauge theory which, in $3d$ $\mathcal{N}=2$ language, contains a $U(2\,k)$ vector multiplet as well as one chiral multiplet $\tilde{S}$ in the symmetric 2-index tensor representation of the gauge group and three chiral multiplets $A_1,\,A_2,\,\tilde{A}$ in the antisymmetric 2-index tensor representation of the gauge group. The corresponding quiver is shown in fig.\ref{OinstantonsCP2}. Note that the total flavor rank corresponds to $N$, while the gauge ranks --as well as the relative configurations of flavor ranks-- correspond to instanton number and other data specifying the instanton.

In turn, the superpotential reads

\begin{equation}
W=\epsilon^{\alpha\beta}\,(A_{\alpha})_{ab}\,\tilde{A}^{bc}\,(A_{\beta})_{cd}\,\tilde{S}^{da}+\tilde{S}^{ab}\,Q^i\,_a\,Q^j\,_b\,M_{ij}\, ,
\end{equation}
being $M$ given by

\begin{eqnarray}
M^{SO(2N)} = \left(\begin{array}{cc} 0 & {\bf 1}_{N\times N} \\ {\bf 1}_{N\times N} & 0 \end{array}\right)\,, \qquad M^{SO(2N+1)} = \left(\begin{array}{ccc} 0 & {\bf 1}_{N\times N} & 0 \\ {\bf 1}_{N\times N} & 0 & 0 \\ 0 & 0 & 1 \end{array}\right)\, .
\end{eqnarray}
As shown in \cite{Mekareeya:2014kca}, the construction of orthogonal instantons on $\mathbb{C}P^2$ can be embedded into that of a parent orthogonal instanton on $\mathbb{C}^2$. As a consequence, the Hilbert series of the instanton on $\mathbb{C}P^2/\mathbb{Z}_n$ matches that of its counterpart on $\mathbb{C}^2$.

\subsection{Resolved moduli space for orthogonal instantons}

The gauge group in the ADHM construction of orthogonal instantons on $\mathbb{C}P^2$ is $U(2\,k)$. However, as shown in \cite{Mekareeya:2014kca}, $k$ can be a half-integer while the Hilbert series is only sensitive to $\lfloor k\rfloor$, that is, the largest integer which is smaller or equal to $k$. In fact, it was conjectured that the instantons are distinguished by their second Stiefel-Whitney class written as $2\,(k-\lfloor k \rfloor)$. From this perspective, it is also natural to expect a notion of ``resolved moduli space" --resolved, as in the unitary case, in the sense that these extra directions associate to other quantum numbers are discerned--.

In order to explore the possibility of such resolved moduli space, following the example set by the unitary case, let us consider the simplest case where such extra directions are present. Since the instanton number was conjectured to be $\lfloor k\rfloor$, the analogous for orthogonal instantons to the case of a unitary instanton with $k_L=0$ as discussed in section \ref{Grassmanian} is $k=\frac{1}{2}$, corresponding to a $U(1)$ gauge theory. Such theory does not have the antisymmetric matter and, on the instanton branch, $\tilde{S}=0$. Therefore the theory only contains the $Q$'s out of which no gauge invariant can be constructed. Hence, very much like the Grassmanian, we find a extra compact manifold associated to the extra directions labelled in this case by the Stiefel-Whitney class. Just like in the unitary case, we can imagine resolving these directions by ungauging the $U(1)$ global symmetry. It is then straightforward to compute the instanton branch Hilbert series, which, upon unrefining the $SO(N)$ labels, reads

\begin{equation}
HS=\frac{1+t}{(1-t)^{N-1}}\ .
\end{equation}
Interestingly, this can be written as

\begin{equation}
HS=\frac{2}{(1-t)^{N-1}}-\frac{1}{(1-t)^{N-2}}\ ,
\end{equation}
which is the Hilbert series for two $\mathbb{C}^{N-1}$ meeting at a $\mathbb{C}^{N-2}$. This is a dimension $N-1$ manifold  analogous to the cone over the Grassmanian in the unitary case. Note that the dimension of the resolved moduli space is $2\,k\,(N-2)$, while that seen by the Hilbert series is $2\,\lfloor k\rfloor\,(N-2)$ \cite{Mekareeya:2014kca}. Hence the difference is $2\,(N-2)\,(k-\lfloor k\rfloor)$. Particularizing to the case $k=\frac{1}{2}$ this is an $N-2$ dimensional compact manifold. Then, the complex cone over it is a $N-1$ complex dimensional manifold, just as we have found.


Note that the case of symplectic instantons does not admit a similar construction. For example, in the quiver in fig.\ref{SpADHM}, the instanton branch appears upon setting to zero an antisymmetric field while keeping the symmetric fields. Hence the theory is never empty of gauge invariant operators, as it happens in the case of unitary and orthogonal instantons, therefore suggesting that no compact directions exist in that case.

\subsection{Constructing $SO(N)$ instantons on $\mathbb{C}P^2/\mathbb{Z}_n$ }

Let us now turn to the construction of orthogonal instantons upon orbifolding the base space. In view of the ALE case, and following the symplectic instanton case in section \ref{sec:sp}, we construct the theories whose instanton branch describes orthogonal instantons on $\mathbb{C}P^2/\mathbb{Z}_n$ by first orbifolding and then orientifolding the unitary instanton case following the rules in \cite{Franco:2007ii} and in \cite{GarciaEtxebarria:2012qx}. As for symplectic instantons, we have qualitatively different situations depending on whether $n$ is even or odd:

\begin{itemize}
\item If \textbf{$n$ is odd} we have only one type of quiver diagram, corresponding to the fact that we have only one inequivalent way to cut the quiver diagram with a line.
\item If \textbf{$n$ is even} we have two types of quiver gauge theories corresponding to two possible inequivalent ways in which we can cut the quiver diagram with a line. Inspired by the ALE case, we will refer to theme as the \textit{VS} case and the \textit{NVS} case respectively.
\end{itemize} 

Also in this case there can be hybrid configurations associated with one choice for the values of the signs implementing the orientifold prescription. As above, we restrict our analysis to the configuration of signs corresponding to the quantum field theory whose instanton branch describes orthogonal instantons on $\mathbb{C}P^2/\mathbb{Z}_n$ which, for the case of even $n$, are either VS or NVS.
Just as in the other cases, the rank of the $SO(N)$ bundle corresponds to the sum of flavor ranks in the ADHM quiver. The rest of the ADHM data corresponds to other data specifying the instanton.

\subsubsection{$SO(N)$ instantons on $\mathbb{C}P^{2}/\mathbb{Z}_{2}$ \ - \textit{VS}}

\begin{figure}[h!]
\centering
\includegraphics[scale=1]{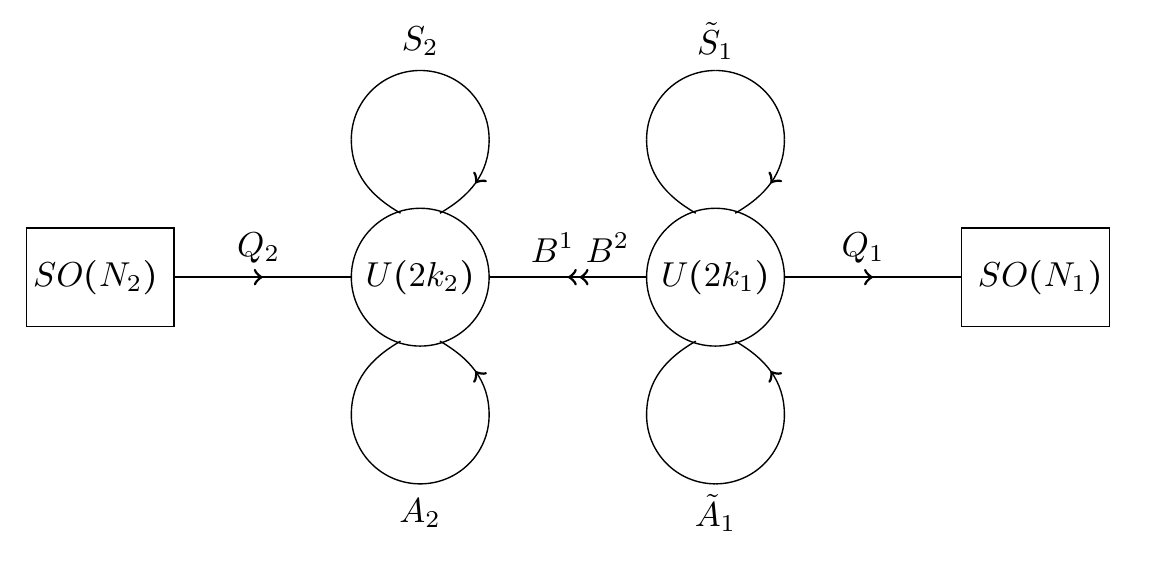}

\caption{Quiver diagram for VS orthogonal instantons on $\mathbb{C}P^{2}/\mathbb{Z}_{2}$. \label{fig:soz2first}}         
        
\end{figure}

Starting from the $\mathbb{C}P^{2}/\mathbb{Z}_{2}$ and applying the rules in \cite{Franco:2007ii} we obtain the theory for $SO(N)$ instantons on $\mathbb{C}P^{2}/\mathbb{Z}_{2}$. The corresponding quiver diagram is reported in fig.\ref{fig:soz2first}, while we summarize the transformations of the fields under the different groups in table \ref{tab:sonz21}

\begin{table}[h!]

\renewcommand\arraystretch{1.3}
\centering
\resizebox{12cm}{!}{
\begin{tabular}{c||cc||cccc||} 
Fields & $U(2k_{1})$ & $U(2k_{2})$ & $SO(N_{1})$ & $SO(N_{2})$ & SU(2) & U(1) \\ 
\hline 
$\tilde{A}_{1}$ & $[0,1,0...,0]_{-2}$ & $[\textbf{0}]$ & $[\textbf{0}]$ & $[\textbf{0}]$ & $[\textbf{0}]$ & 1/2 \\ 
\hline 
$\tilde{S}_{1}$ & $[2,0,...,0]_{-2}$ & $[\textbf{0}]$ & $[\textbf{0}]$ & $[\textbf{0}]$ & $[\textbf{0}]$ & 1/2 \\ 
\hline 
$A_{2}$ & $[\textbf{0}]$ & $[0,1,0,...,0]_{+2}$ & $[\textbf{0}]$ & $[\textbf{0}]$ & $[\textbf{0}]$ & 1/2 \\ 
\hline 
$S_{2}$ & $[\textbf{0}]$ & $[2,0,...,0]_{+2}$ & $[\textbf{0}]$ & $[\textbf{0}]$ & $[\textbf{0}]$ & 1/2 \\ 
\hline 
$B^{j}$ & $[1,0,...,0]_{+1}$ & $[0,...,0,1]_{+1}$ & $[\textbf{0}]$ & $[\textbf{0}]$ & $[1]$ & 1/4 \\ 
\hline 
$Q_{1}$ & $[1,0,...,0]_{+1}$ & $[\textbf{0}]$ & $[1,0,...,0]$ & $[\textbf{0}]$ & $[\textbf{0}]$ & 1/2 \\ 
\hline 
$Q_{2}$ & $[\textbf{0}]$ & $[0,...,0,1]_{+1}$ & $[\textbf{0}]$ & $[1,0,...,0]$ & $[\textbf{0}]$ & 1/2 \\ 
\hline 
$F_{1}$ & $[2,0,...,0]_{+2}$ & $[\textbf{0}]$ & $[\textbf{0}]$ & $[\textbf{0}]$ & $[\textbf{0}]$ & 1 \\ 
\hline 
$F_{2}$ & $[\textbf{0}]$ & $[2,0,...,0]_{-2}$ & $[\textbf{0}]$ & $[\textbf{0}]$ & $[\textbf{0}]$ & 1 \\ 
\end{tabular} 
}

\caption{Transformations of the fields for VS orthogonal instantons on $\mathbb{C}P^2/\mathbb{Z}_2$. \label{tab:sonz21}}
\end{table}

The branch of the moduli space that can be identified with $SO(N)$ instantons on $\mathbb{C}P^{2}/\mathbb{Z}_{2}$ is the one on which $\tilde{S}_{1}=0$ and $S_{2}=0$. The Hilbert series of the instanton branch corresponding the VS theory with flavor symmetry $SO(N_{1}) \times SO(N_{2})$ and gauge ranks $\textbf{k}=(k_{1},k_{2})$ is

\begin{equation}\begin{aligned}
\label{sonhsz2}
& H[\textbf{k},F, \mathbb{C}P^{2}/\mathbb{Z}_{2}](t,x,\textbf{y},\textbf{d})= \int d \mu_{U(2k_{1})}(\textbf{z})\int d \mu_{U(2k_{2})}(\textbf{p}) \times \\
& \textrm{PE}[\chi_{A_{2}}t^{2} +\chi_{\tilde{A}_{1}}t^{2}  + \chi_{B_{j}}t   + \chi_{Q_{1}}t^{2}  + \chi_{Q_{2}}t^{2} -\chi_{F_{1}}t^{4} -\chi_{F_{2}}t^{4}]\, ,
\end{aligned}\end{equation}
where $\textbf{z}$ and $\textbf{p}$ are the fugacities of the $U(2k_{1})$ and $U(2k_{2})$ gauge groups respectively while $\textbf{y}$ and $\textbf{d}$ denote the fugacities of the $SO(N_{1})$ and $SO(N_{2})$ flavor groups. Finally $x$ is the fugacity of the $SU(2)$ symmetry acting on the $B_{j}$ doublet. The contribution of each field is given by
\begin{equation*}
\chi_{F_{1}}=\sum\nolimits_{1 \leq a \leq b \leq 2k_{1}} z_{a}z_{b}, \ \ \chi_{F_{2}}=\sum\nolimits_{1 \leq a \leq b \leq 2k_{2}} p_{a}^{-1}p_{b}^{-1},
\end{equation*}
\begin{equation*}
\chi_{A_{2}} = \sum\nolimits_{1 \leq a < b \leq 2k_{2}} p_{a}p_{b}, \ \ \chi_{\tilde{A}_{1}} = \sum\nolimits_{1 \leq a < b \leq 2k_{1}} z_{a}^{-1}z_{b}^{-1}, \ \ \chi_{B^{j}} = \left(x+\frac{1}{x}\right)\sum_{a=1}^{2k_{1}}\sum_{b=1}^{2k_{2}} z_{a}p_{b}^{-1},
\end{equation*}
\begin{equation*}
\chi_{Q_{1}} = \left(\sum_{a=1}^{2k_{1}}z_{a}\right) \times \left\{
\begin{array}{l}
\sum\limits_{i=1}^{N_{1}/2}\left(y_{i} +\frac{1}{y_{i}}\right) \ N_{1} \ \textrm{even}\, , \\ 1 + \sum\limits_{i=1}^{(N_{1}-1)/2}\left(y_{i} + \frac{1}{y_{i}}\right)  \ N_{1} \ \textrm{odd}\, , 
\end{array}
\right. \ \ \chi_{Q_{2}}=\left(\sum_{b=1}^{2k_{2}}p_{b}^{-1}\right)\times \left\{\begin{array}{l}
\sum\limits_{i=1}^{N_{2}/2}\left(d_{i} +\frac{1}{d_{i}}\right) \ N_{2} \ \textrm{even}\, , \\ 1 + \sum\limits_{i=1}^{(N_{2}-1)/2}\left(d_{i} + \frac{1}{d_{i}}\right)  \ N_{2} \ \textrm{odd}\, . 
\end{array}
\right. 
\end{equation*}

Explicitly computing the Hilbert series with gauge group $G=U(2k_{1})  \times U(2k_{2})$ and flavor group $SO(N_{1}) \times SO(N_{2}) $ for the moduli space of instantons on $\mathbb{C}P^{2}/\mathbb{Z}_{2}$ shows that it is equal to the Hilbert series for $Sp(N)$ instantons on $\mathbb{C}^{2}/\mathbb{Z}_{2}$ with gauge group $G=Sp(K_{1}) \times Sp(K_{2})$ (see \cite{Dey:2013fea} for more details) upon identifying

\begin{equation}
K_{1} = k_{1}\, ,  \qquad K_{2} = k_{2}\, .
\end{equation}
Let us show a few explicit examples.
\\
\\
\textbf{$SO(5)$ instanton: $\mathbf{k} = (1,1)$ and $\mathbf{N}=(2,3)$}\\
\\
Using eq.(\ref{sonhsz2}) and unrefining we find that
\begin{equation*}
H[\mathbf{k}=(1,1),SO(2) \times SO(3), \mathbb{C}P^{2}/\mathbb{Z}_{2}](t,1,1,1) = \frac{1-t^{3} +5t^{6} +4t^{9} +4t^{12} +4t^{15} +5t^{18} -t^{21} +t^{24}}{(1-t^{3})^{6}(1+t^{3})^{2}(1+t^{3}+t^{6})^{3}}\, , 
\end{equation*}
which is the Hilbert series for $SO(5)$ instanton on $\mathbb{C}^{2}/\mathbb{Z}_{2}$ with $\mathbf{K}=(1,1)$ and $\mathbf{N}=(2,3)$.\\
\\
\textbf{$SO(6)$ instanton: $\mathbf{k} = (1,1)$ and $\mathbf{N}=(3,3)$}\\
\\
Using eq.(\ref{sonhsz2}) and unrefining we find that
\begin{equation*}
H[\mathbf{k}=(1,1),SO(3) \times SO(3), \mathbb{C}P^{2}/\mathbb{Z} _{2}](t,1,1,1) = \frac{1-2t^{3} + 8t^{6} + 5t^{12} +12t^{15} +5t^{18} +8t^{24} -2t^{27} +t^{30}}{(1-t^{3})^{8}(1+t^{3})^{2}(1+t^{3}+t^{6})^{4}}\, ,
\end{equation*}
which is the Hilbert series for $SO(6)$ instanton on $\mathbb{C}^{2}/\mathbb{Z}_{2}$ with $\mathbf{K}=(1,1)$ and $\mathbf{N}=(3,3)$.\\
\\

\subsubsection{$SO(N)$ instantons on $\mathbb{C}P^{2}/\mathbb{Z}_{2}$ \ - \textit{NVS}}

\begin{figure}[h!]
\centering
\includegraphics[scale=1]{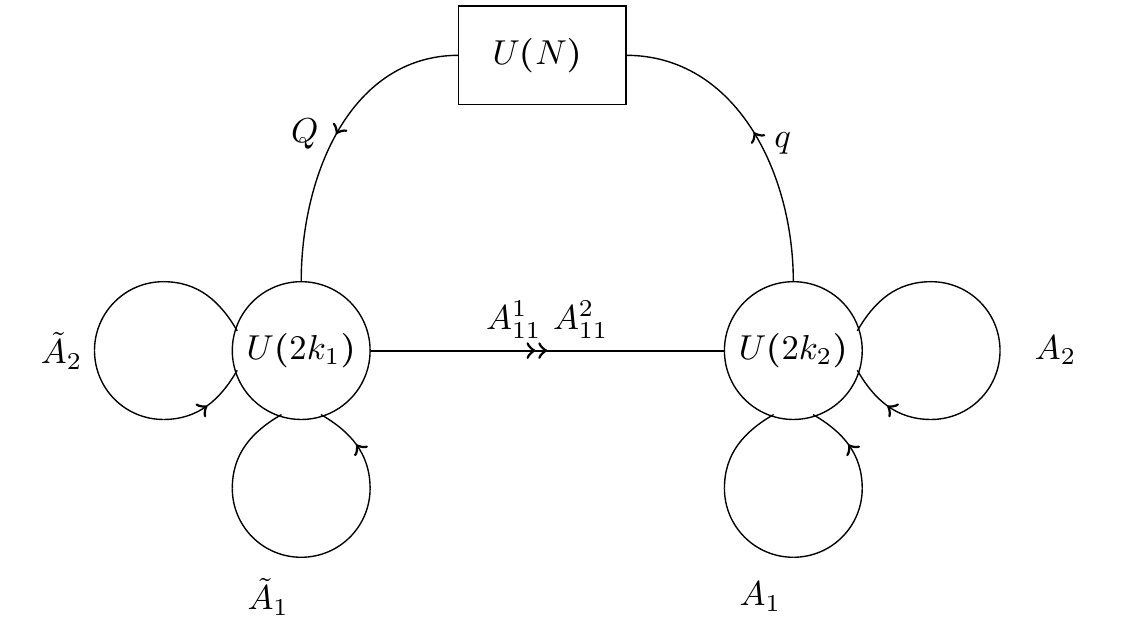}
 \caption{Quiver diagram for NVS orthogonal instantons on $\mathbb{C}P^{2}/\mathbb{Z}_{2}$. \label{fig:sonz2second}}

\end{figure}

Let us now consider the case of orthogonal NVS instantons on $\mathbb{C}P^2/\mathbb{Z}_2$ upon choosing the other, non-equivalent, way to cut the quiver diagram. The quiver diagram of the corresponding theory is reported in fig.\ref{fig:sonz2second}, while the transformations of the fields and of the F-term are summarized in table \ref{tab:sonz22}

\begin{table}[h!]

\renewcommand\arraystretch{1.3}
\centering
\resizebox{12cm}{!}{
\begin{tabular}{c||cc||ccc||} 
Fields & $U(2k_{1})$ & $U(2k_{2})$ & $U(N)$  & $SU(2)$ & $U(1)$ \\ 
\hline 
$\tilde{A}_{1},\tilde{A}_{2}$ & $[0,1,0...,0]_{-2}$ & $[\textbf{0}]$ & $[\textbf{0}]$  & $[1]$ & 1/4 \\ 
\hline 
$A_{1},A_{2}$ & $[\textbf{0}]$ & $[0,1,0,...,0]_{+2}$ & $[\textbf{0}]$  & $[1]$ & 1/4 \\ 
\hline 
$A^{2}_{11}$ & $ [1,0,...,0]_{+1}$ & $[0,0,...,1]_{+1}$ & $[\textbf{0}]$  & $[\textbf{0}]$ & 1/2 \\ 
\hline 
$q$ & $[\textbf{0}]$ & $[1,0,...,0]_{+1}$ & $[0,0,...,1]$ & $[\textbf{0}]$ & 1/2 \\ 
\hline 
$Q$ & $[0,...,0,1]_{+1}$ & $[\textbf{0}]$ & $[1,0,...,0]$ & $[\textbf{0}]$ & 1/2 \\ 
\hline 
$F$ & $[0,...,0,1]_{+1}$ & $[1,0,...,0]_{+1}$ & $[\textbf{0}]$ & $[\textbf{0}]$ & 1 \\ 
\end{tabular} 
}

\caption{Transformations of the fields for NVS orthogonal instantons on $\mathbb{C}P^{2}/\mathbb{Z}_{2}$. \label{tab:sonz22}}
\end{table}

The branch of the moduli space that can be identified with $SO(N)$ instantons on $\mathbb{C}P^{2}/\mathbb{Z}_{2}$ is the one on which $A^{1}_{11}=0$. The Hilbert series of the instanton branch corresponding to the NVS theory with flavor symmetry $U(N)$ and ranks $\textbf{k}=(k_{1},k_{2})$ is
\begin{equation}\begin{aligned}
\label{sonhsz2sec}
& H[\textbf{k},F, \mathbb{C}P^{2}/\mathbb{Z}_{2}](t,x,\textbf{y})= \int d \mu_{U(2k_{1})}(\textbf{z})\int d \mu_{U(2k_{2})}(\textbf{p}) \times \\
& \textrm{PE}[\chi_{A_{i}}t  +\chi_{\tilde{A}_{j}}t  + \chi_{A_{11}^{2}}t^{2}   + \chi_{Q}t^{2}  + \chi_{q}t^{2} -\chi_{F}t^{4} ]\, ,
\end{aligned}\end{equation}
where $\textbf{z}$ and $\textbf{p}$ are the fugacities of the $U(2k_{1})$ and $U(2k_{2})$ gauge groups respectively while $\textbf{y}$  denote the fugacity of the $U(N)$ flavor group. Finally $x$ is the fugacity of the $SU(2)$ acting on the $A_{\beta}$ and on the $\tilde{A}_{\alpha}$ doublets. The contribution of each field is given by
\begin{equation*}
\chi_{A_{j}} = \left(x+\frac{1}{x}\right)\sum\nolimits_{1 \leq a < b \leq 2k_{2}} p_{a}p_{b}, \ \ \chi_{\tilde{A}_{i}} =\left(x+\frac{1}{x}\right) \sum\nolimits_{1 \leq a < b \leq 2k_{1}} z_{a}^{-1}z_{b}^{-1}, \ \  
\end{equation*}
\begin{equation*}
\chi_{A^{2}_{11}} = \sum_{a=1}^{2k_{1}}\sum_{b=1}^{2k_{2}} z_{a}p_{b}^{-1}, \ \
\chi_{Q} = \sum_{i=1}^{N}\sum_{a=1}^{2k_{1}}z_{a}^{-1}y_{i}, \ \ \chi_{q}=\sum_{
j=1}^{N}\sum_{b=1}^{2k_{2}}p_{b}y_{j}^{-1}, \ \ \chi_{F}=\sum_{a=1}^{2k_{1}}\sum_{b=1}^{2k_{2}} z_{a}^{-1}p_{b}\, .
\end{equation*}
The explicit computation of the instanton branch Hilbert series with gauge group $G=U(2k_{1})  \times U(2k_{2})$ and flavor group $U(N)$ shows that it coincides with the Hilbert series for $SO(N)$ instantons on $\mathbb{C}^{2}/\mathbb{Z}_{2}$ with gauge group $G=U(2K_{1})$ (see \cite{Dey:2013fea} for more details regarding the $\mathbb{C}^{2}/\mathbb{Z}_{2}$ Hilbert series) upon setting
\begin{equation}
K_{1} = \textrm{min}(k_{1},k_{2})\, .
\end{equation}
Let us show explicit examples supporting our claim.
\\
\\
\textbf{$SO(6)$ instanton: $\mathbf{k} = (1,1)$ and $N=3$}\\
\\
Using eq.(\ref{sonhsz2sec}) and unrefining we find that
\begin{equation*}\begin{aligned}
& H[\mathbf{k}=(1,1),U(3), \mathbb{C}P^{2}/\mathbb{Z}_{2}](t,1,1,1,1) =  \\
& = \frac{1+2t^{3} +9t^{6} +24t^{9} +50t^{12} +76t^{15} +108t^{18} +120t^{21} + 108t^{24} + \textrm{palindrome} + ... + t^{42}}{(1-t^{3})^{8}(1+t^{3})^{6}(1+t^{3}+t^{6})^{12}}\, ,
\end{aligned}\end{equation*}
which is the Hilbert series for $SO(6)$ instanton on $\mathbb{C}^{2}/\mathbb{Z}_{2}$ with $\mathbf{K}=(1,1)$ and $N=3$.\\
\\
\textbf{$SO(8)$ instanton: $\mathbf{k} = (1,1)$ and $N=4$}\\
\\
Using eq.(\ref{sonhsz2sec}) and unrefining we find that
\begin{equation*}\begin{aligned}
& H[\mathbf{k}=(1,1),U(4), \mathbb{C}P^{2}/\mathbb{Z}_{2}](t,1,1,1,1,1) =  \\
& = \frac{1}{(1-t^{3})^{12}(1+t^{3})^{8}(1+t^{3}+t^{6})^{18}}{(
1+2t^{3} +14t^{6} +44t^{9} +123t^{12} +272t^{15} +546t^{18} +886t^{21} +} \\
& + 1259t^{24}  + 1544t^{27} + 1678t^{30} + \textrm{palindrome} + ... + t^{60})\, ,
\end{aligned}\end{equation*}
which is the Hilbert series for $SO(8)$ instanton on $\mathbb{C}^{2}/\mathbb{Z}_{2}$ with $\mathbf{K}=(1,1)$ and $N=4$.\\
\\

\begin{figure}[h!]
\centering
\includegraphics[scale=1]{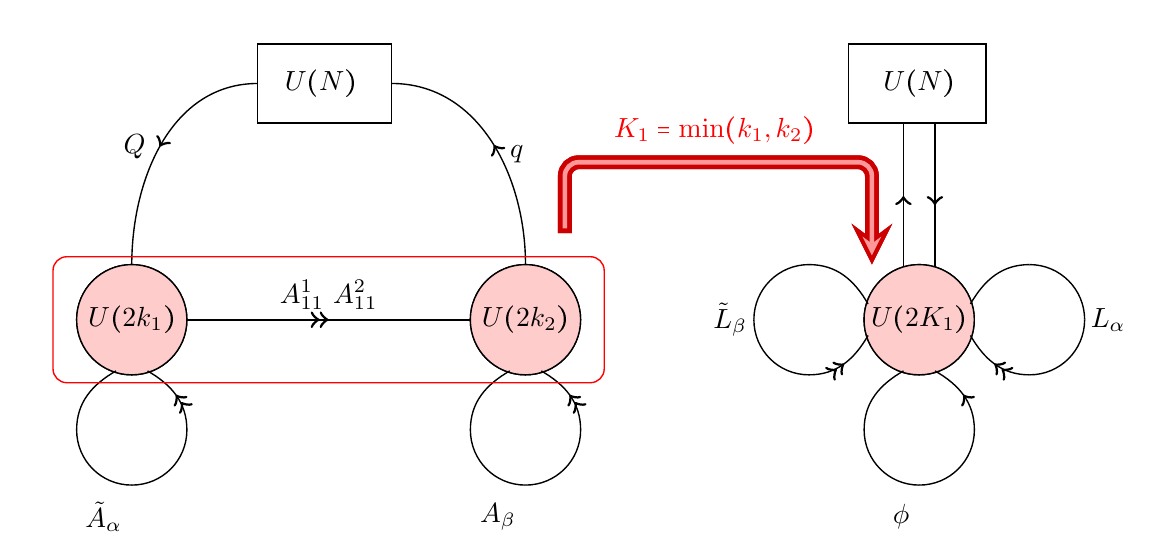}

\caption{Relation between the $\mathbb{C}P^{2}/\mathbb{Z}_{2}$ quiver gauge theory in the NVS case (on the left) and the corresponding $\mathbb{C}^{2}/\mathbb{Z}_{2}$ quiver gauge theory (on the right). Where $\tilde{L}_{\beta}$ are two fields in the antisymmetric conjugate representation of the gauge group $U(2K_{1})$. While $L_{\alpha}$ are two fields in the antisymmetric representation of the gauge group of the gauge group $U(2K_{1})$ (see \cite{Dey:2013fea} for more details).\label{Z2NVSsumm}}

\end{figure}

We graphically summarize in fig.\ref{Z2NVSsumm} the relation between the NVS orthogonal instanton on $\mathbb{C}P^2/\mathbb{Z}_2$ and its cousin on $\mathbb{C}^2/\mathbb{Z}_2$.

\subsubsection{$SO(N)$ instantons on $\mathbb{C}P^{2}/\mathbb{Z}_{3}$ }

\begin{figure}[h!]
\centering
\includegraphics[scale=1]{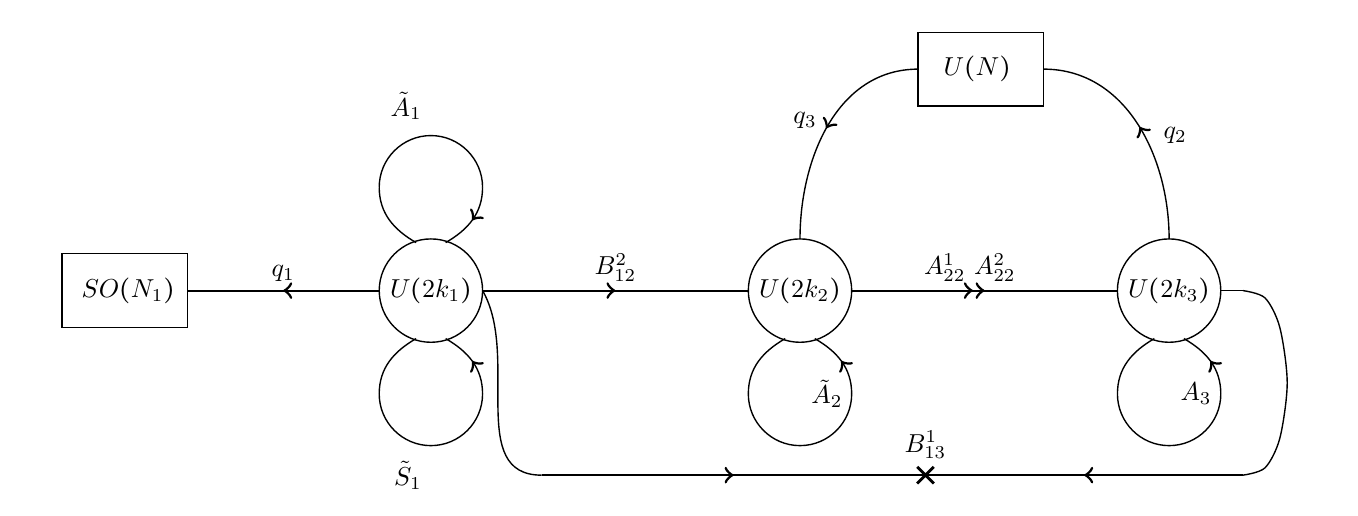}

 \caption{Quiver diagram for $SO(N)$ instantons on $\mathbb{C}P^{2}/\mathbb{Z}_{3}$. \label{fig:orientifoldz3son1}}

\end{figure}

In this case there is only one inequivalent choice of the orientifold action. We report in fig.\ref{fig:orientifoldz3son1} the quiver diagram of the corresponding field theory, while we summarize the fields and F-terms transformations in table \ref{tab:son3}.

\begin{table}\center
\resizebox{13cm}{!}{
\renewcommand\arraystretch{1.3} 
\begin{tabular}{c||ccc||cccc||}

Fields & $U(2k_{1})$ & $U(2k_{2})$ & $U(2k_{3})$ & $SO(N_{1})$ & $U(N_{2})$ & $U(1)_{x}$ & $U(1)$ \\ 
\hline 
$q_{1}$ & $[1,0,...,0]_{+1}$ &  $[\textbf{0}]$ &  $[\textbf{0}]$ & $[1,0,...,0]$ &  $[\textbf{0}]$ &  $[\textbf{0}]$ & 1/2 \\ 
\hline 
$q_{3}$ &  $[\textbf{0}]$ &  $[\textbf{0}]$ & $[1,0,...0]_{+1}$ &  $[\textbf{0}]$ & $[0,...,0,1]_{+1}$ &  $[\textbf{0}]$ & 1/2 \\ 
\hline 
$q_{2}$ &  $[\textbf{0}]$ & $[0,...,0,1]_{+1}$ &  $[\textbf{0}]$ &  $[\textbf{0}]$ & $[1,0,...,0]_{+1}$ &  $[\textbf{0}]$ & 1/2 \\ 
\hline 
$B_{12}^{2}$ & $[1,0,...,0]_{+1}$ & $[0,...,0,1]_{+1}$ &  $[\textbf{0}]$ &  $[\textbf{0}]$ &  $[\textbf{0}]$ &  $[\textbf{0}]$ & 1/4 \\ 
\hline 
$A^{2}_{22}$ &  $[\textbf{0}]$ & $[1,0,...,0]_{+1}$ & $[0,...,0,1]_{+1}$ &  $[\textbf{0}]$ &  $[\textbf{0}]$ &  $[\textbf{0}]$ & 1/2 \\ 
\hline 
$B^{1}_{13}$ & $[1,0,...,0]_{+1}$ &  $[\textbf{0}]$ & $[1,0,...,0]_{+1}$ &  $[\textbf{0}]$ &  $[\textbf{0}]$ &  $[\textbf{0}]$ & 1/4 \\ 
\hline 
$\tilde{A}_{1}$ & $[0,1,0,...,0]_{-2}$ &  $[\textbf{0}]$ &  $[\textbf{0}]$ &  $[\textbf{0}]$ & $[\textbf{0}]$ &  $[\textbf{0}]$ & 1/2 \\ 
\hline 
$\tilde{A}_{2}$ &  $[\textbf{0}]$ & $[0,1,0,...,0]_{-2}$ &  $[\textbf{0}]$ &  $[\textbf{0}]$ &  $[\textbf{0}]$ & $1/x$ & 1/4 \\ 
\hline 
$A_{3}$ &  $[\textbf{0}]$ &  $[\textbf{0}]$ & $[0,1,0,...,0]_{+2}$ &  $[\textbf{0}]$ &  $[\textbf{0}]$ & $x$ & 1/4 \\ 
\hline 
$F_{1}$ & $[2,0,...,0]_{+2}$ &  $[\textbf{0}]$ &  $[\textbf{0}]$ &  $[\textbf{0}]$ &  $[\textbf{0}]$ &  $[\textbf{0}]$ & 1 \\ 
\hline 
$F_{2}$ &  $[\textbf{0}]$ & $[0,...,0,1]_{+1}$ & $[1,0,...,0]_{+1}$ &  $[\textbf{0}]$ &  $[\textbf{0}]$ &  $[\textbf{0}]$ & 1 \\ 
\end{tabular} 
}
\caption{Transformations of the fields for $SO(N)$ instantons on $\mathbb{C}P^{2}/\mathbb{Z}_{3}$. \label{tab:son3}}
\end{table}

The branch of the moduli space that can be identified with $SO(N)$ instantons on $\mathbb{C}P^{2}/\mathbb{Z}_{3}$ is the one on which $A^{1}_{22}=0$ and $\tilde{S}_{1}=0$. The Hilbert series of the instanton branch corresponding to a theory with flavor symmetry $SO(N_{1}) \times U(N_{2})$ and gauge ranks $\textbf{k}=(k_{1},k_{2},k_{3})$ is
\begin{equation}\begin{aligned}
\label{sonhsz3}
& H[\textbf{k},F, \mathbb{C}P^{2}/\mathbb{Z}_{3}](t,x,\textbf{y},\textbf{d})= \int d \mu_{U(2k_{1})}(\textbf{z})\int d \mu_{U(2k_{2})}(\textbf{p}) \int d \mu_{U(2k_{3})}(\textbf{w}) \times \\
& \textrm{PE}[\chi_{q_{1}}t^2  +\chi_{q_{2}}t^{2}  + \chi_{q_{3}}t^{2}   + \chi_{B_{12}^{2}}t  + \chi_{A^{2}_{22}}t^{2} +\chi_{B^{1}_{13}}t + \chi_{\tilde{A}_{1}}t^{2} +\chi_{\tilde{A}_{2}}t +\chi_{A_{3}}t -\chi_{F_{1}}t^{4} -\chi_{F_{2}}t^{4}]\, ,
\end{aligned}\end{equation}
where $\textbf{z}$, $\textbf{p}$ and $\textbf{w}$ are the fugacities of the $U(2k_{1})$, $U(2k_{2})$ and $U(2k_{3})$ gauge groups respectively, while $\textbf{y}$  denote the fugacity of the $SO(N_{1})$ flavor group and $\textbf{d}$ the fugacity of the $U(N_{2})$ gauge group. Finally $x$ is the fugacity of the $U(1)_{x}$ symmetry acting on $\tilde{A}_{2}$ and $A_{3}$. The contribution of each field and of the F-terms are

\begin{equation*}
\chi_{B^{2}_{12}} = \sum_{a=1}^{2k_{1}}\sum_{b=1}^{2k_{2}} z_{a}p_{b}^{-1}, \ \
\chi_{A^{2}_{22}} = \sum_{a=1}^{2k_{2}}\sum_{b=1}^{2k_{3}}p_{a}w_{b}^{-1}, \ \ \chi_{B^{1}_{13}}=\sum_{
a=1}^{2k_{1}}\sum_{b=1}^{2k_{3}}z_{a}w_{b}, \ \ \chi_{F_{1}} = \sum\nolimits_{1\leq a \leq b \leq 2k_{1}} z_{a}z_{b}, 
\end{equation*}
\begin{equation*}
\chi_{q_{1}} = \sum_{a=1}^{2k_{1}}z_{a} \times \left\{
\begin{array}{l}
\sum\limits_{i=1}^{N_{1}/2}\left(y_{i} +\frac{1}{y_{i}}\right) \ N_{1} \ \textrm{even}\, , \\ 1 + \sum\limits_{i=1}^{(N_{1}-1)/2}\left(y_{i} + \frac{1}{y_{i}}\right)  \ N_{1} \ \textrm{odd}\, , 
\end{array}
\right. \ \ \chi_{q_{2}} = \sum_{b=1}^{2k_{3}}\sum_{j=1}^{N_{2}} w_{b}d_{j}^{-1}, \ \ \chi_{q_{3}} = \sum_{a=1}^{2k_{2}}\sum_{j=1}^{N_{2}} p_{a}^{-1}d_{j}, 
\end{equation*}
\begin{equation*}
\chi_{\tilde{A}_{1}} = \sum\nolimits_{1\leq a < b \leq 2k_{1}} z_{a}^{-1}z_{b}^{-1}, \ \ \chi_{\tilde{A}_{2}} = \sum\nolimits_{1\leq a < b \leq 2k_{2}} p_{a}^{-1}p_{b}^{-1}x^{-1}, \ \ \chi_{A_{3}} = \sum\nolimits_{1\leq a < b \leq 2k_{3}} w_{a}w_{b}x, \ \ \chi_{F_{2}} = \sum_{a=1}^{2k_{2}}\sum_{b=1}^{2k_{3}}p_{a}^{-1}w_{b}\, .
\end{equation*}

\begin{figure}[h!]
\centering
\includegraphics[scale=0.8]{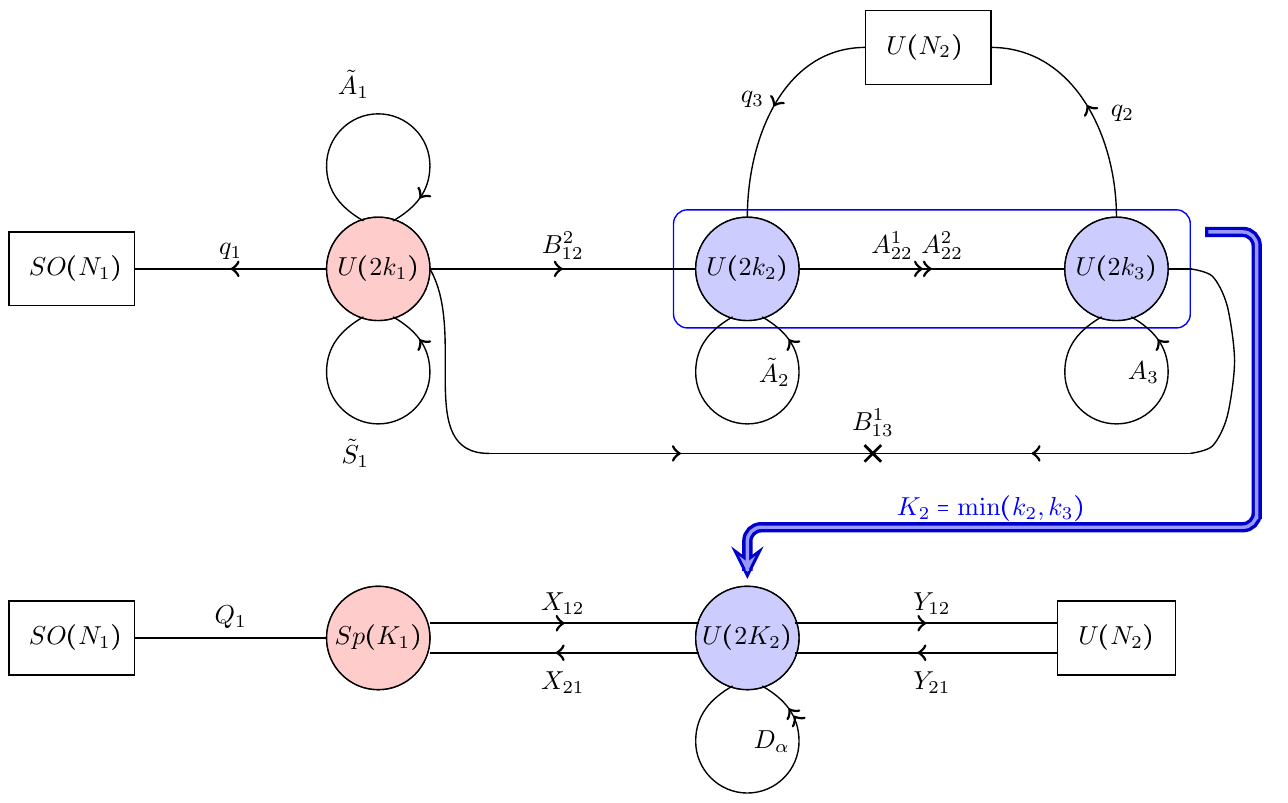}

 \caption{Relation between the quiver diagram for $SO(N)$ instantons on $\mathbb{C}P^{2}/\mathbb{Z}_{3}$ and the quiver diagram for $SO(N)$ instantons on $\mathbb{C}^{2}/\mathbb{Z}_{3}$. Being $D_{\alpha}$ two fields transforming in the antisymmetric representation of $ U(2K_{2})$ gauge group (see \cite{Dey:2013fea} for more details).\label{OZ3rel}}

\end{figure}

By explicitly evaluating the Hilbert series with gauge group $G=U(2k_{1}) \times U(2k_{2}) \times U(2k_{3})$ and flavor group $SO(N_{1}) \times U(N_{2})$ for the moduli space of instantons on $\mathbb{C}P^{2}/\mathbb{Z}_{3}$ we find it to be equal to the Hilbert series for $SO(N)$ instantons on $\mathbb{C}^{2}/\mathbb{Z}_{3}$ with gauge group $G=Sp(K_{1})\times U(2K_{2}) $ and flavor groups $SO(N_{1}) \times U(N_{2})$ (see \cite{Dey:2013fea} for more details) with the identification

\begin{equation}
K_{1} = k_{1}, \ \ K_{2}= \textrm{min}(k_{2},k_{3})\, .
\end{equation}
Supporting our claim, we show a few explicit examples.
\\
\\
\textbf{$SO(5)$ instanton: $\textbf{k}=(1,1,1)$, and $\textbf{N}=(3,1)$} \\
\\
Using eq.(\ref{sonhsz3}) and unrefing we find that
\begin{equation*}\begin{aligned}
 & H[\textbf{k}=(1,1,1),SO(3) \times U(1), \mathbb{C}P^{2}/\mathbb{Z}_{3}](t,1,1,1) =
 \\
& = \frac{1}{(1-t^{3})^{6}(1+t^{3})^{4}(1+t^{6})^{2}(1+t^{3}+t^{6})^{3}}(1+t^{3} + 4t^{6} +9t^{9} + 18t^{12} + 25t^{15} +33t^{18} + \\
& + 30t^{21} +33t^{24} + \textrm{palindrome} + ... + t^{42})\, ,  
\end{aligned}\end{equation*}
which is the Hilbert series for $SO(5)$ instantons on $\mathbb{C}^{2}/\mathbb{Z}_{3}$ with $\textbf{N}=(3,1)$ and $\textbf{K}=(1,1)$.\\
\\
\textbf{$SO(5)$ instanton: $\textbf{k}=(1,1,1)$, and $\textbf{N}=(1,2)$}\\
\\
Using eq.(\ref{sonhsz3}) and unrefing we find that
\begin{equation*}
H[\textbf{k}=(t,1,1,1),SO(2) \times U(1), \mathbb{C}P^{2}/\mathbb{Z}_{3}](t,1,1,1)
= \frac{1-2t^{3}+5t^{6}-2t^{9}+6t^{12}-2t^{15}+5t^{18}-2t^{21}+t^{24}}{(1-t^{3})^{6}(1+t^{6})(1+2t^{3}+2t^{6}+t^{9})^{2}}\, ,
\end{equation*}
which is the Hilbert series for $SO(5)$ instantons on $\mathbb{C}^{2}/\mathbb{Z}_{3}$ with $\textbf{N}=(1,2)$ and $\textbf{K}=(1,1)$.
\\
\\
We can as well graphically summarize the relation between the orthogonal instanton on $\mathbb{C}P^2/\mathbb{Z}_3$ and its cousin on $\mathbb{C}^2/\mathbb{Z}_3$ as in fig. \ref{OZ3rel}.

\subsubsection{$SO(N)$ instantons on $\mathbb{C}P^{2}/\mathbb{Z}_{4}$ \ - \textit{VS}}

\begin{figure}[h!]
\centering
\includegraphics[scale=1]{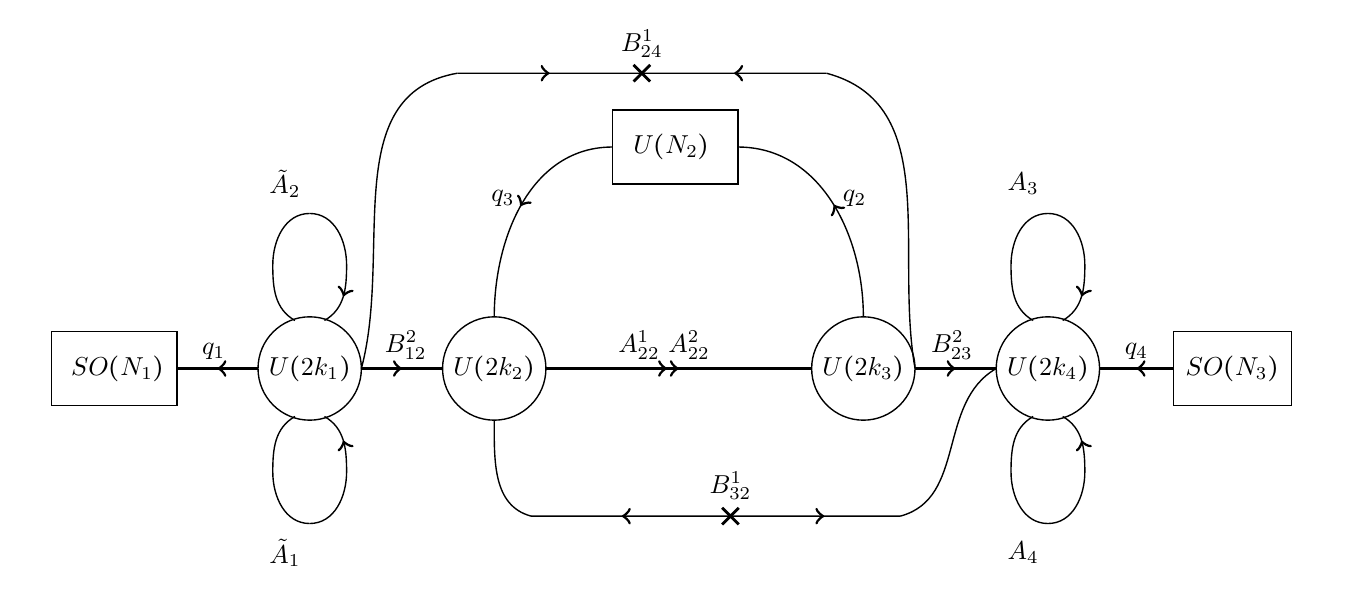}

 \caption{Quiver diagram for \textit{VS} orthogonal instantons on $\mathbb{C}P^{2}/\mathbb{Z}_{4}$. \label{fig:orientifoldz4son}}

\end{figure}

Starting from the theory for unitary instantons on $\mathbb{C}P^{2}/\mathbb{Z}_{4}$ and applying the rules in \cite{Franco:2007ii,GarciaEtxebarria:2012qx} we obtain the theory for $SO(N)$ instantons on $\mathbb{C}P^{2}/\mathbb{Z}_{4}$ in the VS case. The corresponding quiver diagram is reported in fig.\ref{fig:orientifoldz4son}, while we summarize the transformations of the fields under the different groups in table \ref{tab:spoz41}.

\begin{table}[h!]
\renewcommand\arraystretch{1.3} 
\centering
\resizebox{14cm}{!}{
\begin{tabular}{c||cccc||cccc||} 
Fields & $U(2k_{1})$ & $U(2k_{2})$ & $U(2k_{3})$ & $U(2k_{4})$ & $SO(N_{1})$ & $U(N_{2})$ & $SO(N_{3})$ & $U(1)$ \\ 
\hline 
$B^{2}_{12}$ & $[1,0,...,0]_{+1}$ & $[0,...,0,1]_{+1}$ &  $[\textbf{0}]$ &  $[\textbf{0}]$ &  $[\textbf{0}]$ &  $[\textbf{0}]$ &  $[\textbf{0}]$ & 1/4 \\ 
\hline 
$A^{2}_{22}$ &  $[\textbf{0}]$ & $[1,0...,0]_{+1}$ & $[0,...,0,1]_{+1}$ &  $[\textbf{0}]$ &  $[\textbf{0}]$ &  $[\textbf{0}]$ &  $[\textbf{0}]$ & 1/2 \\ 
\hline 
$B^{2}_{23}$ &  $[\textbf{0}]$ &  $[\textbf{0}]$ & $[1,0...,0]_{+1}$ & $[0,...,0,1]_{+1}$ &  $[\textbf{0}]$ &  $[\textbf{0}]$ &  $[\textbf{0}]$ & 1/4 \\ 
\hline 
$\tilde{A}_{2}$ & $[0,1,0...,0]_{-2}$ &  $[\textbf{0}]$ &  $[\textbf{0}]$ &  $[\textbf{0}]$ &  $[\textbf{0}]$ &  $[\textbf{0}]$ &  $[\textbf{0}]$ & 1/2 \\ 
\hline 
$A_{4}$ &  $[\textbf{0}]$ &  $[\textbf{0}]$ &  $[\textbf{0}]$ & $[0,1,0...,0]_{+2}$ &  $[\textbf{0}]$ &  $[\textbf{0}]$ &  $[\textbf{0}]$ & 1/2 \\ 
\hline 
$B^{1}_{24}$ & $[1,0,...,0]_{+1}$ &  $[\textbf{0}]$ & $[1,0,...,0]_{+1}$ &  $[\textbf{0}]$ &  $[\textbf{0}]$ &  $[\textbf{0}]$ &  $[\textbf{0}]$ & 1/4 \\ 
\hline 
$B^{1}_{32}$ &  $[\textbf{0}]$ & $[0,...,0,1]_{+1}$ &  $[\textbf{0}]$ & $[0,...,0,1]_{+1}$ &  $[\textbf{0}]$ &  $[\textbf{0}]$ &  $[\textbf{0}]$ & 1/4 \\ 
\hline 
$q_{1}$ & $[1,0,...,0]_{+1}$ &  $[\textbf{0}]$ &  $[\textbf{0}]$ & $[\textbf{0}]$ & $[1,0,....,0]$ &  $[\textbf{0}]$ &  $[\textbf{0}]$ & 1/2 \\ 
\hline 
$q_{3}$ &  $[\textbf{0}]$ & $[0,...,0,1]_{+1}$ &  $[\textbf{0}]$ &  $[\textbf{0}]$ &  $[\textbf{0}]$ & $[1,0,...,0]_{+1}$ & $[\textbf{0}]$  & 1/2 \\ 
\hline 
$q_{2}$ &  $[\textbf{0}]$ &  $[\textbf{0}]$ & $[1,0,...,0]_{+1}$ &  $[\textbf{0}]$ &  $[\textbf{0}]$ & $[0,...,0,1]_{+1}$ &  $[\textbf{0}]$ & 1/2 \\ 
\hline 
$q_{4}$ &  $[\textbf{0}]$ &  $[\textbf{0}]$ &  $[\textbf{0}]$ & $[0,...,0,1]_{+1}$ &  $[\textbf{0}]$ &  $[\textbf{0}]$ & $[1,0,...,0]$ & 1/2 \\ 
\hline 
$F_{1}$ & $[2,0...,0]_{+2}$ &  $[\textbf{0}]$ &  $[\textbf{0}]$ &  $[\textbf{0}]$ & $[\textbf{0}]$ &  $[\textbf{0}]$ &  $[\textbf{0}]$ & 1 \\ 
\hline 
$F_{2}$ &  $[\textbf{0}]$ & $[0,...,0,1]_{+1}$ & $[1,0,...,0]_{+1}$ &  $[\textbf{0}]$ &  $[\textbf{0}]$ &  $[\textbf{0}]$ &  $[\textbf{0}]$ & 1 \\ 
\hline 
$F_{3}$ &  $[\textbf{0}]$ &  $[\textbf{0}]$ &  $[\textbf{0}]$ & $[2,0,...,,0]_{-2}$ &  $[\textbf{0}]$ &  $[\textbf{0}]$ &  $[\textbf{0}]$ & 1 \\  
\end{tabular} 
}
\caption{Transformation of the fields for  VS orthogonal instantons on $\mathbb{C}P^{2}/\mathbb{Z}_{4}$. \label{tab:spoz41}}
\end{table}

The branch of the moduli space that can be identified with $Sp(N)$ instantons on $\mathbb{C}P^{2}/\mathbb{Z}_{4}$ is the one on which $A^{1}_{22}=0$ , $\tilde{A}_{1}=0$ and $A_{3}=0$. The Hilbert series of the instanton branch corresponding to the VS theory with flavor symmetry $SO(N_{1}) \times U(N_{2}) \times SO(N_{3})$ and gauge ranks $\textbf{k}=(k_{1},k_{2},k_{3},k_{4})$ is
\begin{equation}\begin{aligned}
\label{spohsz4}
& H[\textbf{k},F, \mathbb{C}P^{2}/\mathbb{Z}_{4}](t,x,\textbf{y},\textbf{d},\textbf{u})= \int d \mu_{U(2k_{1})}(\textbf{z})\int d \mu_{U(2k_{2})}(\textbf{p}) \int d \mu_{U(2k_{3})}(\textbf{w}) \times \\ 
& \int d \mu_{U(2k_{4})}(\textbf{v}) \times \textrm{PE}[\chi_{q_{1}}t^2  +\chi_{q_{2}}t^{2}  + \chi_{q_{3}}t^{2}   + \chi_{q_{4}}t^2 + \chi_{B_{12}^{2}}t  + \chi_{A^{2}_{22}}t^{2} +\chi_{B^{2}_{23}}t  + \\ 
& \chi_{B^{1}_{24}}t + \chi_{B^{1}_{32}}t + \chi_{\tilde{A}_{2}}t^{2} +\chi_{A_{4}}t^{2} -\chi_{F_{1}}t^{4} -\chi_{F_{2}}t^{4} -\chi_{F_{3}}t^{4}]\, ,
\end{aligned}\end{equation}
where $\textbf{z}$, $\textbf{p}$, $\textbf{w}$ and $\textbf{v}$  are the fugacities of the $U(2k_{1})$, $U(2k_{2})$, $U(2k_{3})$ and $U(2k_{4})$ gauge groups respectively while $\textbf{y}$ and $\textbf{d}$ denote the fugacities of the $SO(N_{1})$ flavor group ,of the $U(N_{2})$ flavor group and of the $SO(N_{3})$ flavor group respectively. The contributions of the various fields are
\begin{equation*}
\chi_{B^{2}_{12}} = \sum_{a=1}^{2k_{1}}\sum_{b=1}^{2k_{2}}z_{a}p_{b}^{-1}, \ \  \chi_{A_{22}^{2}} = \sum_{a=1}^{2k_{2}}\sum_{b=1}^{2k_{3}}p_{a}w_{b}^{-1}, \ \ \chi_{B^{2}_{23}} = \sum_{a=1}^{2k_{3}}\sum_{b=1}^{2k_{4}}w_{a}v_{b}^{-1},  \ \ \chi_{\tilde{A_{2}}} = \sum\nolimits_{1 \leq a < b \leq 2k_{1}} z_{a}^{-1}z_{b}^{-1},
\end{equation*}
\begin{equation*}
\chi_{q_{1}} = \sum_{a=1}^{2k_{1}}z_{a} \times \left\{
\begin{array}{l}
\sum\limits_{i=1}^{N_{1}/2}\left(y_{i} +\frac{1}{y_{i}}\right) \ N_{1} \ \textrm{even}\, , \\ 1 + \sum\limits_{i=1}^{(N_{1}-1)/2}\left(y_{i} + \frac{1}{y_{i}}\right)  \ N_{1} \ \textrm{odd}\, , 
\end{array}
\right. \ \ \chi_{q_{3}}=\sum_{j=1}^{N_{2}}\sum_{b=1}^{2k_{2}} d_{j}p_{b}^{-1}, \ \ \chi_{B^{1}_{32}} = \sum_{a=1}^{2k_{2}}\sum_{b=1}^{2k_{4}}p_{a}^{-1}v_{b}^{-1},
\end{equation*}
\begin{equation*}
\chi_{B^{1}_{24}} = \sum_{a=1}^{2k_{1}}\sum_{b=1}^{2k_{3}}z_{a}w_{b}, \ \ 
\chi_{q_{2}}=\sum_{a=1}^{2k_{3}}\sum_{i=1}^{N_{2}}w_{a}d_{i}^{-1},  \ \ \chi_{q_{4}}=\sum_{a=1}^{2k_{4}}v_{a}^{-1}\times \left\{
\begin{array}{l}
\sum\limits_{i=1}^{N_{3}/2}\left(y_{i} +\frac{1}{y_{i}}\right) \ N_{3} \ \textrm{even}\, , \\ 1 + \sum\limits_{i=1}^{(N_{3}-1)/2}\left(y_{i} + \frac{1}{y_{i}}\right)  \ N_{3} \ \textrm{odd}\, , 
\end{array}
\right.
\end{equation*}
\begin{equation*}
\chi_{A_{4}} = \sum\nolimits_{1 \leq a < b \leq 2k_{4}}v_{a}v_{b}, \ \ \chi_{F_{1}}= \sum\nolimits_{1 \leq a \leq b \leq 2k_{1}} z_{a}z_{b}, \ \  \chi_{F_{2}} = \sum_{a=1}^{2k_{2}}\sum_{b=1}^{2k_{3}} p_{a}^{-1}w_{b}, \ \ \chi_{F_{3}} = \sum\nolimits_{1 \leq a \leq b \leq 2k_{4}} v_{a}^{-1}v_{b}^{-1}\, . 
\end{equation*}

\begin{figure}[h!]
\centering
\includegraphics[scale=1]{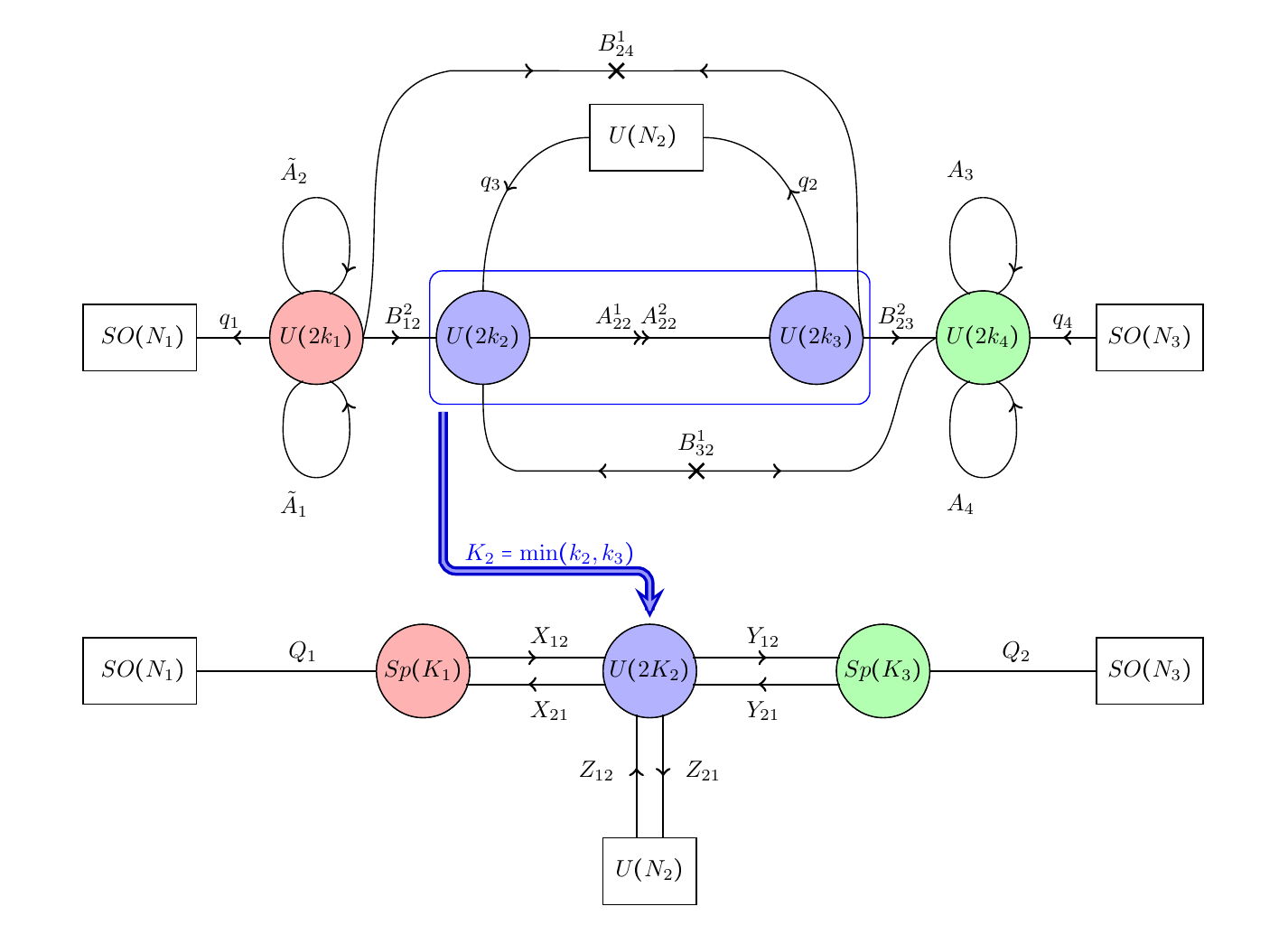}

 \caption{Relation between the $\mathbb{C}P^{2}/\mathbb{Z}_{4}$ quiver gauge theory in the VS case and its relation with the corresponding $\mathbb{C}^{2}/\mathbb{Z}_{4}$ quiver gauge theory.\label{OZ4VSrel}}

\end{figure}

By computing the Hilbert series with gauge group $G=U(2k_{1}) \times U(2k_{2}) \times U(2k_{3}) \times U(2k_{4})$ and flavor group $SO(N_{1}) \times U(N_{2}) \times SO(N_{3})$ we find that it turns out to be equal to the Hilbert series for $SO(N)$ instantons on $\mathbb{C}^{2}/\mathbb{Z}_{4}$ with gauge group $G=Sp(K_{1})\times U(2K_{2}) \times Sp(K_{3}) $ and flavor groups $SO(N_{1}) \times U(N_{2}) \times SO(N_{3})$ (see \cite{Dey:2013fea} for more details) with the identification

\begin{equation}
K_{1} = k_{1}, \ \ K_{2}= \textrm{min}(k_{2},k_{3})\, , \ \ K_{3}=k_{4}\, .
\end{equation}
Let us now show a few explicit examples.
\\
\\
\textbf{ $ SO(6)$ instanton:  $\textbf{k}=(1,1,1,1)$, and  $\textbf{N}=(2,0,4)$}\\
\\
Using eq.(\ref{spohsz4}) and unrefing we find that
\begin{equation*}\begin{aligned}
& H[\textbf{k}=(1,1,1,1),SO(2)\times SO(4), \mathbb{C}P^{2}/\mathbb{Z}_{4}](t,1,1) = \\
& = \frac{1+4t^{6} +22t^{12} +36t^{18} +54t^{24} +36t^{30} +22t^{36} +4t^{42} +t^{48}}{(1-t^{3})^{8}(1+t^{3})^{8}(1+t^{6})^{4}}\, ,
\end{aligned}\end{equation*}
which is the Hilbert series for $SO(6)$ instantons on $\mathbb{C}^{2}/\mathbb{Z}_{4}$ with $\textbf{N}=(2,0,4)$ and $\textbf{K}=(1,1,1)$.\\
\\
\textbf{$ SO(6)$ instanton: $\textbf{k}=(1,1,1,1)$, and $\textbf{N}=(2,1,2)$}\\
\\
Using eq.(\ref{spohsz4}) and unrefing we find that 
\begin{equation*}\begin{aligned}
& H[\textbf{k}=(1,1,1,1),SO(2)\times U(1) \times SO(2), \mathbb{C}P^{2}/\mathbb{Z}_{4}](t,1,1,1) = \\
&  = \frac{1}{(1-t^{3})^{8}(1+t^{3})^{4}(1+t^{6})^{2}(1+t^{3}+t^{6})^{12}(1+t^{3}+t^{6}+t^{9}+t^{12})}(1 + t^{3} +3t^{6} + 7t^{9} +18t^{12} +\\
& 33t^{15} + 51t^{18} + 69t^{21} + 93t^{24} + 110t^{27} + 120t^{30} + 110t^{33} + \textrm{palindrome}+ ... + t^{60})\, ,
\end{aligned}\end{equation*}
which is the Hilbert series for $SO(6)$ instantons on $\mathbb{C}^{2}/\mathbb{Z}_{4}$ with $\textbf{N}=(2,1,2)$ and $\textbf{K}=(1,1,1)$.
\\
\\
Finally, we summarize in fig.\ref{OZ4VSrel} the relation between the theory describing VS orthogonal instantons on $\mathbb{C}P^2/\mathbb{Z}_4$ and its cousin on $\mathbb{C}^2/\mathbb{Z}_4$.

\subsubsection{$SO(N)$ instantons on $\mathbb{C}P^{2}/\mathbb{Z}_{4}$ \ - \textit{NVS}}

\begin{figure}[h!]
\centering
\includegraphics[scale=1]{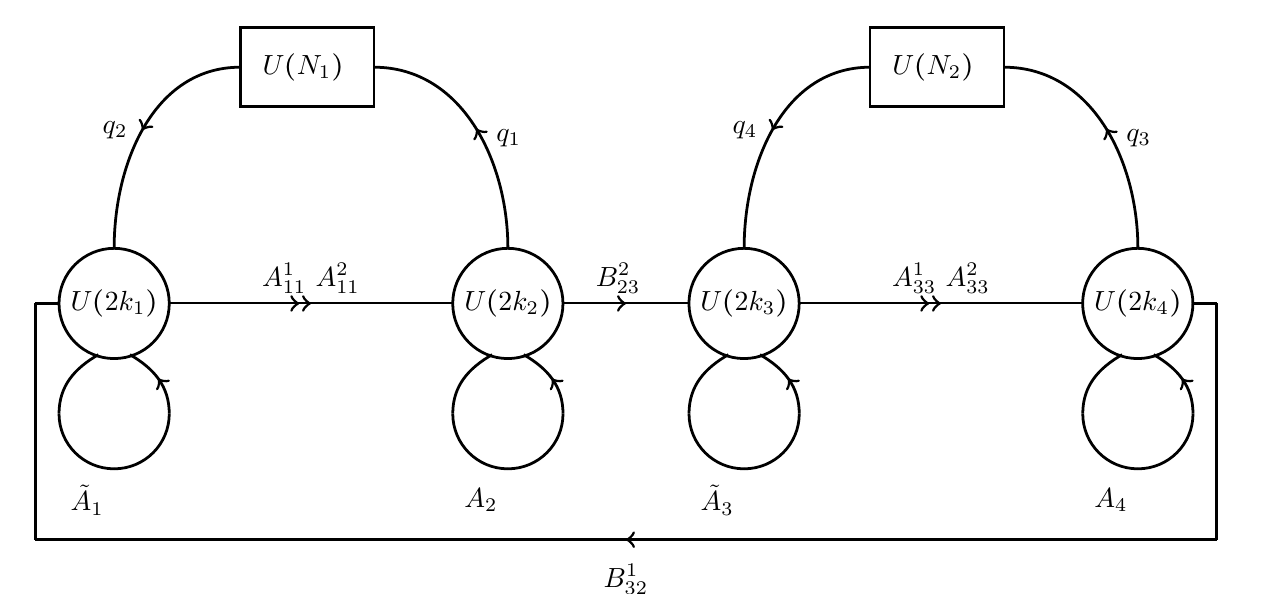}
 
\caption{Quiver diagram for NVS orthogonal instantons on $\mathbb{C}P^{2}/\mathbb{Z}_{4}$. \label{fig:z42son}}

\end{figure}

Let us now consider the second possibility leading to the NVS case. The quiver diagram of the corresponding theory is reported in fig.\ref{fig:z42son}, while the transformations of the fields and of the F-terms are summarized in table \ref{tab:222son}.

\begin{table}[h!]\center

\renewcommand\arraystretch{1.3} 
\resizebox{14cm}{!}{
\begin{tabular}{c||cccc||ccc||}
Fields & $U(2k_{1})$ & $U(2k_{2})$ & $U(2k_{3})$ & $U(2k_{4})$ & $U(N_{1})$ & $U(N_{2})$ & $U(1)$\\ 
\hline 
$A^{2}_{11}$ & $[1,0,...,0]_{+1}$ & $[0,...,0,1]_{+1}$ & $[\textbf{0}]$  & $[\textbf{0}]$  & $[\textbf{0}]$  & $[\textbf{0}]$  & 1/2\\ 
\hline 
$B^{2}_{23}$ & $[\textbf{0}]$  & $[1,0,...,0]_{+1}$ & $[0,...,0,1]_{+1}$ & $[\textbf{0}]$  & $[\textbf{0}]$  & $[\textbf{0}]$  & 1/4\\ 
\hline 
$A^{2}_{33}$ & $[\textbf{0}]$  & $[\textbf{0}]$  & $[1,0,...,0]_{+1}$ & $[0,...,0,1]_{+1}$ & $[\textbf{0}]$  & $[\textbf{0}]$  & 1/2\\ 
\hline 
$B^{1}_{32}$ & $[0,...,0,1]_{+1}$ & $[\textbf{0}]$  & $[\textbf{0}]$  & $[1,0,...,0]_ {+1}$ & $[\textbf{0}]$  & $[\textbf{0}]$  & 1/4\\ 
\hline 
$\tilde{A}_{1}$ & $[0,1,0,...,0]_{-2}$ & $[\textbf{0}]$  & $[\textbf{0}]$  & $[\textbf{0}]$  & $[\textbf{0}]$  & $[\textbf{0}]$  & 1/4\\ 
\hline 
$A_{2}$ & $[\textbf{0}]$  & $[0,1,0,...,0]_{+2}$ & $[\textbf{0}]$  & $[\textbf{0}]$  & $[\textbf{0}]$  & $[\textbf{0}]$  & 1/4\\ 
\hline 
$\tilde{A}_{3}$ & $[\textbf{0}]$  & $[\textbf{0}]$  & $[0,1,0,...,0]_{-2}$  &$[\textbf{0}]$  & $[\textbf{0}]$  & $[\textbf{0}]$  & 1/4\\ 
\hline 
$A_{4}$ & $[\textbf{0}]$  & $[\textbf{0}]$  & $[\textbf{0}]$  & $[0,1,0,...,0]_{+2}$ & $[\textbf{0}]$  & $[\textbf{0}]$  & 1/4\\ 
\hline 
$q_{1}$ & $[\textbf{0}]$  & $[1,0,...,0]_{+1}$ & $[\textbf{0}]$  & $[\textbf{0}]$  & $[0,...,0,1]_{+1}$ & $[\textbf{0}]$  & 1/2\\ 
\hline 
$q_{2}$ & $[0,...,0,1]_{+1}$ & $[\textbf{0}]$  & $[\textbf{0}]$  & $[\textbf{0}]$  & $[1,0,...,0]_{+1}$ & $[\textbf{0}]$  & 1/2\\ 
\hline 
$q_{3}$ & $[\textbf{0}]$  & $[\textbf{0}]$  & $[\textbf{0}]$  & $[1,0,...,0]_{+1}$ & $[\textbf{0}]$  & $[0,...,0,1]_{+1}$ & 1/2\\ 
\hline 
$q_{4}$ & $[\textbf{0}]$  & $[\textbf{0}]$  & $[0,...,0,1]_{+1}$ & $[\textbf{0}]$  & $[\textbf{0}]$ & $[1,0,...,0]_{+1}$  & 1/2 \\ 
\hline 
$F_{1}$ & $[0,...,0,1]_{+1}$ & $[1,0,...,0]_{+1}$ &  $[\textbf{0}]$  &  $[\textbf{0}]$  &  $[\textbf{0}]$  &  $[\textbf{0}]$  & 1 \\ 
\hline 
$F_{2}$ &  $[\textbf{0}]$  &  $[\textbf{0}]$  & $[0,...,0,1]_{+1}$ & $[1,0,...,0]_{+1}$ &  $[\textbf{0}]$  &  $[\textbf{0}]$  & 1\\ 
\end{tabular} 
}
\caption{Transformation of the fields for NVS orthogonal instantons on $\mathbb{C}P^{2}/\mathbb{Z}_{4}$. \label{tab:222son}}
\end{table}

The branch of the moduli space that can be identified with $SO(N)$ instantons on $\mathbb{C}P^{2}/\mathbb{Z}_{4}$ is the one on which $A^{1}_{11}=0$  and $A_{33}^{1}=0$. The Hilbert series of the instanton branch corresponding to the NVS theory with flavor symmetry $U(N_{1}) \times U(N_{2})$ and gauge ranks $\textbf{k}=(k_{1},k_{2},k_{3},k_{4})$ is
\begin{equation}\begin{aligned}
\label{sohsz42}
& H[\textbf{k},F, \mathbb{C}P^{2}/\mathbb{Z}_{4}](t,x,\textbf{y},\textbf{d})= \int d \mu_{U(2k_{1})}(\textbf{z})\int d \mu_{U(2k_{2})}(\textbf{p}) \int d \mu_{U(2k_{3})}(\textbf{w}) \times \\ 
& \int d \mu_{U(2k_{4})}(\textbf{v}) \times 
\textrm{PE}[\chi_{q_{1}}t^2  +\chi_{q_{2}}t^{2}  + \chi_{q_{3}}t^{2}   + \chi_{q_{4}}t^2 + \chi_{B_{23}^{2}}t  + \chi_{A^{2}_{11}}t^{2} +\chi_{A^{2}_{33}}t^{2} + \\ 
& \chi_{B^{1}_{32}}t  +\chi_{\tilde{A}_{1}}t + \chi_{A_{2}}t +
+\chi_{\tilde{A}_{3}}t +\chi_{A_{4}}t -\chi_{F_{1}}t^{4} -\chi_{F_{2}}t^{4}]\, ,
\end{aligned}\end{equation}
where $\textbf{z}$, $\textbf{p}$, $\textbf{w}$ and $\textbf{v}$  are the fugacities of the $U(2k_{1})$, $U(2k_{2})$, $U(2k_{3})$ and $U(2k_{4})$ gauge groups respectively while $\textbf{y}$ and $\textbf{d}$ denote the fugacities of the $U(N_{1})$ flavor group and the $U(N_{2})$ flavor group respectively. The contributions of the various fields are given by
\begin{equation*}
\chi_{\tilde{A}_{3}} = \sum\nolimits_{1 \leq a < b \leq 2k_{3}} w_{a}^{-1}w_{b}^{-1}, \ \ \chi_{A_{4}} = \sum\nolimits_{ 1 \leq a < b \leq 2k_{4}}  v_{a}v_{b}, \ \ \chi_{q_{1}} = \sum_{a=1}^{2k_{2}}\sum_{i=1}^{N_{1}} p_{a}y_{i}^{-1}\, ,
\end{equation*}
\begin{equation*}
\chi_{B^{1}_{32}} = \sum_{a=1}^{2k_{1}}\sum_{b=1}^{2k_{4}} v_{b}z_{a}^{-1}, \ \ \chi_{\tilde{A}_{1}} = \sum\nolimits_{ 1 \leq a < b \leq 2k_{1}} z_{a}^{-1}z_{b}^{-1}, \ \ \chi_{A_{2}}=\sum\nolimits_{ 1 \leq a < b \leq 2k_{2}}  p_{a}p_{b}\, ,
\end{equation*}
\begin{equation*}
\chi_{q_{2}} = \sum_{a=1}^{2k_{1}}\sum_{i=1}^{N_{1}} z_{a}^{-1}y_{i}, \ \ \chi_{q_{3}} = \sum_{a=1}^{2k_{4}}\sum_{j=1}^{N_{2}} v_{a}d_{j}^{-1}, \ \ \chi_{q_{4}} = \sum_{a=1}^{2k_{3}}\sum_{j=1}^{N_{2}} w_{a}^{-1}d_{j}, \ \ \chi_{F_{1}} = \sum_{a=1}^{2k_{1}}\sum_{b=1}^{2k_{2}} p_{b}z_{a}^{-1}\, ,
\end{equation*}
\begin{equation*}
\chi_{A^{2}_{11}} = \sum_{a=1}^{2k_{1}}\sum_{b=1}^{2k_{2}}z_{a}p_{b}^{-1}, \ \ \chi_{B^{2}_{23}}=\sum_{a=1}^{2k_{2}}\sum_{b=1}^{2k_{3}} p_{a}w_{b}^{-1}, \ \ \chi_{A^{2}_{33}}=\sum_{a=1}^{2k_{3}}\sum_{b=1}^{2k_{4}} w_{a}v_{b}^{-1}, \ \ \chi_{F_{2}} = \sum_{a=1}^{2k_{3}}\sum_{b=1}^{2k_{4}} w_{a}^{-1}v_{b}\, .
\end{equation*}

\begin{figure}[h!]
\centering
\includegraphics[scale=0.8]{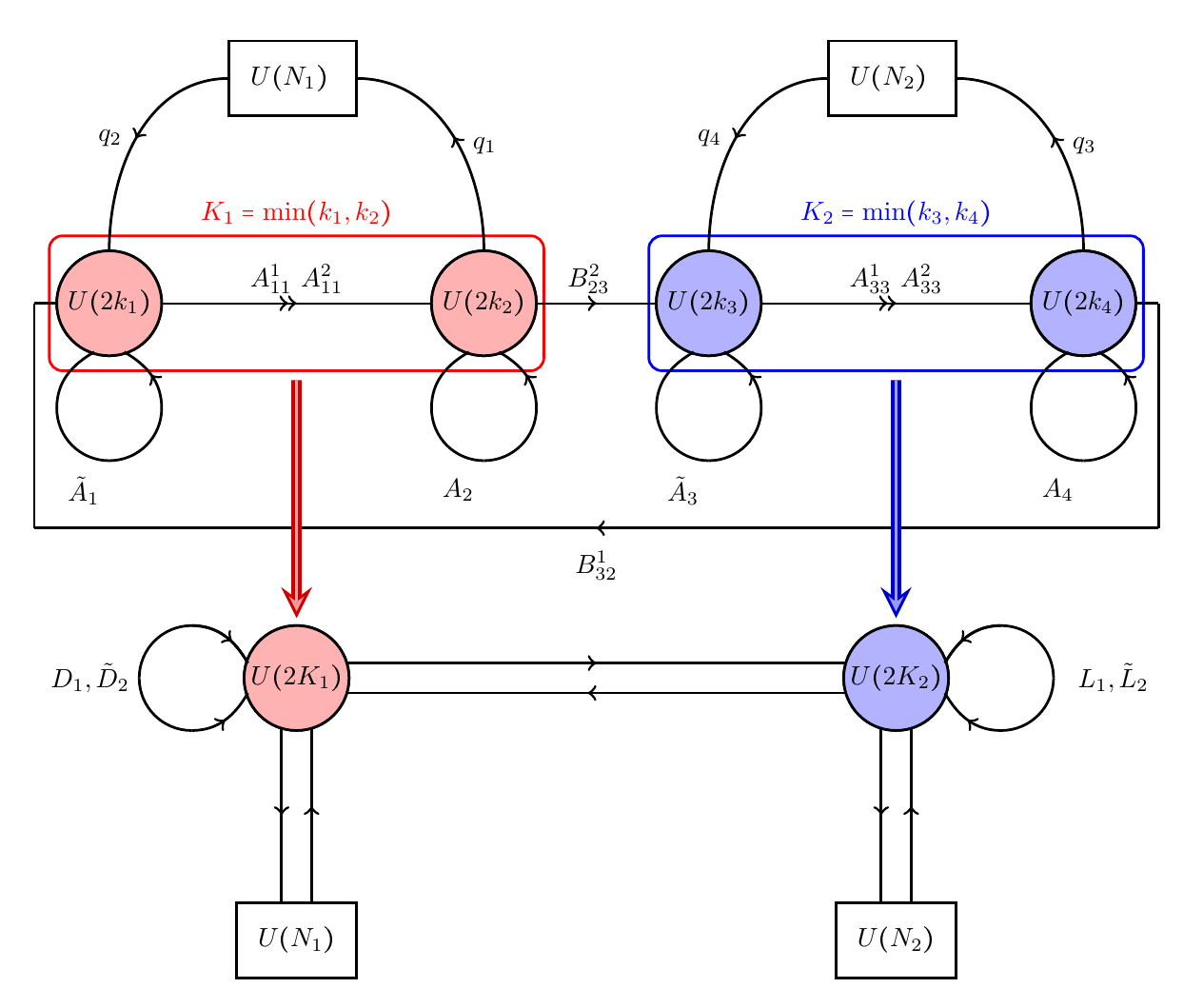}

\caption{Relation between the $\mathbb{C}P^{2}/\mathbb{Z}_{4}$ quiver gauge theory in the NVS case and the corresponding $\mathbb{C}^{2}/\mathbb{Z}_{4}$ quiver gauge theory. Where $D_{1},\tilde{D}_{2}$ are two fields  in the antisymmetric representation of the gauge group $U(2K_{1})$, while $L_{1},\tilde{L}_{2}$ are two fields in the antisymmetric representation of the gauge group $U(2K_{2})$. \label{fig:z42relation}}

\end{figure}

Performing the computation of the Hilbert series with gauge group $G=U(2k_{1}) \times U(2k_{2}) \times U(2k_{3}) \times U(2k_{4})$ and flavour group $U(N_{1}) \times U(N_{2})$ we find that it coincides with the Hilbert series for $SO(N)$ instantons on $\mathbb{C}^{2}/\mathbb{Z}_{4}$ with gauge group $G=U(2K_{1})\times U(2K_{2}) $ and flavor groups $U(N_{1}) \times U(N_{2})$ (see \cite{Dey:2013fea} for more details) with the identification

\begin{equation}
K_{1} = \textrm{min}(k_{1},k_{2})\, ,  \qquad K_{2}= \textrm{min}(k_{3},k_{4})\, .
\end{equation}
Let us show an explicit example of our claim.
\\
\\
\textbf{$SO(6)$ instanton: $\textbf{k}=(1,1,1,1)$, and  $\textbf{N}=(2,1)$}\\
\\
Using eq.(\ref{sohsz42}) and unrefining we obtain
\begin{equation*}\begin{split}
& H\left[\textbf{k}=\left(1,1,1,1\right),U(2)\times U(1), \mathbb{C}P^{2}/\mathbb{Z}_{4}\right](t,1,1,1) = \\ 
& = \frac{1}{(1-t^3)^8(1+t^3)^6(1+t^6)^3(1+t^3+t^6)^3(1+t^3+t^6+t^9+t^{12})^2}(1 +3t^3 +9t^{6} +  22t^{9} + \\ 
& + 54t^{12} + 114t^{15} + 219t^{18} + 371t^{21} + 582t^{24} + 827t^{27} + 1092t^{30} + 1323t^{33} + 1493t^{36} + \\
& + 1548t^{39}  + 1493t^{42} + \textrm{palindrome} + t^{72})\, ,
\end{split}\end{equation*}
which is the Hilbert series for $SO(6)$  instantons on $\mathbb{C}^{2}/\mathbb{Z}_{4}$ with $\textbf{N}=(2,1)$ and $\textbf{K}=(1,1)$.
\\ 
\\
Finally, we graphically summarize the relation between the theory describing the NVS orthogonal instantons on $\mathbb{C}P^2/\mathbb{Z}_4$ and its cousin on $\mathbb{C}^2/\mathbb{Z}_4$ in fig.\ref{fig:z42relation}.

\subsubsection{$SO(N)$ instantons on $\mathbb{C}P^{2}/\mathbb{Z}_{n}$ with $n>4$}
\label{subsec:genarlso}

Let us now consider the generic case of instantons on $\mathbb{Z}_n$ orbifolds of $\mathbb{C}P^2$ with $n >4$. Based on the previous examples above, we can extract the generic pattern of both the quiver as well as the relation between the orthogonal instanton on $\mathbb{C}P^2/\mathbb{Z}_n$ with its relative on $\mathbb{C}^2/\mathbb{Z}_n$.

Recall that $N$ is the sum of the ranks of the flavor groups in the ADHM quiver, while the ranks of the gauge groups is related to instanton number and, together with the relative flavor ranks, to other possible quantum numbers labelling the instanton. Unfortunately also in this case the precise identification between quiver data and instanton data is not know.
\\
\\
\textbf{$SO(N)$ instantons on $\mathbb{C}P^{2}/\mathbb{Z}_{2n+1}$}
\\
\\
Elaborating on the previous examples, we conjecture that the theory describing orthogonal instantons on $\mathbb{C}P^2/\mathbb{Z}_{2n+1}$ is related to its counterpart on $\mathbb{C}^2/\mathbb{Z}_{2n+1}$ as in fig.\ref{fig:cp22n+1o}. Moreover, the gauge ranks are related by
\begin{equation}
K_{1} = k_{1}, \ \ K_{2}=\textrm{min}(k_{2},k_{3}), \ \ K_{3}=\textrm{min}(k_{4},k_{5}), \ \ ... \ \ K_{n+1}=\textrm{min}(k_{2n},k_{2n+1})\, . 
\end{equation}

\textbf{$SO(N)$ instantons on $\mathbb{C}P^{2}/\mathbb{Z}_{2n}$- \textit{VS}}
\\
\\
In this case, based on the lowest $n$ examples, the relation between the theory describing VS instantons on $\mathbb{C}P^2/\mathbb{Z}_{2n}$ and their VS counterparts on $\mathbb{C}^2/\mathbb{Z}_{2n}$ is summarized in fig.\ref{fig:cp22n+1firsto}. In addition, we find the gauge rank identification

\begin{equation}
 K_{1}=k_{1}, \ \ K_{2}=\textrm{min}(k_{2},k_{3}), \ \ ... \ \ K_{n-1}=\textrm{min}(k_{2n-2},k_{2n-1}),  \ \ K_{n}=k_{2n}\, . 
\end{equation}

\textbf{$SO(N)$ instantons on $\mathbb{C}P^{2}/\mathbb{Z}_{2n}$- \textit{NVS}}
\\
\\
Elaborating on the previous examples, we conjecture that the theory describing NVS orthogonal instantons on $\mathbb{C}P^2/\mathbb{Z}_{2n+1}$ is related to its NVS counterpart on $\mathbb{C}^2/\mathbb{Z}_{2n+1}$ as in fig.\ref{fig:cp22n+1secondo}. In addition, the gauge rank assignation is

\begin{equation}
 K_{1}=\textrm{min}(k_{1},k_{2}), \ \ K_{2}=\textrm{min}(k_{3},k_{4}), \ \ ... \ \ K_{n}=\textrm{min}(k_{2n-1},k_{2n})\, . 
\end{equation}
\\
\\
Note that, as in the symplectic case, the ``merging nodes" are those going over to unitary nodes in the parent $\mathbb{C}^2/\mathbb{Z}_n$ theory. It would be very interesting to deeper understand this feature, as well as the topological data classifying orthogonal instantons.

\section{Conclusions}
\label{conclusions}

In this paper we have analyzed several aspects of the moduli space of instantons on $\mathbb{C}P^2$. Since $\mathbb{C}P^2$ is a K\"ahler manifold, its K\"ahler form  naturally induces an orientation which, in particular, intrinsically distinguishes ASD and SD 2-forms. This is very relevant for the construction of gauge bundles whose curvature has definite duality properties, as such construction will be different depending on whether we are interested in the SD or ASD case. In this paper we have been interested on SD connections. In turn, these are the ones which admit an ADHM-like construction recently embedded into a $3d$ $\mathcal{N}=2$ gauge theory arising from a brane construction in \cite{Mekareeya:2014kca}. 

Since $\mathbb{C}P^2$ is a topologically non-trivial manifold, the gauge bundles of interest are classified by more than simply the instanton number. Indeed, they admit a non-zero first Chern class. As a consequence, the moduli space of instantons on $\mathbb{C}P^2$ typically has compact submanifolds associated to these extra directions. In turn, the Hilbert series of the moduli space --that is, the generating function of holomorphic functions on the instanton moduli space or, equivalently, the generating function of gauge-invariant operators in the ADHM description of the instanton moduli space--, which coincides with the Nekrasov instanton partition function and it's therefore a very interesting quantity; is not sensible to these compact directions. Hence, in retrospect, it is natural to expect that it would coincide with the Hilbert series for a parent instanton on $\mathbb{C}^2$, as it was explicitly shown in \cite{Mekareeya:2014kca}. In this paper we have provided evidence of this picture by probing the compact directions in a slightly indirect way.  Focusing on the simplest case admitting such directions, and following \cite{Hanany:2014hia}, we considered the master space of the gauge theory describing these instantons. This amounts to ungauging a $U(1)$, which allows to construct extra gauge-invariants otherwise not present. These precisely reproduce a moduli space which is a complex cone over the non-compact directions. By using this strategy we have been able to understand the extra directions in the unitary and orthogonal cases. In turn, the case of symplectic instantons does not admit a similar construction, consistent with the observation in \cite{Mekareeya:2014kca} that it does not seem to involve quantum numbers other than the instanton number. Note however that we explicitly checked this picture for the lowest instanton numbers. It would be worth exploring this further to all instanton numbers, including studying the geometry of the moduli space with extra directions, which is not simply a direct product of the non-compact times the compact directions (this can be checked already in the simplest cases by studying the relations among operators in the moduli space). 

The case of unitary instantons is particularly interesting, as its AHDM construction is in terms of the gauge theory dual to M2 branes probing a certain $CY_4$ cone \cite{Benini:2009qs}. Hence, it is natural to guess that, at least partially, the instanton moduli space can be read from the $AdS/CFT$ duality. Typically, fundamental degrees of freedom --that is, open string-like-- are not captured by the geometry alone in $AdS/CFT$. Hence, it is natural to expect that the backgrounds in \cite{Benini:2009qs} can only capture the part of the instanton moduli space which does not involve fundamental fields. We have explicitly checked this proposal, finding a complete agreement between field theory results and gravity computations. Turning things around, we can think of our results as a non-trivial check of the proposed $AdS_4/CFT_3$ duality in \cite{Benini:2009qs}, where we explicitly match charges in field theory with geometrical data in $AdS$.

The ambient manifold where our instantons live is $\mathbb{C}P^2$, which is in particular a toric manifold. Being acted by a $\mathbb{T}^2$, it is natural to consider quotienting by a discrete subgroup --that is, orbifolding--. In turn, by means of the standard methods we can orbifold the $\mathbb{C}P^2$ ADHM construction as a field theory to find the ADHM construction of instantons on $\mathbb{C}P^2/\mathbb{Z}_n$. This way we have constructed the ADHM construction for unitary, symplectic and orthogonal instantons on $\mathbb{C}P^2/\mathbb{Z}_n$. Note that the orbifolded space has a non-trivial topology containing 2-cycles of a somewhat different origin. On one hand we originally had a 2-cycle in the $\mathbb{C}P^2$ which gets mirrored by the orbifold. On the other hand, the orbifold introduces extra (vanishing) 2-cycles at the orbifold fixed point. It is natural to expect that the cycles originating from the original one in $\mathbb{C}P^2$ are invisible to the Hilbert series --just as the original one was-- while the others, introduced by the orbifold, are indeed visible. In fact, it is natural to guess that the Hilbert series for instantons on $\mathbb{C}P^2/\mathbb{Z}_n$ coincides with the Hilbert series of a parent instanton on $\mathbb{C}^2/\mathbb{Z}_n$ just as in the unorbifolded case. Note that, consistently, the Hilbert series of instantons on $\mathbb{C}^2/\mathbb{Z}_n$ is indeed sensible to the 2-cycles associated to the orbifold fixed point \cite{Dey:2013fea}. \footnote{Strictly speaking, this applies to unitary instantons. The case of orthogonal and symplectic instantons is more involved, as the ADHM construction does not allow for enough FI parameters so as to blow-up all cycles (see \cite{Tachikawa:2014qaa} for related discussions).} In this paper we have indeed confirmed this picture, in particular by explicitly showing the matching of the $\mathbb{C}P^2/\mathbb{Z}_n$ Hilbert series with that of a parent $\mathbb{C}^2/\mathbb{Z}_n$ one. As shown in the text, the process suggests a certain ``folding" of the $\mathbb{C}P^2/\mathbb{Z}_n$ quiver by ``node merging" into that of $\mathbb{C}^2/\mathbb{Z}_n$. In fact, since at least for unitary instantons on $\mathbb{C}^2/\mathbb{Z}_n$ the matching between quiver data and instanton data is known, this naturally suggests, at least partially, an identification of the quiver data with the instanton data in the $\mathbb{C}P^2/\mathbb{Z}_n$ case.  Unfortunately, the full identification with the ADHM quiver data of the relevant quantum numbers specifying instantons on the orbifolded $\mathbb{C}P^2$ space is not known. Nevertheless we have provided --at least for the case of unitary instantons-- certain conjectures based on the mapping into $\mathbb{C}^2/\mathbb{Z}_n$. As a check, the expected compact directions can be recovered upon appropriate ungaugings of $U(1)$'s. Of course, a more comprehensive study of these aspects would be very interesting.

\appendix

\section{Hybrid configuration (an example)}
\label{app:A}

\begin{figure}[h!]\center
\includegraphics{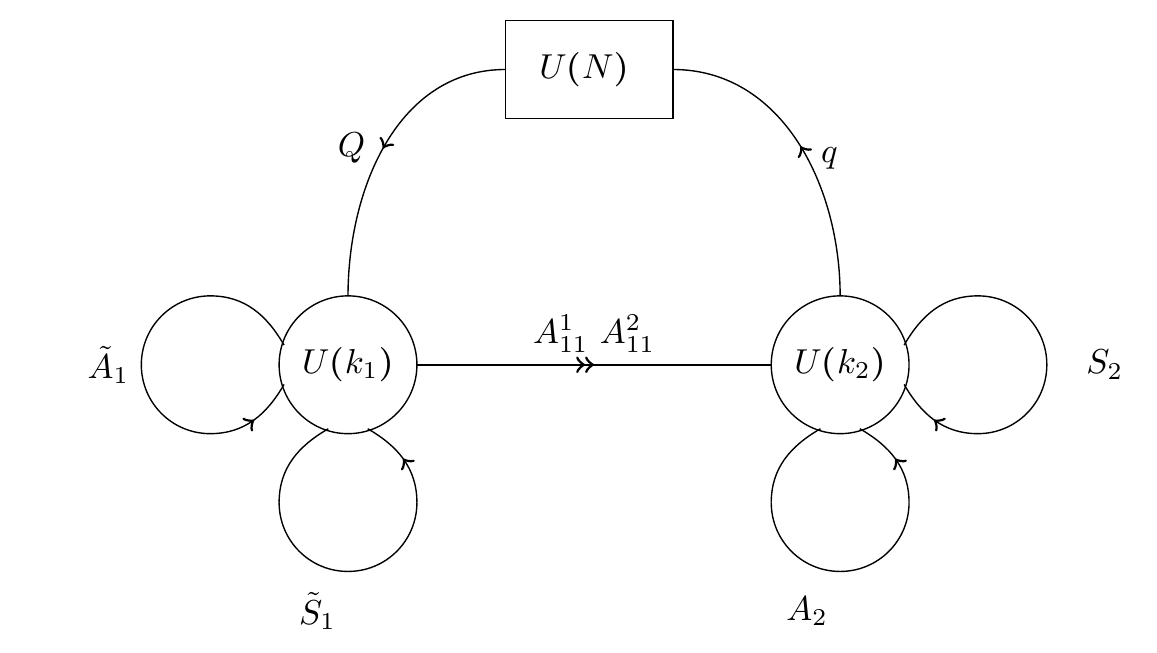}
 
\caption{Quiver diagram for instantons of the hybrid configuration on $\mathbb{C}P^{2}/\mathbb{Z}_{2}$. \label{fig:cp2hybrid}}
\end{figure}

In this appendix we study an example of hybrid configuration, making the following choice for the charges of the orientifolds plane in fig.\ref{fig:orientifolds} $(I,II,III,IV)=(+,-,+,-)$. The corresponding quiver is reported in fig.\ref{fig:cp2hybrid}, while the transformations of the fields are summarized in table \ref{tab:hybrid}.

\begin{table}[h!]
 \renewcommand\arraystretch{1.3} 

\centering
\resizebox{12cm}{!}{
\begin{tabular}{c||cc||cccc||} 
Fields & $U(k_{1})$ & $U(k_{2})$ & $U(N)$  & $U(1)_{s}$ &$U(1)_{a}$&  $U(1)$ \\ 
\hline 
$\tilde{S}_{1}$ & $[2,0,...,0]_{-2}$ & $[\textbf{0}]$ & $[\textbf{0}]$  & $1/s$ & $[0]$ & 1/4 \\ 
\hline 
$S_{2}$ & $[\textbf{0}]$ & $[2,0,...,0]_{+2}$ & $[\textbf{0}]$  & $s$ & $[\textbf{0}]$ & 1/4 \\ 
\hline 
$\tilde{A}_{1}$ & $[0,1,0...,0]_{-1}$ & $[\textbf{0}]$ & $[\textbf{0}]$  & $[\textbf{0}]$ & $1/a$ & 1/4 \\ 
\hline 
$A_{2}$ & $[\textbf{0}]$ & $[0,1,0,...,0]_{+1}$ & $[\textbf{0}]$  & $[\textbf{0}]$ &  $a$ & 1/4 \\ 
\hline
$A^{2}_{11}$ & $ [1,0,...,0]_{+1}$ & $[0,0,...,1]_{+1}$ & $[\textbf{0}]$ & $[\textbf{0}]$ & $[\textbf{0}]$ & 1/2 \\ 
\hline 
$q$ & $[\textbf{0}]$ & $[1,0,...,0]_{+1}$ & $[0,...,0,1]_{+1}$ & $[\textbf{0}]$ & $[\textbf{0}]$ & 1/2 \\ 
\hline 
$Q$ & $[0,...,0,1]_{+1}$ & $[\textbf{0}]$ & $[1,0,...,0]_{+1}$ & $[\textbf{0}]$ & $[\textbf{0}]$ &1/2 \\ 
\hline 
$F$ & $[0,...,0,1]_{+1}$ & $[1,0,...,0]_{+1}$ & $[\textbf{0}]$ & $[\textbf{0}]$ & [$\textbf{0}$]& 1 \\ 
\end{tabular} 
}

\caption{Transformations of the fields for instantons of the hybrid configuration on $\mathbb{C}P^2/\mathbb{Z}_{2}$. \label{tab:hybrid}}
\end{table}

The Hilbert series of the hybrid configuration is given by

\begin{equation}\begin{aligned}
\label{hybridhs}
& H[\textbf{k},F,\mathbb{C}P^{2}/Z_{2}](t,a,s,\textbf{y})= \int d \mu_{U(k_{1})}(\textbf{z})\int d \mu_{U(k_{2})}(\textbf{p}) \times \\
& \textrm{PE}[\chi_{\tilde{S_{1}}}t  +\chi_{S_{2}}t  +\chi_{\tilde{A_{1}}}t  +\chi_{A_{2}}t+ \chi_{A_{11}^{2}}t^{2}   + \chi_{Q}t^{2}  + \chi_{q}t^{2} -\chi_{F}t^{4} ]\, ,
\end{aligned}\end{equation}
where $\textbf{z}$ and $\textbf{p}$ are the fugacities of the $U(k_{1})$ and $U(k_{2})$ gauge groups respectively, $\textbf{y}$  denote the fugacity of the $U(N)$ flavor group, $s$ denote the fugacity of the global $U(1)_{s}$ symmetry acting $\tilde{S}_{1}$ and $S_{2}$, while $a$ denote the fugacity of the global $U(1)_{a}$ symmetry acting on $\tilde{A}_{1}$ and $A_{2}$.  The contribution of each field is given by
\begin{equation*}
\chi_{A^{2}_{11}} = \sum_{a=1}^{k_{1}}\sum_{b=1}^{k_{2}} z_{a}p_{b}^{-1}, \ \
\chi_{Q} = \sum_{i=1}^{N}\sum_{a=1}^{k_{1}}z_{a}^{-1}y_{i}, \ \ \chi_{q}=\sum_{
j=1}^{N}\sum_{b=1}^{k_{2}}p_{b}y_{j}^{-1}, \ \ \chi_{F}=\sum_{a=1}^{k_{1}}\sum_{b=1}^{k_{2}} z_{a}^{-1}p_{b}\, ,
\end{equation*}
\begin{equation*}
\chi_{S_{2}} = s\sum\nolimits_{1 \leq a \leq b \leq k_{2}} p_{a}p_{b}, \ \ \chi_{\tilde{S}_{1}} = \frac{1}{s}\sum\nolimits_{1 \leq a \leq b \leq k_{1}} z_{a}^{-1}z_{b}^{-1}, \ \  \chi_{A_{2}} = a\sum\nolimits_{1 \leq a < b \leq k_{2}} p_{a}p_{b}, \ \ \chi_{\tilde{A}_{1}} = \frac{1}{a}\sum\nolimits_{1 < a < b \leq k_{1}} z_{a}^{-1}z_{b}^{-1}\, .
\end{equation*}

In this case, by explicit computation of the Hilbert series for the hybrid configuration with gauge group $G=U(k_{1})  \times U(k_{2})$ and flavor group $U(N)$ we find it to be equal to the Hilbert series for the \textit{SA} hybrid configuration on $\mathbb{C}^{2}/\mathbb{Z}_{2}$ with gauge group $G=U(K_{1})$ (see \cite{Dey:2013fea} for more details). The two theories share the same flavor group and the gauge groups are related in the following way
\begin{equation}
K_{1} = \textrm{min}(k_{1},k_{2})\, .
\end{equation}
Let us explicitly show a few examples supporting our claim
\\
\\
\textbf{$\textbf{k}=(1,1)$, and $N=1$}\\
\\
Using eq.(\ref{hybridhs}) and unrefining we find that
\begin{equation*}
 H[\textbf{k}=(1,1),U(1), \mathbb{C}P^{2}/\mathbb{Z}_{2}](t,1,1)= \frac{1-t^{18}}{(1 - t^6)(1-t^9)^2}\, ,
\end{equation*}
which is the Hilbert series for the \textit{SA} hybrid configuration on $\mathbb{C}^{2}/\mathbb{Z}_{2}$ with $N=1$ and $K_{1}=1$.\\
\\
\textbf{$\textbf{k}=(1,1)$, and $N=2$}\\
\\
Using eq.(\ref{hybridhs}) and unrefining we find that
\begin{equation*}
 H[\textbf{k}=(1,1),U(2),\mathbb{C}P^{2}/\mathbb{Z}_{2}](t,1,1)= \frac{1 + 2 t^6 + 4 t^9 + 
 2 t^{12} + t^{18}}{(1 - t^3)^4 (1 + 2 t^3 + 2 t^6 + t^9)^2}\, ,
\end{equation*}
which is the Hilbert series for the \textit{SA} hybrid configuration on $\mathbb{C}^{2}/\mathbb{Z}_{2}$ with $N=2$ and $K_{1}=1$.\\
\\
\textbf{$\textbf{k}=(1,2)$, and $N=2$}\\
\\
Using eq.(\ref{hybridhs}) and unrefining we find that
\begin{equation*}
 H[\textbf{k}=(1,2),U(2),\mathbb{C}P^{2}/\mathbb{Z}_{2}](t,1,1)= \frac{1 + 2 t^6 + 4 t^9 + 
 2 t^{12} + t^{18}}{(1 - t^3)^4 (1 + 2 t^3 + 2 t^6 + t^9)^2}\, ,
\end{equation*}
which is again the Hilbert series for the \textit{SA} hybrid configuration on $\mathbb{C}^{2}/\mathbb{Z}_{2}$ with $N=2$ and $K_{1}=1$.\\
\\
\textbf{$\textbf{k}=(1,1)$, and $N=3$}\\
\\
Using eq.(\ref{hybridhs}) and unrefining we find that
\begin{equation*}
 H[\textbf{k}=(1,1),U(3), \mathbb{C}P^{2}/\mathbb{Z}_{2}](t,1,1)=
 \frac{1 + t^3 + 6 t^6 + 15 t^9 + 21 t^{12} + 18 t^{15} + 21 t^{18} + 15 t^{21} + 
 6 t^{24} + t^{27} + t^{30}}{(1 - t^3)^6 (1 + t^3)^4 (1 + t^3 + t^6)^3}\, ,
\end{equation*}
which is the Hilbert series for the \textit{SA} hybrid configuration on $\mathbb{C}^{2}/\mathbb{Z}_{2}$ with $N=3$ and $K_{1}=1$.\\
\\
\textbf{$\textbf{k}=(1,2)$, and $N=3$}\\
\\
Using eq.(\ref{hybridhs}) and unrefining we find that
\begin{equation*}
 H[\textbf{k}=(1,1),U(3), \mathbb{C}P^{2}/\mathbb{Z}_{2}](t,1,1)=
 \frac{1 + t^3 + 6 t^6 + 15 t^9 + 21 t^{12} + 18 t^{15} + 21 t^{18} + 15 t^{21} + 
 6 t^{24} + t^{27} + t^{30}}{(1 - t^3)^6 (1 + t^3)^4 (1 + t^3 + t^6)^3}\, ,
\end{equation*}
which is again the Hilbert series for the \textit{SA} hybrid configuration on $\mathbb{C}^{2}/\mathbb{Z}_{2}$ with $N=3$ and $K_{1}=1$.\\
\\
\textbf{$\textbf{k}=(1,1)$, and $N=4$}\\
\\
Using eq.(\ref{hybridhs}) and unrefining we find that
\begin{equation*}\begin{split}
 & H[\textbf{k}=(1,1),U(4), \mathbb{C}P^{2}/\mathbb{Z}_{2}](t,1,1) = \\
& = \frac{1 + 2 t^3 + 13 t^6 + 40 t^9 + 86 t^{12} + 132 t^{15} + 194 t^{18} + 
 220 t^{21} + 194 t^{24} + \textrm{palindrome} + t^{42}}{(1 - t^3)^8 (1 + t^3)^6 (1 + t^3 + t^6)^4}\, , 
\end{split}\end{equation*}
which is the Hilbert series for the \textit{SA} hybrid configuration on $\mathbb{C}^{2}/\mathbb{Z}_{2}$ with $N=4$ and $K_{1}=1$.\\
\\
\textbf{$\textbf{k}=(2,2)$, and $N=1$}\\
\\
Using eq.(\ref{hybridhs}) and unrefining we find that
\begin{equation*}
H[\textbf{k}=(2,2),U(1), \mathbb{C}P^{2}/\mathbb{Z}_{2}](t,1,1,1)= \frac{1 - t^3 + 
 2 t^9 - t^{15} + t^{18}}{(1 - t^3)^4 (1 + t^3)^2 (1 + t^3 + t^6 + t^9 +
    t^{12})}\, ,
\end{equation*}
which is the Hilbert series for the \textit{SA} hybrid configuration on $\mathbb{C}^{2}/\mathbb{Z}_{2}$ with $N=1$ and $K_{1}=2$.\\
\\

\section{Quivers and relations for $Sp(N)$ and $SO(N)$ instantons on $\mathbb{C}P^{2}/\mathbb{Z}_{n}$ with $n>4$}
In this appendix we collect the quiver diagrams for $Sp(N)$ and $SO(N)$ instantons on $\mathbb{C}P^{2}/\mathbb{Z}_{n}$ (with $n >4$) showing their relations with the corresponding quiver diagrams of the corresponding $\mathbb{C}^{2}/\mathbb{Z}_{n}$ theory.

\begin{figure}[h!]
\centering
\includegraphics[scale=1]{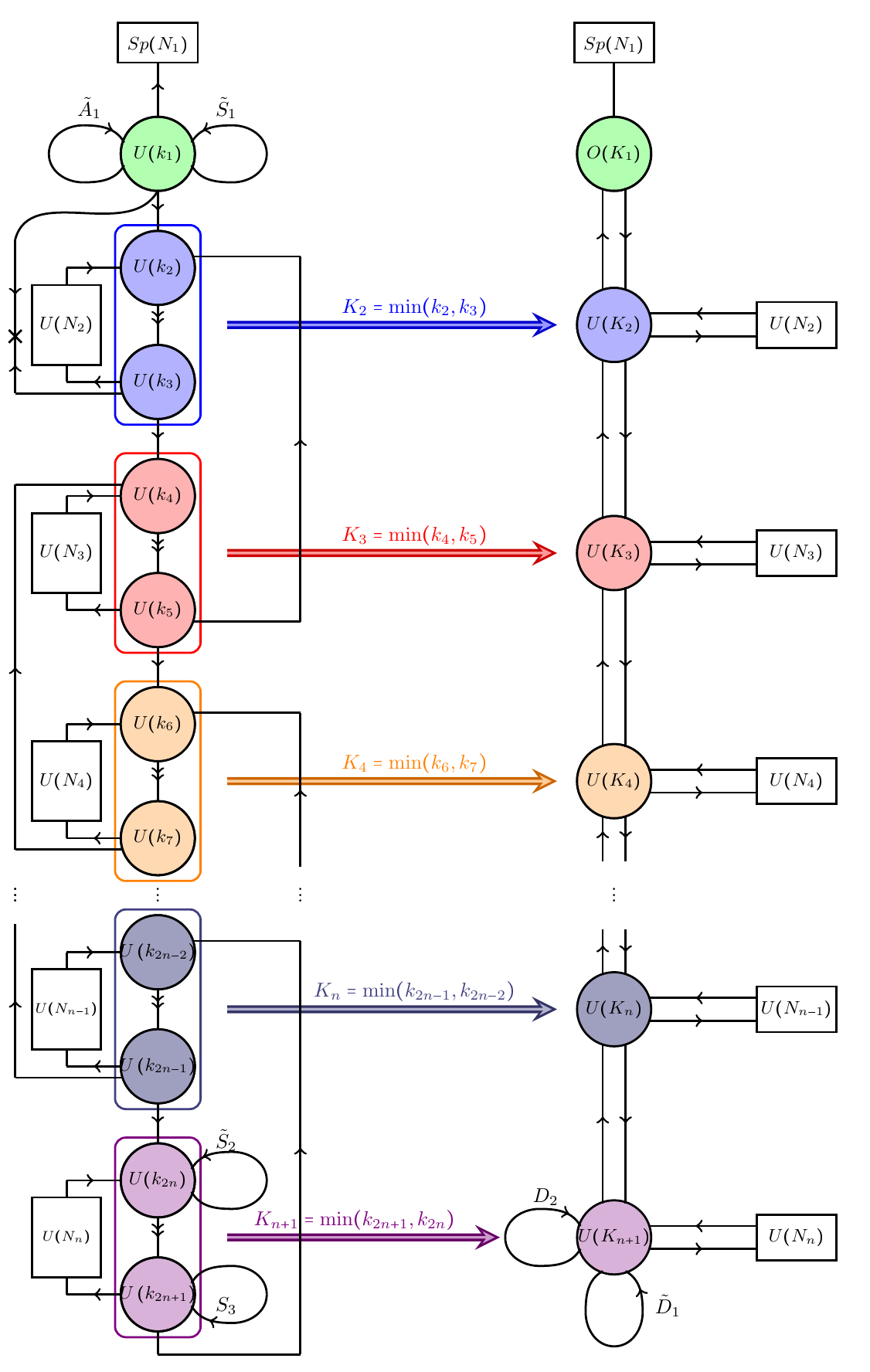}

\caption{Relation between the quiver diagram for $Sp(N)$ instantons on $\mathbb{C}P^{2}/\mathbb{Z}_{2n+1}$ (on the left) and the quiver diagram for $Sp(N)$ instantons on $\mathbb{C}^{2}/\mathbb{Z}_{2n+1}$ (on the right). Where $\tilde{D}_{1}$ and $D_{2}$ are two fields in the symmetric representation of the gauge group $U(K_{n+1})$. \label{fig:cp22n+1}}
     
\end{figure}

\begin{figure}[h!]
\centering
\includegraphics[scale=1]{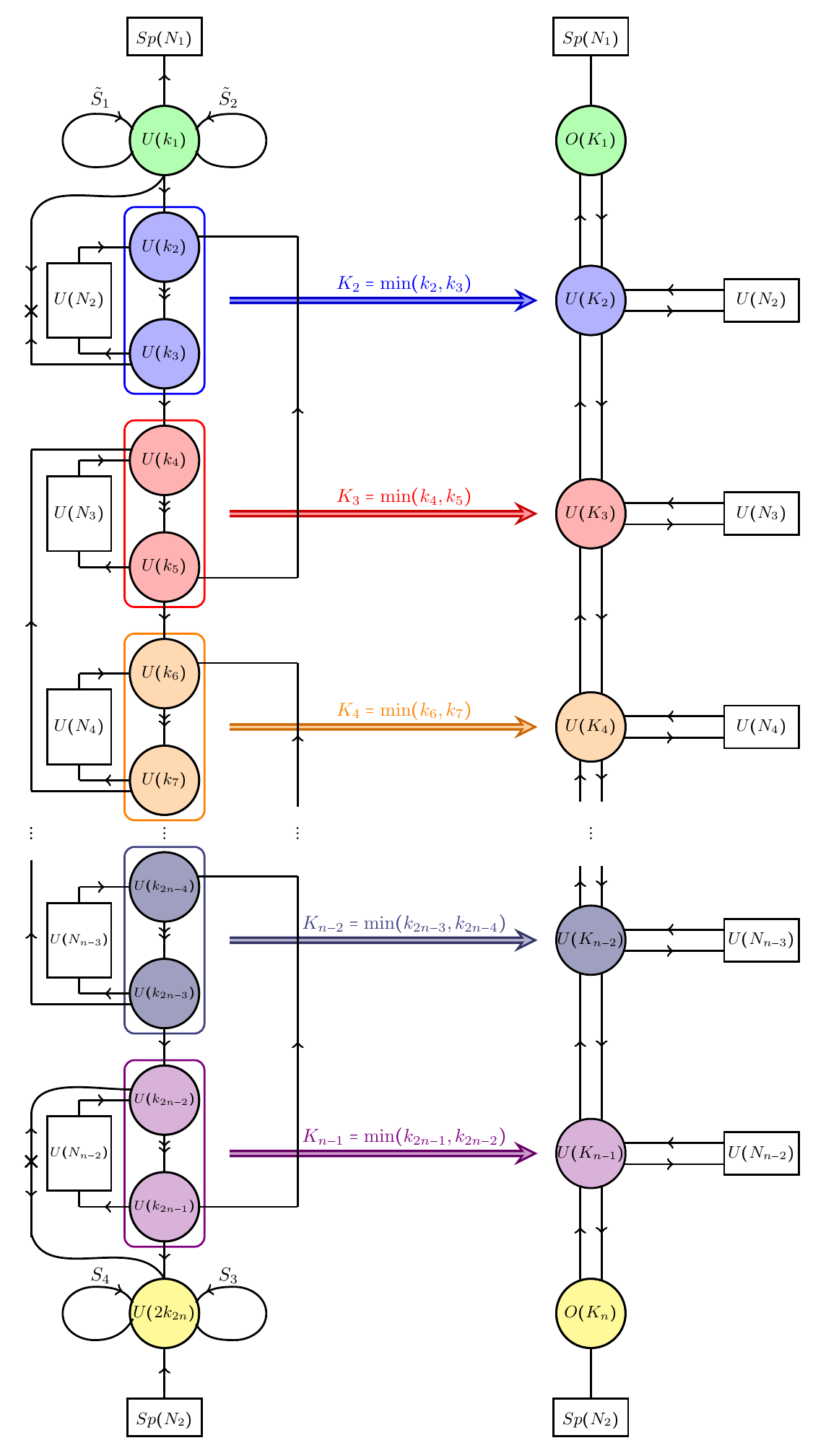}

\caption{Relation between the quiver diagram for VS $Sp(N)$ instantons on $\mathbb{C}P^{2}/\mathbb{Z}_{2n}$ (on the left) and the quiver diagram for VS $Sp(N)$ instantons on $\mathbb{C}^{2}/\mathbb{Z}_{2n}$ (on the right). \label{fig:cp22n+1first}}
\end{figure}

\begin{figure}[h!]
\centering
\includegraphics[scale=1]{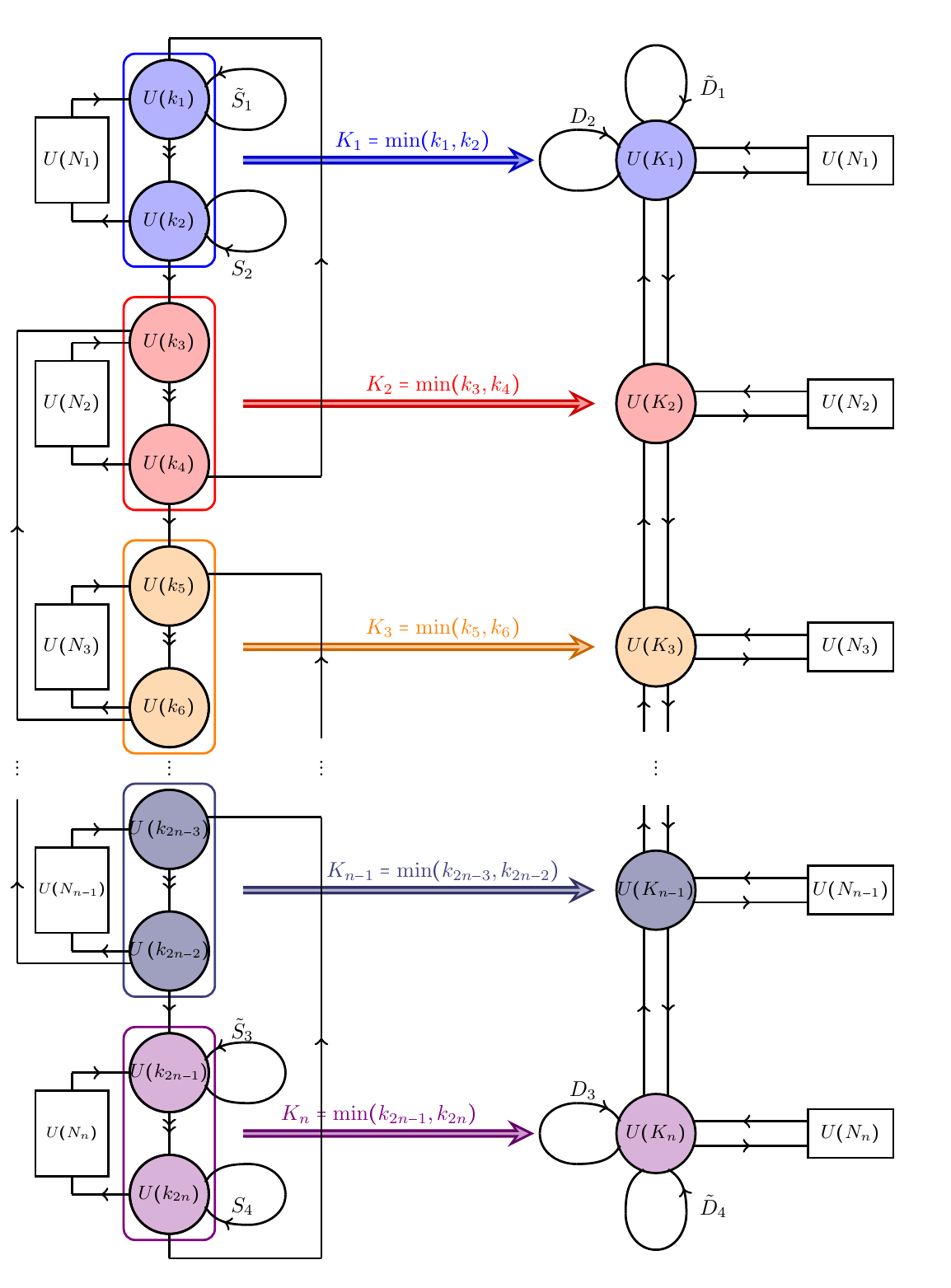}

\caption{Relation between the quiver diagram for NVS $Sp(N)$ instantons on $\mathbb{C}P^{2}/\mathbb{Z}_{2n}$ (on the left) and the quiver diagram for NVS $Sp(N)$ instantons on $\mathbb{C}^{2}/\mathbb{Z}_{2n}$ (on the right). Where $\tilde{D}_{1}$ and $D_{2}$ are two fields in the symmetric representation of the gauge group $U(K_{1})$, while $D_{3}$ and $\tilde{D}_{4}$  are two fields in the symmetric representation of the gauge group  $U(K_{n})$. \label{fig:cp22n+1second}}
     
     \end{figure}

\begin{figure}[h!]
\centering
\includegraphics[scale=1]{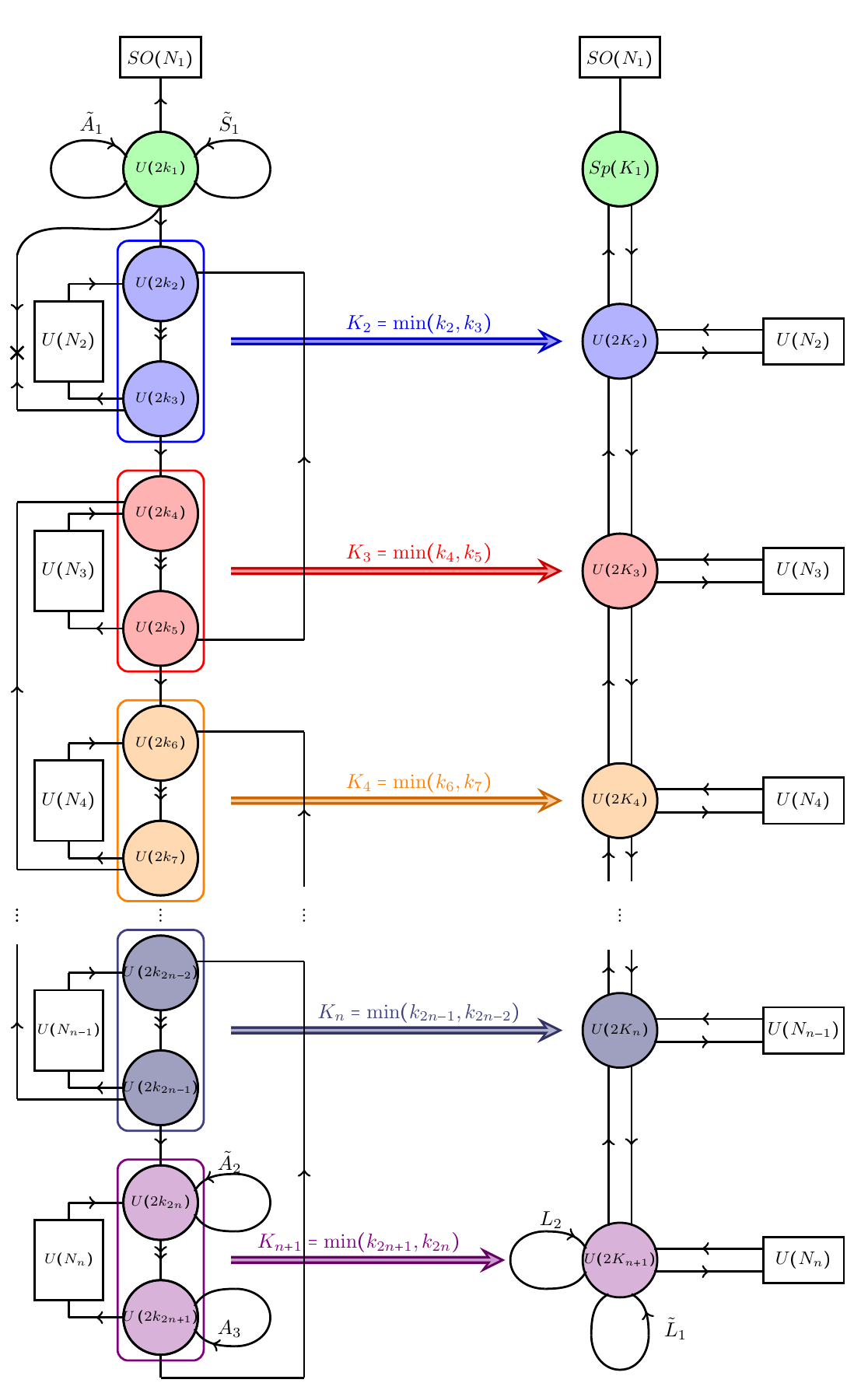}

\caption{Relation between the quiver diagram for $SO(N)$ instantons on $\mathbb{C}P^{2}/\mathbb{Z}_{2n+1}$ (on the left) and the quiver diagram for $SO(N)$ instantons on $\mathbb{C}^{2}/\mathbb{Z}_{2n+1}$ (on the right). Where $\tilde{L}_{1}$ and $L_{2}$ are two fields in the antisymmetric representation of the gauge group $U(2K_{n+1})$. \label{fig:cp22n+1o}}
     
     \end{figure}

\begin{figure}[h!]
\centering
\includegraphics[scale=0.99]{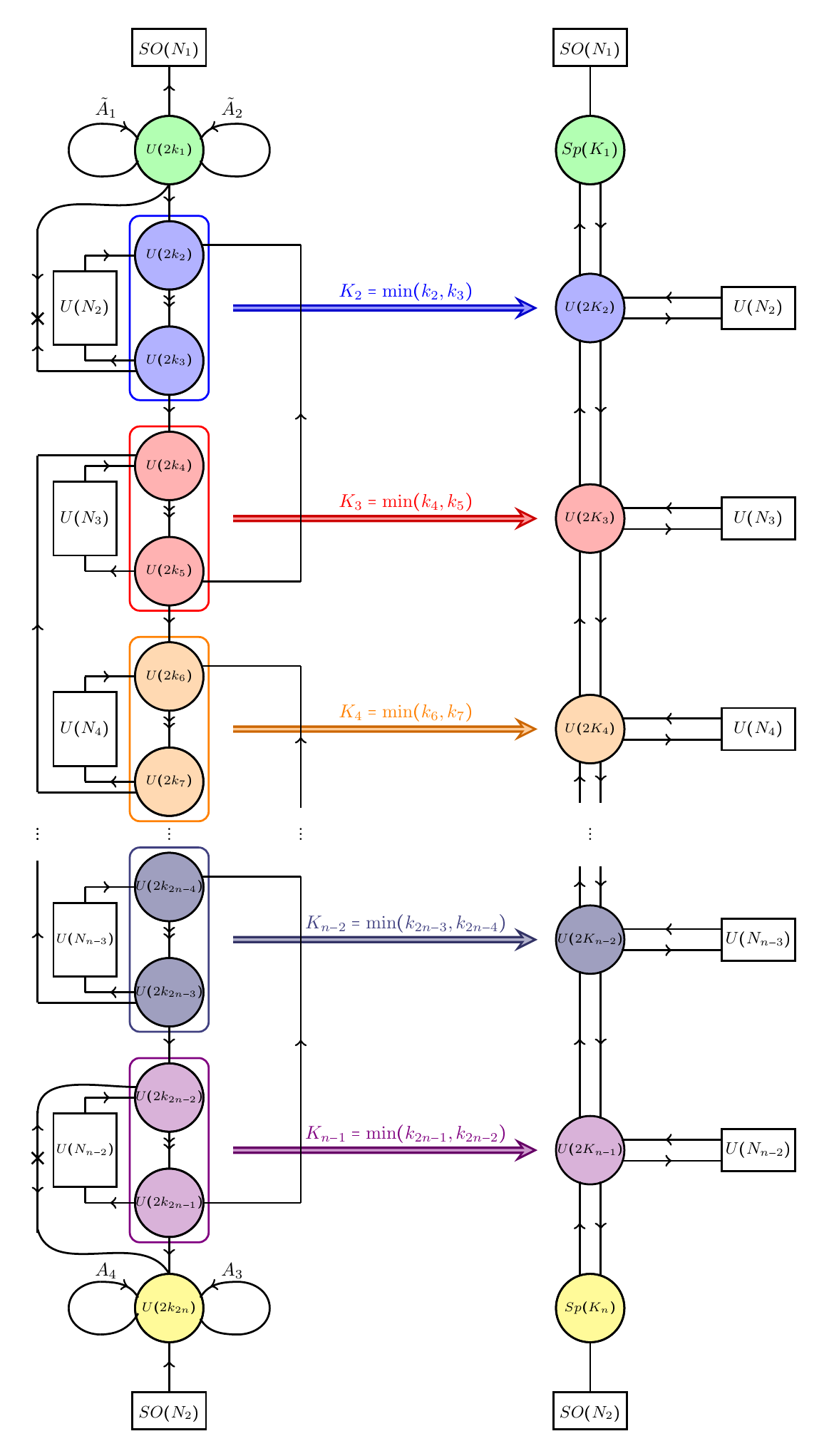}

\caption{Relation between the quiver diagram for VS $SO(N)$ instantons on $\mathbb{C}P^{2}/\mathbb{Z}_{2n}$ (on the left) and the quiver diagram for VS $SO(N)$ instantons on $\mathbb{C}^{2}/\mathbb{Z}_{2n}$ (on the right). \label{fig:cp22n+1firsto}}
     
     \end{figure}

\begin{figure}[h!]
\centering
\includegraphics[scale=1]{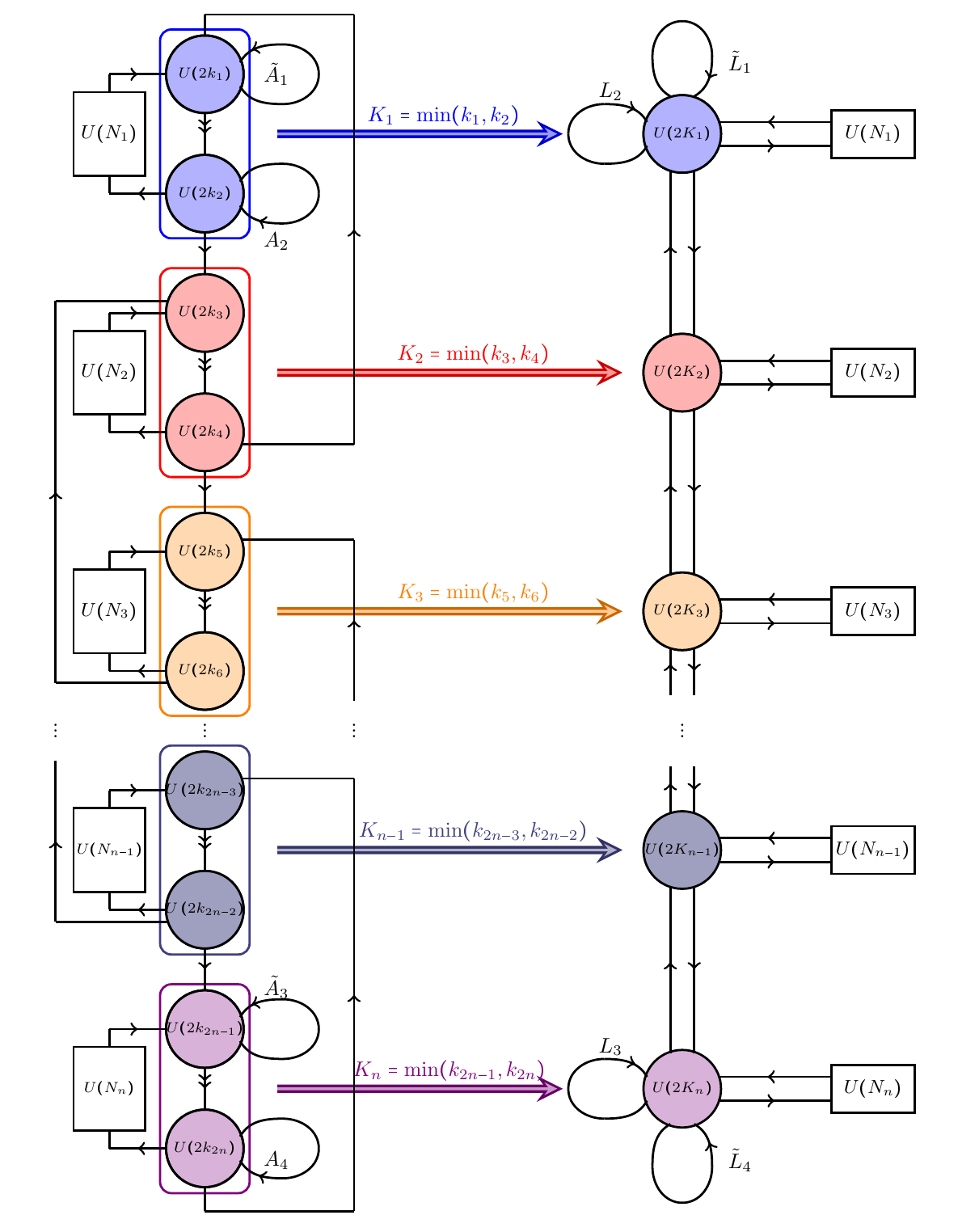}

\caption{Relation between the quiver diagram for NVS $SO(N)$ instantons on $\mathbb{C}P^{2}/\mathbb{Z}_{2n}$ (on the left) and the quiver diagram for NVS $SO(N)$ instantons on $\mathbb{C}^{2}/\mathbb{Z}_{2n}$ (on the right).  Where $\tilde{L}_{1}$ and $L_{2}$ are two fields in the antisymmetric representation of the gauge group $U(2K_{1})$, while $L_{3}$ and $\tilde{L}_{4}$  are two fields in the antisymmetric representation of the gauge group $U(2K_{n})$. \label{fig:cp22n+1secondo}}
     
     \end{figure}

\newpage 
\mbox{}
\newpage 
\mbox{}
\newpage
\mbox{}
\newpage
\mbox{}

\section*{Acknowledgements}

The authors would like to thank Amihay Hanany, Noppadol Mekareeya and Johannes Schmude for numerous useful discussions. The authors are partly supported by the spanish grant MINECO-13-FPA2012-35043-C02-02. In addition, they acknowledge financial support from the Ramon y Cajal grant RYC-2011-07593 as well as the EU CIG grant UE-14-GT5LD2013-618459. A.P. is supported by the Asturias government's Severo Ochoa FICYT grant BP14-003.


\begin{thebibliography}{99}


\bibitem{Pestun:2007rz} 
  V.~Pestun,
  ``Localization of gauge theory on a four-sphere and supersymmetric Wilson loops,''
  Commun.\ Math.\ Phys.\  {\bf 313}, 71 (2012)
  [arXiv:0712.2824 [hep-th]].


\bibitem{Keller:2011ek} 
  C.~A.~Keller, N.~Mekareeya, J.~Song and Y.~Tachikawa,
  ``The ABCDEFG of Instantons and W-algebras,''
  JHEP {\bf 1203}, 045 (2012)
  [arXiv:1111.5624 [hep-th]].
  
\bibitem{Rodriguez-Gomez:2013dpa} 
  D.~Rodriguez-Gomez and G.~Zafrir,
  ``On the 5d instanton index as a Hilbert series,''
  Nucl.\ Phys.\ B {\bf 878}, 1 (2014)
  [arXiv:1305.5684 [hep-th]].




\bibitem{adhm} 
  M. Atiyah,V. Drinfeld, N. Hitchin and Y. Manin,
  ``Constructions of instantons,'' \textit{Physics Letter A} \textbf{65} (\textbf{3}) 185-187.
  
  
  
  
\bibitem{Witten:1994tz}
  E.~Witten,
  ``Sigma models and the ADHM construction of instantons,''
  J.\ Geom.\ Phys.\  {\bf 15} (1995) 215
  [hep-th/9410052].
    
 
  
\bibitem{Witten:1995gx} 
  E.~Witten,
  ``Small instantons in string theory,''
  Nucl.\ Phys.\ B {\bf 460}, 541 (1996)
  [hep-th/9511030].
  
  
\bibitem{Douglas:1995bn} 
  M.~R.~Douglas,
  ``Branes within branes,''
  In *Cargese 1997, Strings, branes and dualities* 267-275
  [hep-th/9512077].
  
\bibitem{Douglas:1996sw} 
  M.~R.~Douglas and G.~W.~Moore,
  ``D-branes, quivers, and ALE instantons,''
  hep-th/9603167.
  
  
  



\bibitem{Buchdal1} 
  N. Buchdal,
  ``Instantons on CP2,''
\textit{ J.Diff.Geom.} \textbf{24} (1986) 19-52. 

\bibitem{Buchdal2} 
  N. Buchdal,
  ``Stable 2-bundles on Hirzebruch surfaces,''
\textit{Math. Z.} \textbf{194} (1987) 143-152. 


\bibitem{Buchdal3} 
  N. Buchdal,
  ``Hermitiean-Einstein Connections and Stable Vector Bundles  Over Compact Algebraic Surfaces,''
\textit{Math. Ann.} \textbf{280} (1988) 625-648. 



\bibitem{King} 
  A. King,
  ``Instantons and holomorphic bundles on the blown-up plane,''
  PhD-thesis, Thesis, Oxford University, 1989.


\bibitem{Sanders} 
  J. Bryan and M. Sanders,
  ``Instantons on $\left \{ S4 \right \}$ and cp2, rank stabilization, and bott periodicity,''


\bibitem{Mekareeya:2014kca} 
  N.~Mekareeya and D.~Rodriguez-Gomez,
  ``The ADHM-like constructions for instantons on $\mathbb{CP}^2$ and three-dimensional gauge theories,''
  Nucl.\ Phys.\ B {\bf 891}, 346 (2015)
  [arXiv:1404.3738 [hep-th]].

\bibitem{NY1}
H.~Nakajima and K.~Yoshioka, ``Perverse coherent sheaves on blow-up. I. A Quiver description," arXiv:0802.3120 [math.AG]

\bibitem{NY2}
H.~Nakajima and K.~Yoshioka, ``Perverse coherent sheaves on blow-up. II. K-theoretic parition function," arXiv:math/0505553 [math.AG].

\bibitem{NY3}
H. Nakajima and K. Yoshioka, ``Instanton counting on blowup. I. 4-dimensional pure gauge theory," Invent.Math. 162 (2005) 313-355, arXiv:math/0306198 [math.AG].



\bibitem{Hanany:1997vm} 
  A.~Hanany and K.~Hori,
  ``Branes and N=2 theories in two-dimensions,''
  Nucl.\ Phys.\ B {\bf 513}, 119 (1998)
  [hep-th/9707192].


\bibitem{Festuccia:2011ws} 
  G.~Festuccia and N.~Seiberg,
  ``Rigid Supersymmetric Theories in Curved Superspace,''
  JHEP {\bf 1106}, 114 (2011)
  [arXiv:1105.0689 [hep-th]].


\bibitem{Klare:2013dka} 
  C.~Klare and A.~Zaffaroni,
  ``Extended Supersymmetry on Curved Spaces,''
  JHEP {\bf 1310}, 218 (2013)
  [arXiv:1308.1102 [hep-th]].
  
\bibitem{Rodriguez-Gomez:2014eza} 
  D.~Rodriguez-Gomez and J.~Schmude,
  ``Partition functions for equivariantly twisted $\mathcal{N}=2$ gauge theories on toric K\"ahler manifolds,''
  arXiv:1412.4407 [hep-th].

\bibitem{Karlhede:1988ax} 
  A.~Karlhede and M.~Rocek,
  ``Topological Quantum Field Theory and $N=2$ Conformal Supergravity,''
  Phys.\ Lett.\ B {\bf 212}, 51 (1988).


\bibitem{Witten:1994ev} 
  E.~Witten,
  ``Supersymmetric Yang-Mills theory on a four manifold,''
  J.\ Math.\ Phys.\  {\bf 35}, 5101 (1994)
  [hep-th/9403195].
 
\bibitem{Davey:2009sr} 
  J.~Davey, A.~Hanany, N.~Mekareeya and G.~Torri,
  JHEP {\bf 0906}, 025 (2009)
  [arXiv:0903.3234 [hep-th]].

\bibitem{Benini:2009qs}
  F.~Benini, C.~Closset and S.~Cremonesi,
  ``Chiral flavors and M2-branes at toric CY4 singularities,''
  JHEP {\bf 1002} (2010) 036
  [arXiv:0911.4127 [hep-th]].


\bibitem{Benvenuti:2010pq} 
  S.~Benvenuti, A.~Hanany and N.~Mekareeya,
  ``The Hilbert Series of the One Instanton Moduli Space,''
  JHEP {\bf 1006}, 100 (2010)
  [arXiv:1005.3026 [hep-th]].
  
\bibitem{Hanany:2012dm} 
  A.~Hanany, N.~Mekareeya and S.~S.~Razamat,
  ``Hilbert Series for Moduli Spaces of Two Instantons,''
  JHEP {\bf 1301}, 070 (2013)
  [arXiv:1205.4741 [hep-th]].


\bibitem{Dey:2013fea} 
  A.~Dey, A.~Hanany, N.~Mekareeya, D.~Rodríguez-Gómez and R.~-K.~Seong,
  ``Hilbert Series for Moduli Spaces of Instantons on $\mathbb{C}$$^{2}$/$\mathbb{Z}$$_{n}$,''
  JHEP {\bf 1401}, 182 (2014)
  [arXiv:1309.0812 [hep-th]].

\bibitem{Nibbelink:2008qf}
  S.~Groot Nibbelink, F.~P.~Correia and M.~Trapletti,
  ``Non-Abelian bundles on heterotic non-compact K3 orbifold blowups,''
  JHEP {\bf 0811} (2008) 044
  [arXiv:0809.4430 [hep-th]].



\bibitem{Hanany:2014hia} 
  A.~Hanany and R.~K.~Seong,
  ``Hilbert Series and Moduli Spaces of k U(N) Vortices,''
  arXiv:1403.4950 [hep-th].
  
 
  
  
\bibitem{Forcella:2008bb} 
  D.~Forcella, A.~Hanany, Y.~H.~He and A.~Zaffaroni,
  ``The Master Space of N=1 Gauge Theories,''
  JHEP {\bf 0808}, 012 (2008)
  [arXiv:0801.1585 [hep-th]].


\bibitem{Kitao:1998mf} 
  T.~Kitao, K.~Ohta and N.~Ohta,
  ``Three-dimensional gauge dynamics from brane configurations with (p,q) - five-brane,''
  Nucl.\ Phys.\ B {\bf 539}, 79 (1999)
  [hep-th/9808111].

\bibitem{Bergman:1999na} 
  O.~Bergman, A.~Hanany, A.~Karch and B.~Kol,
  ``Branes and supersymmetry breaking in three-dimensional gauge theories,''
  JHEP {\bf 9910}, 036 (1999)
  [hep-th/9908075].
  

\bibitem{Bergman:2012qh} 
  O.~Bergman and D.~Rodriguez-Gomez,
  ``Probing the Higgs branch of 5d fixed point theories with dual giant gravitons in AdS(6),''
  JHEP {\bf 1212}, 047 (2012)
  [arXiv:1210.0589 [hep-th]].


\bibitem{RodriguezGomez:2009ae} 
  D.~Rodriguez-Gomez,
  ``M5 spikes and operators in the HVZ membrane theory,''
  JHEP {\bf 1003}, 039 (2010)
  [arXiv:0911.0008 [hep-th]].
  
\bibitem{Klebanov:2007us} 
  I.~R.~Klebanov and A.~Murugan,
  ``Gauge/Gravity Duality and Warped Resolved Conifold,''
  JHEP {\bf 0703}, 042 (2007)
  [hep-th/0701064].
  

\bibitem{Aharony:2008gk} 
  O.~Aharony, O.~Bergman and D.~L.~Jafferis,
  ``Fractional M2-branes,''
  JHEP {\bf 0811}, 043 (2008)
  [arXiv:0807.4924 [hep-th]].


\bibitem{Butti:2007jv}
  A.~Butti, D.~Forcella, A.~Hanany, D.~Vegh and A.~Zaffaroni,
  ``Counting Chiral Operators in Quiver Gauge Theories,''
  JHEP {\bf 0711} (2007) 092
  [arXiv:0705.2771 [hep-th]].


\bibitem{Hanany:2008fj}
  A.~Hanany, D.~Vegh and A.~Zaffaroni,
  ``Brane Tilings and M2 Branes,''
  JHEP {\bf 0903} (2009) 012
  [arXiv:0809.1440 [hep-th]].



\bibitem{Franco:2007ii} 
  S.~Franco, A.~Hanany, D.~Krefl, J.~Park, A.~M.~Uranga and D.~Vegh,
  ``Dimers and orientifolds,''
  JHEP {\bf 0709}, 075 (2007)
  [arXiv:0707.0298 [hep-th]].




\bibitem{GarciaEtxebarria:2012qx}
  I.~Garcia-Etxebarria, B.~Heidenreich and T.~Wrase,
  ``New N=1 dualities from orientifold transitions. Part I. Field Theory,''
  JHEP {\bf 1310} (2013) 007
  [arXiv:1210.7799 [hep-th]].



\bibitem{Tachikawa:2014qaa} 
  Y.~Tachikawa,
  ``Moduli spaces of SO(8) instantons on smooth ALE spaces as Higgs branches of 4d N = 2 supersymmetric theories,''
  JHEP {\bf 1406}, 056 (2014)
  [arXiv:1402.4200 [hep-th]].
  




\end{thebibliography}
\end{document}